\documentclass[a4paper,12pt]{article}

\usepackage{amsmath}
\usepackage{amsfonts}
\usepackage[position=b]{subcaption}
\usepackage{tikz}
\usetikzlibrary{decorations.pathmorphing,decorations.markings,trees,arrows,patterns}
\usepackage{comment}
\usepackage{color}
\usepackage{multirow}
\usepackage{enumitem}
\usepackage[utf8]{inputenc}
\RequirePackage{graphicx}
\RequirePackage[numbers,sort&compress]{natbib}
\RequirePackage[colorlinks,citecolor=blue,urlcolor=blue,linkcolor=blue]{hyperref}
\newcommand{\fig}[1]{Fig.~\ref{#1}}

\newcommand{\eq}[1]{Eq.~(\ref{#1})}

\newcommand{\Tp}{\mathrm{T}}

\numberwithin{equation}{section}
\usepackage{enumitem}
\usepackage{authblk} 
\usepackage{textpos} 

\usepackage{times}

\title{Phases of QCD$_3$ with three families of fundamental flavors}

\author[1,2,3]{Abdullah Khalil\thanks{khabdu@liverpool.ac.uk}}
\author[1]{Radu Tatar\thanks{radu.tatar@liverpool.ac.uk}}
\affil[1]{Department of Mathematical Sciences, University of Liverpool, Liverpool L69 3BX, United Kingdom}
\affil[2]{Department of Physics, National Tsing Hua University, Hsinchu City 30013, Taiwan}
\affil[3]{Department of Physics, Cairo University, Giza 12613, Egypt}

\date{}

\begin{document} 



\maketitle 
  \begin{textblock*}{3cm}(13.6cm,-8.2cm)
  LTH 1226
\end{textblock*}

\begin{abstract}
We explore the phase diagram for an $SU(N)$ gauge theory in $2 + 1$ dimensions with three families of fermions with different masses, all in the fundamental representation. The phase diagram is three dimensional and contains cuboid, planar and linear quantum regions, depending on the values of the fermionic masses. Among other checks, we consider the consistency with boson/fermion dualities and verify the reduction of the phase diagram to the one and two-family diagrams.
\end{abstract}

\newpage

\tableofcontents

\section{Introduction}\label{sec:1}
In recent years, a sustained effort has been dedicated to studying the phases of 2+1 dimensional gauge theories, inspired by applications in condensed matter systems and supersymmetric dualities. One important outcome was the proposal of infrared dualities between non-supersymmetric Chern-Simons gauge theories with fundamental matter \cite{Aharony:2015mjs,Hsin:2016blu} where certain fermion-boson dualities were conjectured. 
This avenue included theories with various gauge groups and matter representations \cite{Hsin:2016blu,Aharony:2016jvv,Karch:2016aux,Seiberg:2016gmd,Karch:2016sxi,Aitken:2017nfd,Benini:2017dus,Benini:2017aed,Jensen:2017bjo,Aitken:2018cvh}. A specific fermion-boson duality that is of interest for our work is 
\begin{equation}
SU(N)_k+ F\text{ Fermions} \longleftrightarrow U(k+F/2)_{-N}+ F\text{ Scalars}\ ,\label{aduality}
\end{equation}
and its time-reversal version. 

However, these dualities do not describe the full phase diagram of the gauge theory in $2+1$ dimensions for any number of flavors, colors, and level. Komargodski and Seiberg considered such a general case and presented the complete phase diagram of three dimensional $SU(N)_k$ gauge theory with Chern-Simons level $k$ coupled to $F$ fundamental Dirac fermions with mass $m$. The diagram was drawn as a function of $m$ \cite{Komargodski:2017keh}. The main new feature of the phase diagram is that the infrared (IR) description has a quantum phase that is hidden semiclassically for a large number of flavors. The discussion was extended in \cite{Gomis:2017ixy,Bashmakov:2018wts,Choi:2018ohn,Choi:2019eyl}.

The next step is to consider fermions in fundamental representation with different masses. One step in this direction was to introduce two sets of fundamental fermions with different masses $m_1 \ne m_2$, where the phase diagram acquires a two-dimensional structure and includes planes of quantum phases in the IR \cite{Argurio:2019tvw, Baumgartner:2019frr}. Our goal in this work is to consider a theory with three sets of fundamental fermions with three different masses and consider the corresponding three-dimensional diagram. We will see that there are three types of quantum regions of dimensions 1, 2, or 3 and will perform various consistency checks on the perturbations of infrared descriptions.

In the rest of this section, we review the description of the one and two-family cases to explain how the flavor symmetry breaking mechanism helps to understand the phase diagram for small masses when the semiclassical description fails. In section \ref{sec:2}, we give the details of the IR description for the three-family case for various ranges of $k$ and draw the full phase diagram. In section \ref{sec:3}, we perform various checks which, on the one hand, verify our proposal and, on the other hand, confirm the consistency of the results for one family and two families of fermions \cite{Komargodski:2017keh, Argurio:2019tvw, Baumgartner:2019frr}. Section \ref{sec:4} contains the conclusions.
\subsection{Review of the one-family case}\label{subsec:1.1}
The work \cite{Komargodski:2017keh} considered the phase diagram of $SU(N)$ gauge theory in $2+1$ dimensions with Chern-Simons level $k$ and $F$ fermions in the fundamental representation. The theory has a global flavor symmetry $U(F)$, which is spontaneously broken for small values of $k$. The spontaneous breaking of the flavor symmetry allows them to split the phase diagram into two cases as follows:$\footnote{We use the notations and conventions of \cite{Komargodski:2017keh, Argurio:2019tvw, Baumgartner:2019frr}}$.
\begin{itemize}
\item[1.] \underline{$k\geq F/2$:} In this case, the theory in the IR is semiclassically accessible, and the phase diagram is described by the asymptotic theories obtained after integrating the fermions out when their mass $m$ is positive or negative. The two phases are then $SU(N)_{k+F/2}$ for positive $m$, which has a level-rank dual $U(k+F/2)_{-N}$ and $SU(N)_{k-F/2}$ for negative $m$ with a level-rank dual $U(k-F/2)_{-N}$. The two phases are separated by a phase transition described by the critical theory $SU(N)_k+F\, \psi^0$, where the superscript $0$ refers to the fermions being massless. This phase transition could be first or second order, in some cases, it could be a series of phase transitions \cite{Bashmakov:2018wts}. The asymptotic phases are gapped and are described by pure topological quantum field theories (TQFT). The phase diagram is as in figure \ref{fig1}, where the transition between the two asymptotic phases occurs at the blue point. 
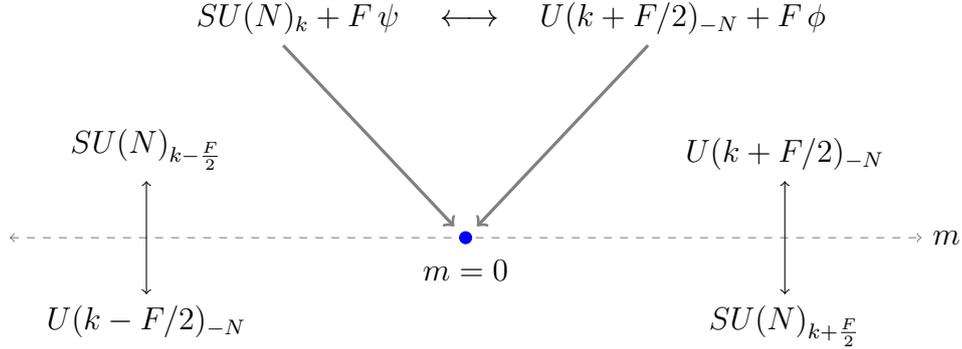
\begin{figure}[h!]
       \centering
 \begin{tikzpicture}[scale=1.5]
\draw[<->,gray, dashed] (-4,0) -- (4,0) node[right,black]{$m$};
\draw  (.4,1.7) node[align=left,   above] {$SU(N)_k+F\, \psi\quad \longleftrightarrow\quad U(k+F/2)_{-N}+F\, \phi$};
\draw[very thick,->,gray] (-1.6,1.7)--(-.1,0.1);
\draw[very thick,->,gray] (1.6,1.7)--(.1,0.1);
\filldraw[blue] (0,0)circle (1.5pt);
\draw[<->] (2.8,0.5) node[align=right,   above] {$U(k+F/2)_{-N}$}--(2.8,-0.5)node[align=right,below]{$SU(N)_{k+\frac{F}{2}}$};
\draw[<->] (-2.8,0.5) node[align=right,   above] {$SU(N)_{k-\frac{F}{2}}$}--(-2.8,-0.5)node[align=right,below]{$U(k-F/2)_{-N}$};
\draw (0,-.1)node[below]{$m=0$};
\end{tikzpicture}
\caption{ Phase diagram of $SU(N)_k+F\, \psi$ with $k\geq F/2$.}        
        \label{fig1}
    \end{figure}

      \item[2.] \underline{$k<F/2$:} In this case, the theory is semiclassically accessible only for large mass $(m\rightarrow \pm \infty)$. The asymptotic theories are $SU(N)_{k+F/2}\longleftrightarrow U(k+F/2)_{-N} $ for large positive $m$ and $SU(N)_{k-F/2}\longleftrightarrow U(F/2-k)_{N} $ for large negative $m$. However, integrating out the scalars from the dual bosonic theory $U(k+F/2)_{-N}+F\, \phi$ for large negative mass squared leads to a sigma-model phase. In \cite{Komargodski:2017keh}, the authors suggested that for the fermionic theory there is some value of the number of flavors $F^\ast$ at which the $U(F)$ symmetry is spontaneously broken into $U(F/2+k)\times U(F/2-k)$, leading to a sigma-model $\sigma$ in the IR that matches the bosonic phase and is given by the Grassmannian
      \begin{equation}
      Gr(F/2+k,F) = \frac{U(F)}{U(F/2+k)\times U(F/2-k)}\ .
      \label{sigma}
\end{equation}   
The sigma-model is a purely quantum gapless phase that does not appear semiclassically. The phase diagram now consists of the two asymptotic topological phases and a new quantum region for small $m$, which separates the two topological phases. The quantum region lies between two transition points that are described by some positive mass $m^+$ on the right and some negative mass $m^-$ on the left, as shown in figure \ref{fig2}. The authors also conjectured the existence of a new duality in the form $SU(N)_k+F\, \psi \longleftrightarrow U(F/2-k)_N+F\,\phi$ to cover the phase diagram for negative $m$.

 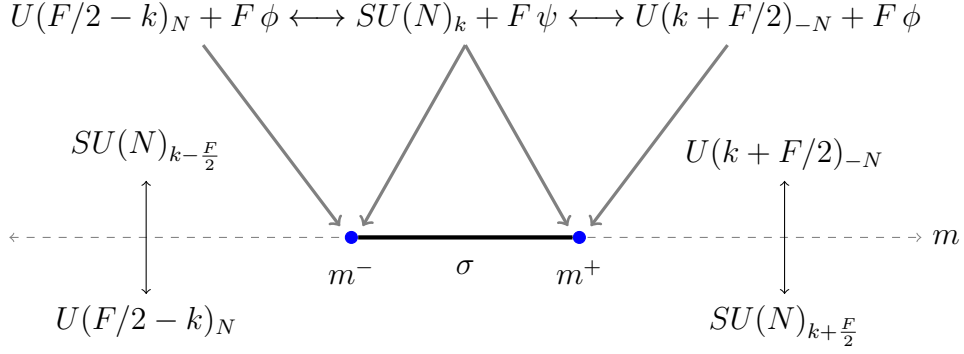
\begin{figure}[h]
     \centering
      \begin{tikzpicture}[scale=1.5]
\draw[<->,gray, dashed] (-4,0) -- (4,0) node[right,black]{$m$};
\draw  (0,1.7) node[align=left,   above]{ $ U(F/2-k)_N+F\,\phi\longleftrightarrow SU(N)_k+F\, \psi \longleftrightarrow U(k+F/2)_{-N}+F\, \phi$};
\draw[very thick,gray,->] (-2.3,1.7)--(-1.1,0.1);
\draw[very thick,gray,->] (0,1.7)--(-.9,0.1);
\draw[very thick,gray,->] (2.3,1.7)--(1.1,0.1);
\draw[very thick,gray,->] (0,1.7)--(.9,0.1);
\filldraw[blue] (-1,0)circle (1.5pt);
\filldraw[blue] (1,0)circle (1.5pt);
\draw[<->] (2.8,0.5) node[align=right,   above] {$U(k+F/2)_{-N}$}--(2.8,-0.5)node[align=right,below]{$SU(N)_{k+\frac{F}{2}}$};
\draw[<->] (-2.8,0.5) node[align=right,   above] {$SU(N)_{k-\frac{F}{2}}$}--(-2.8,-0.5)node[align=right,below]{$U(F/2-k)_{N}$};
\draw (1,-.1)node[below]{$m^+$};
\draw (-1,-.1)node[below]{$m^-$};
\draw (0,-.1)node[below,align=left]{$\sigma $};
\draw[ultra thick](-.94,0)--(.94,0);
\end{tikzpicture}
       \caption{Phase diagram of $SU(N)_k+F\, \psi$ with $ k<F/2$.}
        \label{fig2}
    \end{figure}  
\end{itemize} 
\subsection{Review of the two-family phase diagrams}\label{subsec:1.2}
We now move to \cite{Argurio:2019tvw,Baumgartner:2019frr} where the authors considered the case when the $F$ fermions are split into two sets of fermions: $p\, \psi_1$ fermions with mass $m_1$ and $(F-p)\, \psi_2$ fermions with mass $m_2$. The flavor symmetry is now explicitly broken into $U(p)\times U(F-p)$ and the phase diagram looks different for three particular cases: $k\geq F/2$, $F/2-p\leq k<F/2$, and $0\leq k<F/2-p$, where the range of $p$ is such that $0\leq p\leq F/2$.  In analogy to the one-family case, there are two dual bosonic theories $U(k+F/2)_{-N}+p\, \phi_1+(F-p)\, \phi_2$ and $U(F/2-k)_{N}+p\, \phi_1+(F-p)\, \phi_2$ for small values of $k$ to cover the full phase diagram.
\begin{itemize}
\item[1.] $\underline{k\geq F/2:}$ There is no flavor symmetry breaking, and the four topological theories describe the phase diagram. Integrating out both $\psi_1$ and $\psi_2$ when their masses are both positive and negative, one obtains various topological phases $\Tp_a (m_1>0, m_2>0)$, $\Tp_b (m_1<0, m_2 >0), 
\Tp_c(m_1<0, m_2 <0), \Tp_d (m_1 > 0, m_2 <0)$:
\begin{equation}
  \begin{aligned}
&\Tp_a:SU(N)_{k+\frac{F}{2}}\quad\longleftrightarrow \ U(k+F/2)_{-N}\ ,\\
&\Tp_b: SU(N)_{k+\frac{F}{2}-p}\longleftrightarrow U(k+F/2-p)_{-N}\ ,\\
&\Tp_c: SU(N)_{k-\frac{F}{2}}\quad\longleftrightarrow U(k-F/2)_{-N}\ ,\\
&\Tp_d: SU(N)_{k-\frac{F}{2}+p}\longleftrightarrow U(k-F/2+p)_{-N}\ .\ 
\label{t2kgf}
  \end{aligned}
\end{equation}

These topological phases are separated by critical theories represented as red lines in figure \ref{fig:3a}. The critical theories are $SU(N)_{k\pm \frac{p}{2}}+(F-p)\, \psi_2$ on the horizontal red line and $SU(N)_{k\pm\frac{F-p}{2}} +p\, \psi_1$ on the vertical red line. The blue point is a transition point that separates the four different topological phases. We will use the term type $\mathrm{I}$ phase diagram to label this case.

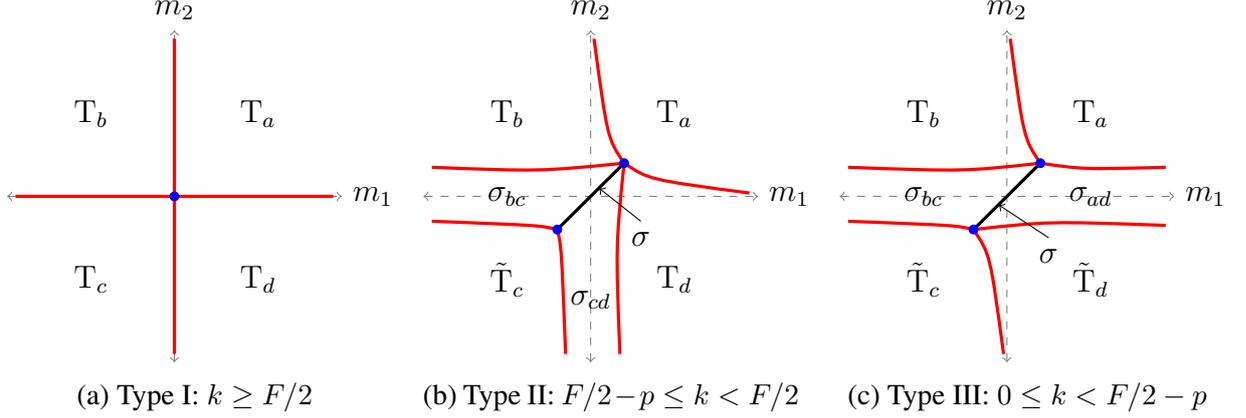
\begin{figure*}[h]
    \centering
    \begin{subfigure}[b]{0.3\textwidth}
\begin{tikzpicture}[scale=.55]
\draw[<->,gray,dashed](-4,0)--(4,0);
\draw (4,0) node[right]{$m_1$};
\draw[<->,gray,dashed](0,-4)--(0,4);
\draw (0,4)node[above]{$m_2$};
\draw (2,2) node{$\Tp_a$};
\draw (-2,2) node{$\Tp_b$};
\draw (2,-2) node{$\Tp_d$};
\draw (-2,-2) node{$\Tp_c$};
\draw[ very thick, red] (-3.8,0)--(3.8,0);
\draw[very thick, red] (0,-3.8)--(0,3.8);
\filldraw[blue] (0,0) circle (3pt);
\end{tikzpicture}
\caption{Type I: $k\geq F/2$}        
        \label{fig:3a}
    \end{subfigure}
    \quad
    \begin{subfigure}[b]{0.3\textwidth}
      \begin{tikzpicture}[scale=.55]
\draw[<->,gray,dashed](-4,0)--(4,0);
\draw (4,0) node[right]{$m_1$};
\draw[<->,gray,dashed](0,-4)--(0,4);
\draw (0,4)node[above]{$m_2$};
\draw (2,2) node{$\Tp_a$};
\draw (-2,2) node{$\Tp_b$};
\draw (2,-2) node{$\Tp_d$};
\draw (-2,-2) node{$\tilde{\Tp}_c$};
\draw[very thick] (-.8,-.8)--(.8,.8);
\draw[very thick, red] (-3.8,-.6)..controls(-1.2,-.7)..(-.8,-.8)..controls (-0.7,-1.2)..(-.6,-3.8);
\draw[very thick, red] (-3.8,.7)..controls(-1.2,.6)..(.8,.8)..controls (0.6,-1.2)..(.7,-3.8); 
\draw[very thick, red] (3.8,.08)..controls(1.4,.4)..(.8,.8)..controls(0.4,1.4)..(0.08,3.8); 
\filldraw[blue] (0.8,0.8) circle (3pt);
\filldraw[blue] (-.8,-.8) circle (3pt);
\draw (-2,0) node{$\sigma_{bc}$};
\draw (0,-2) node[below]{$\sigma_{cd}$};
\draw[->] (1.2,-.6)node[below]{$\sigma$}--((.21,.19);
\end{tikzpicture}
        \caption{Type II: $F/2-p\leq k<F/2$}
        \label{fig:3b}
    \end{subfigure}
  \quad
    \begin{subfigure}[b]{0.3\textwidth}
      \begin{tikzpicture}[scale=.55]
\draw[<->,gray,dashed](-4,0)--(4,0);
\draw (4,0) node[right]{$m_1$};
\draw[<->,gray,dashed](0,-4)--(0,4);
\draw (0,4)node[above]{$m_2$};
\draw (2,2) node{$\Tp_a$};
\draw (-2,2) node{$\Tp_b$};
\draw (2,-2) node{$\tilde{\Tp}_d$};
\draw (-2,-2) node{$\tilde{\Tp}_c$};
\draw[very thick] (-.8,-.8)--(.8,.8);
\draw[very thick, red] (-3.8,-.6)..controls(-1.2,-.7)..(-.8,-.8)..controls (-0.4,-1.4)..(-.08,-3.8);
\draw[very thick, red] (-3.8,.7)..controls(-1.,.6)..(.8,.8)..controls (1.8,.65)..(3.8,.7); 
\draw[very thick, red] (.08,3.8)..controls(.4,1.4)..(.8,.8); 
\draw[very thick, red] (-.8,-.8)..controls(1,-.6)..(3.8,-.7); 
\filldraw[blue] (0.8,0.8) circle (3pt);
\filldraw[blue] (-.8,-.8) circle (3pt);
\draw (-2,0) node{$\sigma_{bc}$};
\draw (2,0) node{$\sigma_{ad}$};
\draw[->] (1.,-1)node[below]{$\sigma$}--(-.19,-.21);
\end{tikzpicture}
        \caption{Type III: $0\leq k<F/2-p$}
        \label{fig:3c}
    \end{subfigure}
    \caption{Phases of $SU(N)_k+p\, \psi_1+(F-p)\, \psi_2$.}\label{fig3}
\end{figure*}
\item[2.] \underline{$F/2-p\leq k<F/2:$} The asymptotic phases can be found similarly by sending both masses to $\pm \infty$. The topological phases $\Tp_a$, $\Tp_b$, and $\Tp_d$ remain as in \eq{t2kgf} while $\Tp_c$ becomes
\begin{equation}
\Tp_c\rightarrow \tilde{\Tp}_c: SU(N)_{k-\frac{F}{2}}\longleftrightarrow U(F/2-k)_{N}\ .
\label{t2klf1}
\end{equation}
There are quantum regions in this case where the flavor symmetry is spontaneously broken. The strategy is to check the asymptotic phases where we send only one of the masses to $\pm \infty$, and the theory becomes one-family with a shifted level. The theory is strongly coupled and has sigma-model descriptions $\sigma_{bc}$ and $\sigma_{cd}$ for small and/or negative $m_1$ and $m_2$, respectively. The sigma-models have target spaces with the following Grassmannians
\begin{align}
&\sigma_{bc}: Gr(F/2-k,F-p)=\frac{U(F-p)}{U(F/2-k)\times U(k+F/2-p)}\ ,
\label{sigma23}\\
&\sigma_{cd}: Gr(F/2-k,p)\hspace{.75cm}=\frac{U(p)}{U(F/2-k)\times U(k-F/2+p)}\ .
\label{sigma34}
\end{align} 
The theory also has a sigma-model $\sigma$ on the diagonal line $m_1=m_2$, where it is reduced to the one-family case. In this case, $\sigma$ acts as a phase transition between $\sigma_{bc}$ and $\sigma_{cd}$. The red line that separates $\Tp_a$ and $\Tp_d$ corresponds to the critical theory $SU(N)_{k+\frac{p}{2}}+(F-p)\, \psi_2$ while the red line separating $\Tp_a$ and $\Tp_b$ corresponds to the critical theory $SU(N)_{k+\frac{F-p}{2}}+p\, \psi_1$. There also exist critical theories separating the topological theories and the quantum phases, which are described via the dual bosonic theories. For example, the red line that separates $\Tp_b$ and $\sigma_{bc}$ is given by the critical theory $U(F/2-k)_N+(F-p)\, \phi_2$, and this applies similarly for the remaining red lines. The full phase diagram for this case is shown in figure \ref{fig:3b}, and we label it as a type $\mathrm{II}$ phase diagram.

\item[3.] \underline{$0\leq k<F/2-p$:} In this range, the phase diagram is similar to the previous case in the sense that there exist topological phases that can be found asymptotically plus expected quantum regions due to the symmetry breaking scenario. The topological phases $\Tp_a$, $\Tp_b$, and $\tilde{\Tp}_c$ remain the same while $\Tp_d$ becomes
\begin{align}
\Tp_d\rightarrow \tilde{\Tp}_d: SU(N)_{k-\frac{F}{2}+p}\longleftrightarrow U(F/2-p-k)_{N}\ .
\label{t2klf2}
\end{align}
In this case, the quantum phases are $\sigma$ models on the diagonal line with a Grassmannian given by \eq{sigma}. They are  $\sigma_{bc}$ which is a plane describing the quantum region for small and negative $m_2$ with a Grassmannian given by \eq{sigma23} and $\sigma_{ad}$ for small and positive $m_2$ with a Grassmannian
\begin{align}
\sigma_{ad}: Gr(F/2+k,F-p)=\frac{U(F-p)}{U(F/2+k)\times U(F/2-p-k))}\ .
\label{sigma14}
\end{align} 

The phase diagram is shown in figure \ref{fig:3c} wihch is a type $\mathrm{III}$ phase diagram. As in type $\mathrm{II}$, the $\sigma$ line acts as a transition between $\sigma_{bc}$ and $\sigma_{ad}$.
\end{itemize}
To summarize, the two-family theory has the following features:
\begin{itemize}
\item[$\bullet$] For $k\geq F/2$, the theory in the IR is described by the four topological field theories in \eq{t2kgf}.
\item[$\bullet$] For small $k$, the IR description is given by four topological field theories and a line of sigma-model $\sigma$ separating two planes of sigma-models $(\sigma_{bc},\, \sigma_{cd})$ or $(\sigma_{bc},\, \sigma_{ad})$. 
\end{itemize}
The theory passes various consistency checks such as matching the phases of the dual bosonic theory near the transition points as well as perturbing the $\sigma$ model to make sure the description is valid when both masses are small \cite{Argurio:2019tvw}.

\section{The theory with three families}\label{sec:2}
We now turn to the main work of this paper where we split the $F$ fermions into three sets of fermions $p$ fermions $\psi_1$ with mass $m_1$, $q$ fermions $\psi_2$ with mass $m_2$ and $F-p-q$ fermions $\psi_3$ with mass $m_3$. The global flavor symmetry $U(F)$ is explicitly broken to $U(p)\times U(q)\times U(F-p-q)$. There are two possible scenarios to cover the full phase diagram of the three-family case. The two scenarios depend on the value of the number of flavors for the extra family; The first is for $F-p-q \leq F/2$ and the second is when $F-p-q>F/2$, each scenario is split into five different cases depending on the value of $k$. However, most of these cases are the same in the two different situations, while the difference appears when we check the phases of $k$ with a boundary of $|F/2-p-q|$ flavors.
\subsection{$F-p-q\leq F/2$ Scenario}\label{sec:2a}

In order to keep the analysis under control, we consider the choice of $p$ and $q$ such that $0<q\leq p\leq F-(p+q) \leq F/2$. The range of $k$ diagram is divided into five cases: $k\geq F/2$, $F/2-q\leq k<F/2$, $F/2-p\leq k<F/2-q$, $(p+q)-F/2\leq k<F/2-p$, and $0\leq k<(p+q)-F/2$, each one with a different phase diagram.

Our strategy for finding the topological phases is by sending the three masses to $\pm\infty$. For small values of $k$, we look for the quantum regions in two steps; the first is to find the asymptotic limits when two of the masses are sent to $\pm\infty$ and then when we send only one of the three masses to $\pm\infty$. The theory on the three-dimensional diagonal line $m_1=m_2=m_3$ is reduced to the one-family case with the phase diagram described by Figs.~\ref{fig1} and \ref{fig2}, which means that we always expect $\sigma$ to appear as a phase for all the ranges with $k<F/2$. 
\subsubsection*{\underline{Case 1: $k\geq F/2$}}
In this case, The theory in the IR is weakly coupled with no symmetry breaking scenario. We describe the phase diagram by eight topological field theories and their level-rank dual descriptions; six critical lines separate these phases. The topological phases are determined by finding the asymptotic limits of the three masses and they appear in the following ranges of the masses $m_1,\, m_2,\,\text{and } m_3$ as
$\Tp_1(m_1>0,m_2>0,m_3>0),\Tp_2(m_1>0,m_2>0,m_3<0), \Tp_3(m_1>0,m_2<0,m_3>0) \Tp_4(m_1<0,m_2>0,m_3>0),\Tp_5(m_1>0,m_2<0,m_3>0),\Tp_6(m_1<0,m_2>0,m_3<0)$, $\Tp_7(m_1<0,m_2<0,m_3>0), \Tp_8(m_1<0,m_2<0,m_3<0)$. They are
\begin{equation}
  \begin{aligned}
&\Tp_1:SU(N)_{k+\frac{F}{2}}\qquad\longleftrightarrow \ U(k+F/2)_{-N}\\
&\Tp_2:SU(N)_{k+p+q-\frac{F}{2}}\longleftrightarrow\ U(k+p+q-F/2)_{-N}\\
&\Tp_3:SU(N)_{k+\frac{F}{2}-q}\quad \longleftrightarrow\ U(k+F/2-q)_{-N}\\
&\Tp_4:SU(N)_{k+\frac{F}{2}-p}\quad \longleftrightarrow\ U(k+F/2-p)_{-N}\\
&\Tp_5: SU(N)_{k+p-\frac{F}{2}}\quad \longleftrightarrow\ U(k+p-F/2)_{-N}\\
&\Tp_6: SU(N)_{k+q-\frac{F}{2}}\quad\longleftrightarrow\ U(k+q-F/2)_{-N}\\
&\Tp_7: SU(N)_{k+\frac{F}{2}-p-q}\longleftrightarrow\ U(k+F/2-p-q)_{-N}\\
&\Tp_8:SU(N)_{k-\frac{F}{2}}\qquad \longleftrightarrow\ U(k-F/2)_{-N}
  \end{aligned}
  \label{t3kgf}
\end{equation}

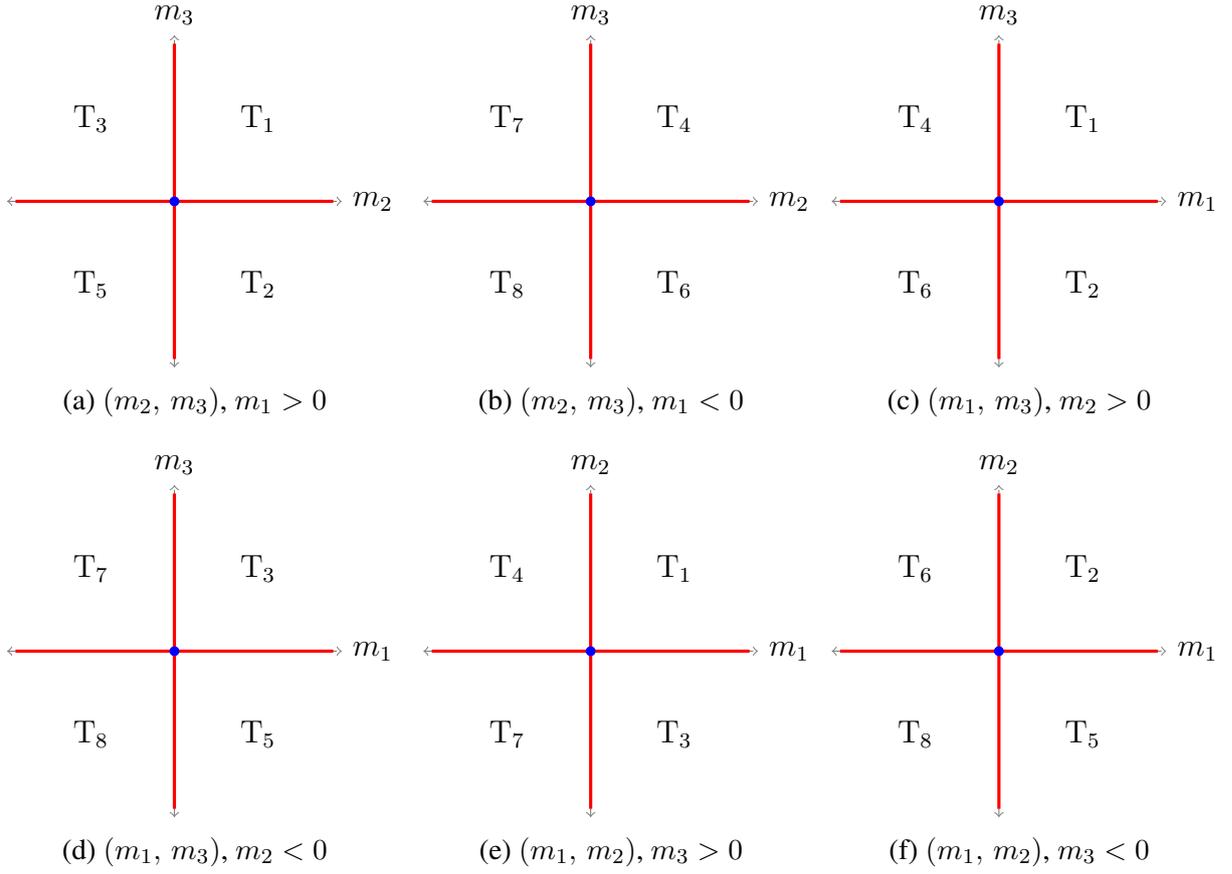
\begin{figure*}[h]
    \centering
    \begin{subfigure}[b]{0.3\textwidth}
\begin{tikzpicture}[scale=.55]
\draw[<->,gray,dashed](-4,0)--(4,0);
\draw (4,0) node[right]{$m_2$};
\draw[<->,gray,dashed](0,-4)--(0,4);
\draw (0,4)node[above]{$m_3$};
\draw (2,2) node{$\Tp_1$};
\draw (-2,2) node{$\Tp_3$};
\draw (2,-2) node{$\Tp_2$};
\draw (-2,-2) node{$\Tp_5$};
\draw[ very thick, red] (-3.8,0)--(3.8,0);
\draw[very thick, red] (0,-3.8)--(0,3.8);
\filldraw[blue] (0,0) circle (3pt);
\end{tikzpicture}
\caption{$(m_2,\,m_3)$, $m_1>0$}        
        \label{fig:4a}
    \end{subfigure} \quad
        \begin{subfigure}[b]{0.3\textwidth}
\begin{tikzpicture}[scale=.55]
\draw[<->,gray,dashed](-4,0)--(4,0);
\draw (4,0) node[right]{$m_2$};
\draw[<->,gray,dashed](0,-4)--(0,4);
\draw (0,4)node[above]{$m_3$};
\draw (2,2) node{$\Tp_4$};
\draw (-2,2) node{$\Tp_7$};
\draw (2,-2) node{$\Tp_6$};
\draw (-2,-2) node{$\Tp_8$};
\draw[ very thick, red] (-3.8,0)--(3.8,0);
\draw[very thick, red] (0,-3.8)--(0,3.8);
\filldraw[blue] (0,0) circle (3pt);
\end{tikzpicture}
\caption{$(m_2,\,m_3)$, $m_1<0$}        
        \label{fig:4b}
    \end{subfigure}\quad
            \begin{subfigure}[b]{0.3\textwidth}
\begin{tikzpicture}[scale=.55]
\draw[<->,gray,dashed](-4,0)--(4,0);
\draw (4,0) node[right]{$m_1$};
\draw[<->,gray,dashed](0,-4)--(0,4);
\draw (0,4)node[above]{$m_3$};
\draw (2,2) node{$\Tp_1$};
\draw (-2,2) node{$\Tp_4$};
\draw (2,-2) node{$\Tp_2$};
\draw (-2,-2) node{$\Tp_6$};
\draw[ very thick, red] (-3.8,0)--(3.8,0);
\draw[very thick, red] (0,-3.8)--(0,3.8);
\filldraw[blue] (0,0) circle (3pt);
\end{tikzpicture}
\caption{$(m_1,\,m_3)$, $m_2>0$}        
        \label{fig:4c}
    \end{subfigure}\\ \vspace*{.3 cm}
        \begin{subfigure}[b]{0.3\textwidth}
\begin{tikzpicture}[scale=.55]
\draw[<->,gray,dashed](-4,0)--(4,0);
\draw (4,0) node[right]{$m_1$};
\draw[<->,gray,dashed](0,-4)--(0,4);
\draw (0,4)node[above]{$m_3$};
\draw (2,2) node{$\Tp_3$};
\draw (-2,2) node{$\Tp_7$};
\draw (2,-2) node{$\Tp_5$};
\draw (-2,-2) node{$\Tp_8$};
\draw[ very thick, red] (-3.8,0)--(3.8,0);
\draw[very thick, red] (0,-3.8)--(0,3.8);
\filldraw[blue] (0,0) circle (3pt);
\end{tikzpicture}
\caption{$(m_1,\,m_3)$, $m_2<0$}        
        \label{fig:4d}
    \end{subfigure} \quad
    \begin{subfigure}[b]{0.3\textwidth}
\begin{tikzpicture}[scale=.55]
\draw[<->,gray,dashed](-4,0)--(4,0);
\draw (4,0) node[right]{$m_1$};
\draw[<->,gray,dashed](0,-4)--(0,4);
\draw (0,4)node[above]{$m_2$};
\draw (2,2) node{$\Tp_1$};
\draw (-2,2) node{$\Tp_4$};
\draw (2,-2) node{$\Tp_3$};
\draw (-2,-2) node{$\Tp_7$};
\draw[ very thick, red] (-3.8,0)--(3.8,0);
\draw[very thick, red] (0,-3.8)--(0,3.8);
\filldraw[blue] (0,0) circle (3pt);
\end{tikzpicture}
\caption{$(m_1,\,m_2)$, $m_3>0$}        
        \label{fig:4e}
    \end{subfigure}\quad
        \begin{subfigure}[b]{0.3\textwidth}
\begin{tikzpicture}[scale=.55]
\draw[<->,gray,dashed](-4,0)--(4,0);
\draw (4,0) node[right]{$m_1$};
\draw[<->,gray,dashed](0,-4)--(0,4);
\draw (0,4)node[above]{$m_2$};
\draw (2,2) node{$\Tp_2$};
\draw (-2,2) node{$\Tp_6$};
\draw (2,-2) node{$\Tp_5$};
\draw (-2,-2) node{$\Tp_8$};
\draw[ very thick, red] (-3.8,0)--(3.8,0);
\draw[very thick, red] (0,-3.8)--(0,3.8);
\filldraw[blue] (0,0) circle (3pt);
\end{tikzpicture}
\caption{$(m_1,\,m_2)$, $m_3<0$}        
        \label{fig:4f}
    \end{subfigure}
        \caption{Phases of $SU(N)_k+p\, \psi_1+q\, \psi_2+(F-p-q)\,\psi_3$ with $k\geq F/2$.}
    \label{fig4}
\end{figure*}
\begin{figure}[h!]
\centering
\includegraphics[scale=.2]{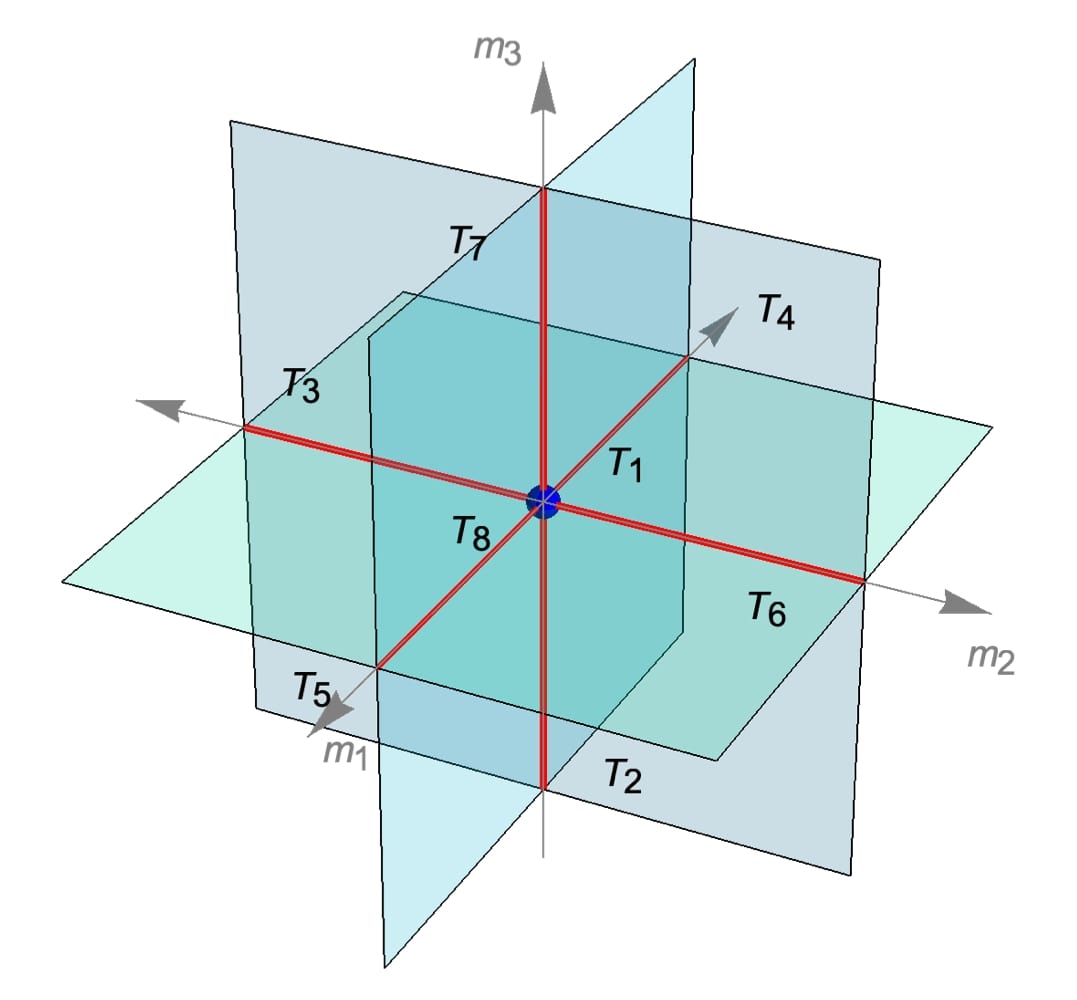}
\caption{The three-dimensional phase diagram of $SU(N)_k+p\,\psi_1+q\,\psi_2+(F-p-q)\,\psi_3$ with $k\geq F/2$. The blue ball represents the critical point, the red lines are the critical lines given by \eq{crit} while the planes in cyan are the critical planes, each plane is described by one of the critical theories in \eq{critplanes}.}
\label{G1}
\end{figure}

The phase diagram is three-dimensional in the space $(m_1,m_2,m_3)$ as shown in figure \ref{G1} and the six different projections are shown in figure \ref{fig4}. The separation between the eight topological theories occurs via point, lines, and planes of critical theories. The blue point is described by a critical theory given by $SU(N)_k+p\,  \psi^0_1+q\, \psi^0_2+(F-p-q)\, \psi^0_3$. Each point on the critical lines belong to one of the following critical theories

\begin{equation}
  \begin{aligned}
&C_1^{\pm}:SU(N)_{k\pm\frac{p}{2}}+q\,\psi^0_2+(F-p-q)\,\psi^0_3\ ,\\
&C_2^{\pm}:SU(N)_{k\pm\frac{q}{2}}+p\,\psi^0_1+(F-p-q)\,\psi^0_3\ ,\\
&C_3^{\pm}:SU(N)_{k\pm\frac{F-p-q}{2}}+p\,\psi^0_1+q\,\psi^0_2\ .
  \end{aligned}
  \label{crit}
\end{equation}
 $C_1^{\pm}$, $C_2^{\pm}$, and $C_3^{\pm}$ describe the lines on the mass axes $m_1$, $m_2$, and $m_3$, respectively. The plus or minus sign indicates whether the theory is along the positive or negative axes. The critical planes are given by any of the following theories
 \begin{equation}
  \begin{aligned}
&C_{12}^{\pm}:SU(N)_{k\pm\frac{p+q}{2}}+(F-p-q)\,\psi^0_3\ ,\\
&C_{21}^{\pm}:SU(N)_{k\pm\frac{p-q}{2}}+(F-p-q)\,\psi^0_3\ ,\\
&C_{13}^{\pm}:SU(N)_{k\pm\frac{F-q}{2}}+q\,\psi^0_2\ ,\\
&C_{31}^{\pm}:SU(N)_{k\pm\left(\frac{F-q}{2}-p\right)}+q\,\psi^0_2\ ,\\
&C_{23}^{\pm}:SU(N)_{k\pm\frac{F-p}{2}}+p\,\psi^0_1 ,\\
&C_{32}^{\pm}:SU(N)_{k\pm(\frac{F-p}{2}-q)}+p\,\psi^0_1\ .
  \end{aligned}
  \label{critplanes}
\end{equation}
$C_{ij}^{\pm}$ describes the theories on the planes $(m_i,m_j)$ with both of the masses being positive or negative, while $C_{ji}^{\pm}$ for the planes when one of the masses is positive and the other is negative.
 \subsubsection*{\underline{Case 2: $F/2-q\leq k<F/2$}}
In this range of $k$ and all the remaining cases, we need two bosonic dual descriptions to fill in the full phase diagram as conjectured in \cite{Komargodski:2017keh}. These bosonic theories are $U(F/2+k)_{-N}+p\,\phi_1+q\, \phi_2+(F-p-q)\, \phi_3$ and $U(F/2-k)_{N}+p\,\phi_1+q\, \phi_2+(F-p-q)\, \phi_3$, where $\phi_1$, $\phi_2$, and $\phi_1$ are scalars in the fundamental representation of $SU(N)$. The three masses asymptotic limits yield the same topological theories as in \eq{t3kgf} except for $\Tp_8$, which becomes
\begin{equation}
\Tp_8\rightarrow \tilde{\Tp}_8: SU(N)_{k-\frac{F}{2}}\longleftrightarrow U(F/2-k)_{N}\ .
\label{t3klf1}
\end{equation}

The two mass asymptotic limits produce a one-family theory with a shifted level. In some limits, the shifted level becomes lower than the remaining number of flavors divided by two, and the theory is strongly coupled. The three-dimensional phase diagram includes quantum regions described by sigma-models, but this time the quantum regions appear as cuboids in the three-dimensional picture. The cuboid quantum phases are signature of the three-family theory as the planes of sigma-models were signatures of the two-family case. The theory that has a level within this range experience a cuboid sigma-model in its phase diagram in the following cases:

\begin{itemize}
\item[$\bullet$] $m_2$, $m_3 \, \rightarrow -\infty$: the theory becomes $SU(N)_{k-F/2+p/2}$ $+p \, \psi_1$. The cuboid quantum region is described by a sigma-model $\sigma_1^c$. From now on we shorthand the cuboid Grassmannains by their corresponding sigma-models, the Grassmannian for this region is
\begin{equation}
\sigma_1^c= \frac{U(p)}{U(F/2-k)\times U(k-F/2+p)}\ .
\label{sigma1c}
\end{equation} 

\item[$\bullet$] $m_1$, $m_3 \, \rightarrow -\infty$: the theory becomes $SU(N)_{k-F/2+q/2}$ $+q \, \psi_2$ which has a sigma-model $\sigma_2^c$ given by
\begin{equation}
\sigma_2^c= \frac{U(q)}{U(F/2-k)\times U(k-F/2+q)}\ .
\label{sigma2c}
\end{equation} 
\item[$\bullet$] $m_1$, $m_2 \, \rightarrow -\infty$: the theory is $SU(N)_{k-p/2-q/2}+(F-p-q) \, \psi_3$ with a sigma-model $\sigma_3^c$ given by
\begin{equation}
\sigma_3^c= \frac{U(F-p-q)}{U(F/2-k)\times U(F/2-p-q+k)}\ .
\label{sigma3c}
\end{equation} 
\end{itemize}
Now let us perform a consistency check by sending one mass to infinity where the theory with the remaining flavors is reduced to a two-family case with shifted level. In each limit, we check the relation between the remaining number of flavors and the shifted level, which determines whether the phase diagram is type $\mathrm{I}$, $\mathrm{II}$, or $\mathrm{III}$. We summarize the check as follows:

\begin{enumerate}[label=(\roman*)]
\item $m_1 \rightarrow +\infty$: we integrate $\psi_1$ out, and the theory is reduced to 
\begin{align}
SU(N)_{k+\frac{p}{2}}+q\, \psi_2+ (F-p-q)\, \psi_3 \equiv SU(N)_{k_1^+}  +q\, \psi_2+ (F_1-q)\, \psi_3\ ,
\end{align}
 where $k_1^+= k+p/2$ and $F_1 = F-p$ with $k_1^+>F_1/2$.  Due to the range of $k_1^+$,  this region of the three-dimensional phase diagram has a type $\mathrm{I}$ phase diagram with the topological theories $\Tp_1$, $\Tp_2$, $\Tp_3$, and $\Tp_5$. 
  \item $m_1 \rightarrow -\infty$: integrating $\psi_1$ out leads to 
\begin{align}
SU(N)_{k-\frac{p}{2}}+q\, \psi_2+ (F-p-q)\, \psi_3 \equiv SU(N)_{k_1^-}  +q\, \psi_2+ (F_1-q)\, \psi_3\ ,
\end{align}
 where $k_1^-= k-p/2$ and $F_1/2-q<k_1^-<F_1$. Due to the range of $k_1^-$, the phase diagram is then of type $\mathrm{II}$ with the topological theories $\Tp_4$, $\Tp_6$, $\Tp_7$, and $\tilde{\Tp}_8$. Alongside these topological phases, we have a sigma-model on the diagonal given by 
 \begin{align}
 \sigma_{23}^d: Gr(F_1/2+k_1^-,F_1)=\frac{U(F-p)}{U(F/2-k)\times U(F/2-p+k)}\ .
 \label{sigma23d}
 \end{align}
 This diagonal sigma-model is not a line but rather a plane region in the three-dimensional picture. Only one side appears here, which will become clear in section \ref{sec:4}. The horizontal and vertical sigma-models are $\sigma_{78}$ which separates $\Tp_7$ and $\tilde{\Tp}_8$ as well as $\sigma_{68}$ separating $\Tp_6$ and $\tilde{\Tp}_8$. They are given by
 \begin{align}
\sigma_{78}&:Gr(F_1/2-k_1^-,F_1-q)=\frac{U(F-p-q)}{U(F/2-k)\times U(F/2-p-q+k)}\equiv \sigma_3^c\ ,
\label{sigma78}\\
\sigma_{68}&:Gr(F_1/2-k_1^-,q)=\frac{U(q)}{U(F/2-k)\times U(k-F/2+q)}\equiv \sigma_2^c\ .
\label{sigma68}
\end{align} 
We notice that $\sigma_{78}\equiv \sigma_3^c$ which means that $\sigma_{78}$ is just one side of the three-dimensional quantum region $\sigma_3^c$, the same thing applies for $\sigma_{68}$ which is equivalent to $\sigma_2^c$. 
\item $m_2 \rightarrow +\infty$: we integrate $\psi_2$ out, and the theory becomes
\begin{align}
SU(N)_{k+\frac{q}{2}}+p\, \psi_1+ (F-p-q)\, \psi_3 \equiv SU(N)_{k_2^+}  +p\, \psi_1+ (F_1-p)\, \psi_3\ ,
\end{align}
 where $k_2^+= k+p/2$ and $F_2 = F-q$ with $k_2^+>F_2/2$ and the phase diagram is of type $\mathrm{I}$ with the topological phases $\Tp_1$, $\Tp_2$, $\Tp_4$, and $\Tp_6$. 
 \begin{figure*}[h]
    \centering
     \begin{subfigure}[b]{0.3\textwidth}
\begin{tikzpicture}[scale=.55]
\draw[<->,gray,dashed](-4,0)--(4,0);
\draw (4,0) node[right]{$m_2$};
\draw[<->,gray,dashed](0,-4)--(0,4);
\draw (0,4)node[above]{$m_3$};
\draw (2,2) node{$\Tp_1$};
\draw (-2,2) node{$\Tp_3$};
\draw (2,-2) node{$\Tp_2$};
\draw (-2,-2) node{$\Tp_5$};
\draw[ very thick, red] (-3.8,0)--(3.8,0);
\draw[very thick, red] (0,-3.8)--(0,3.8);
\filldraw[blue] (0,0) circle (3pt);
\end{tikzpicture}
\caption{$(m_2,\,m_3)$, $m_1>0$}        
        \label{fig:5a}
    \end{subfigure}\quad
        \begin{subfigure}[b]{0.3\textwidth}
      \begin{tikzpicture}[scale=.55]
\draw[<->,gray,dashed](-4,0)--(4,0);
\draw (4,0) node[right]{$m_2$};
\draw[<->,gray,dashed](0,-4)--(0,4);
\draw (0,4)node[above]{$m_3$};
\draw (2,2) node{$\Tp_4$};
\draw (-2,2) node{$\Tp_7$};
\draw (2,-2) node{$\Tp_6$};
\draw (-2,-2) node{$\tilde{\Tp}_8$};
\draw[very thick] (-.8,-.8)--(.8,.8);
\draw[very thick, red] (-3.8,-.6)..controls(-1.2,-.7)..(-.8,-.8)..controls (-0.7,-1.2)..(-.6,-3.8);
\draw[very thick, red] (-3.8,.7)..controls(-1.2,.6)..(.8,.8)..controls (0.6,-1.2)..(.7,-3.8); 
\draw[very thick, red] (3.8,.08)..controls(1.4,.4)..(.8,.8)..controls(0.4,1.4)..(0.08,3.8); 
\filldraw[blue] (0.8,0.8) circle (3pt);
\filldraw[blue] (-.8,-.8) circle (3pt);
\draw (-2,0) node{$\sigma_{78}$};
\draw (0,-2) node[below]{$\sigma_{68}$};
\draw[->] (1.2,-.6)node[right]{$\sigma_{23}^d$}--((.21,.19);
\end{tikzpicture}
\caption{$(m_2,\,m_3)$, $m_1<0$}        
        \label{fig:5b}
    \end{subfigure}\quad
        \begin{subfigure}[b]{0.3\textwidth}
\begin{tikzpicture}[scale=.55]
\draw[<->,gray,dashed](-4,0)--(4,0);
\draw (4,0) node[right]{$m_1$};
\draw[<->,gray,dashed](0,-4)--(0,4);
\draw (0,4)node[above]{$m_3$};
\draw (2,2) node{$\Tp_1$};
\draw (-2,2) node{$\Tp_4$};
\draw (2,-2) node{$\Tp_2$};
\draw (-2,-2) node{$\Tp_6$};
\draw[ very thick, red] (-3.8,0)--(3.8,0);
\draw[very thick, red] (0,-3.8)--(0,3.8);
\filldraw[blue] (0,0) circle (3pt);
\end{tikzpicture}
\caption{$(m_1,\,m_3)$, $m_2>0$}        
        \label{fig:5c}
    \end{subfigure}\\ \vspace*{.3 cm}
        \begin{subfigure}[b]{0.3\textwidth}
      \begin{tikzpicture}[scale=.55]
\draw[<->,gray,dashed](-4,0)--(4,0);
\draw (4,0) node[right]{$m_1$};
\draw[<->,gray,dashed](0,-4)--(0,4);
\draw (0,4)node[above]{$m_3$};
\draw (2,2) node{$\Tp_3$};
\draw (-2,2) node{$\Tp_7$};
\draw (2,-2) node{$\Tp_5$};
\draw (-2,-2) node{$\tilde{\Tp}_8$};
\draw[very thick] (-.8,-.8)--(.8,.8);
\draw[very thick, red] (-3.8,-.6)..controls(-1.2,-.7)..(-.8,-.8)..controls (-0.7,-1.2)..(-.6,-3.8);
\draw[very thick, red] (-3.8,.7)..controls(-1.2,.6)..(.8,.8)..controls (0.6,-1.2)..(.7,-3.8); 
\draw[very thick, red] (3.8,.08)..controls(1.4,.4)..(.8,.8)..controls(0.4,1.4)..(0.08,3.8); 
\filldraw[blue] (0.8,0.8) circle (3pt);
\filldraw[blue] (-.8,-.8) circle (3pt);
\draw (-2,0) node{$\sigma_{78}$};
\draw (0,-2) node[below]{$\sigma_{58}$};
\draw[->] (1.2,-.6)node[right]{$\sigma_{13}^d$}--((.21,.19);
\end{tikzpicture}
\caption{$(m_1,\,m_3)$, $m_2<0$}        
        \label{fig:5d}
    \end{subfigure}  \quad
     \begin{subfigure}[b]{0.3\textwidth}
\begin{tikzpicture}[scale=.55]
\draw[<->,gray,dashed](-4,0)--(4,0);
\draw (4,0) node[right]{$m_1$};
\draw[<->,gray,dashed](0,-4)--(0,4);
\draw (0,4)node[above]{$m_2$};
\draw (2,2) node{$\Tp_1$};
\draw (-2,2) node{$\Tp_4$};
\draw (2,-2) node{$\Tp_3$};
\draw (-2,-2) node{$\Tp_7$};
\draw[ very thick, red] (-3.8,0)--(3.8,0);
\draw[very thick, red] (0,-3.8)--(0,3.8);
\filldraw[blue] (0,0) circle (3pt);
\end{tikzpicture}
\caption{$(m_1,\,m_2)$, $m_3>0$}        
        \label{fig:5e}
    \end{subfigure}\quad
        \begin{subfigure}[b]{0.3\textwidth}
      \begin{tikzpicture}[scale=.55]
\draw[<->,gray,dashed](-4,0)--(4,0);
\draw (4,0) node[right]{$m_1$};
\draw[<->,gray,dashed](0,-4)--(0,4);
\draw (0,4)node[above]{$m_2$};
\draw (2,2) node{$\Tp_2$};
\draw (-2,2) node{$\Tp_6$};
\draw (2,-2) node{$\Tp_5$};
\draw (-2,-2) node{$\tilde{\Tp}_8$};
\draw[very thick] (-.8,-.8)--(.8,.8);
\draw[very thick, red] (-3.8,-.6)..controls(-1.2,-.7)..(-.8,-.8)..controls (-0.7,-1.2)..(-.6,-3.8);
\draw[very thick, red] (-3.8,.7)..controls(-1.2,.6)..(.8,.8)..controls (0.6,-1.2)..(.7,-3.8); 
\draw[very thick, red] (3.8,.08)..controls(1.4,.4)..(.8,.8)..controls(0.4,1.4)..(0.08,3.8); 
\filldraw[blue] (0.8,0.8) circle (3pt);
\filldraw[blue] (-.8,-.8) circle (3pt);
\draw (-2,0) node{$\sigma_{68}$};
\draw (0,-2) node[below]{$\sigma_{58}$};
\draw[->] (1.2,-.6)node[right]{$\sigma_{12}^d$}--((.21,.19);
\end{tikzpicture}
\caption{$(m_1,\,m_2)$, $m_3<0$}        
        \label{fig:5f}
    \end{subfigure}
    \caption{Phases of $SU(N)_k+p\, \psi_1+q\, \psi_2+(F-p-q)\,\psi_3$ with $F/2-q\leq k<F/2$.}
    \label{fig5}
\end{figure*}
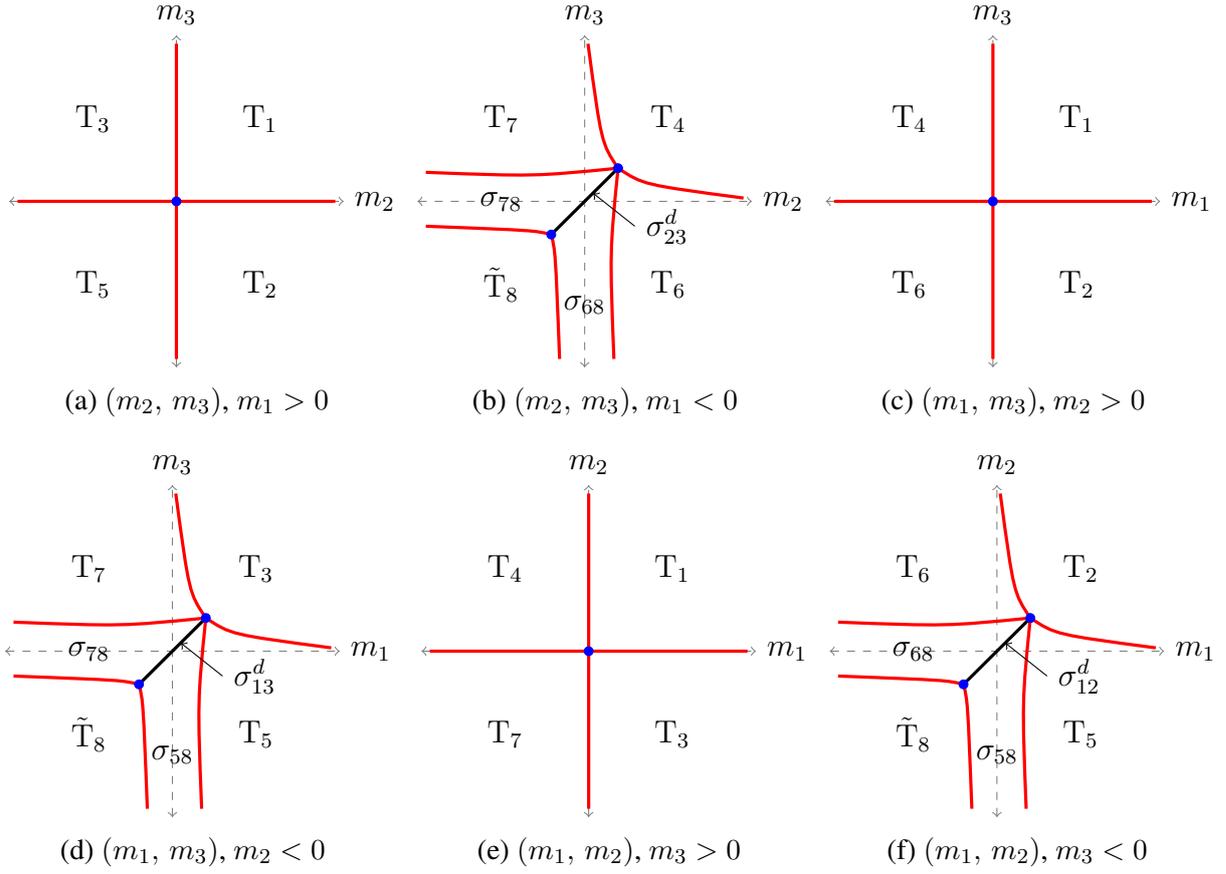
\begin{figure}[h!]
\centering
\includegraphics[scale=.2]{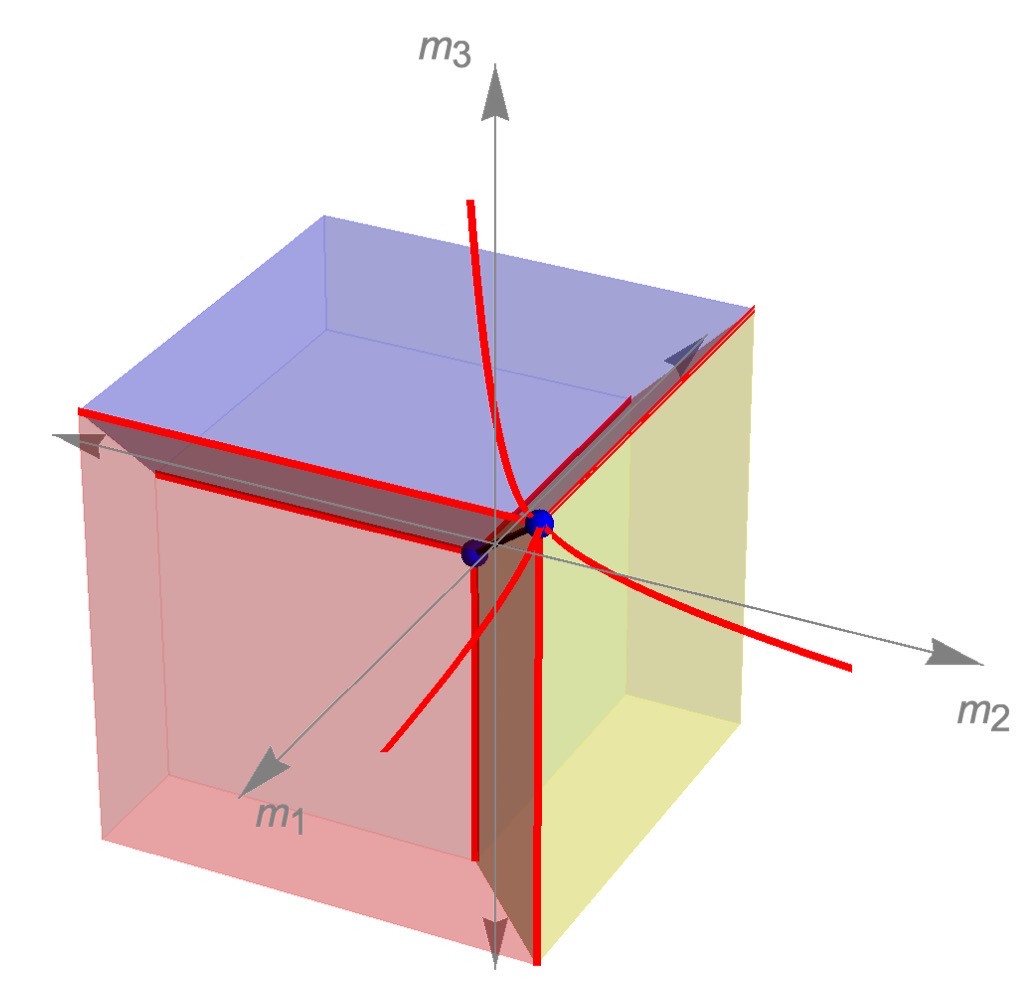}
\caption{The three-dimensional phase diagram of $SU(N)_k+p\,\psi_1+q\,\psi_2+(F-p-q)\,\psi_3$ with $F/2-q\leq k< F/2$. $\sigma$ is represented by the thick black line between the two critical points. $\sigma_1^c$, $\sigma_2^c$, and $\sigma_3^c$ are represented by the red, yellow, and blue regions, respectively. The diagonal sigma-models are the dark brown planes separating the cuboid quantum regions.}
\label{G2}
\end{figure}
  \item $m_2 \rightarrow -\infty$: integrating $\psi_2$ gives 
\begin{align}
SU(N)_{k-\frac{q}{2}}+p\, \psi_1+ (F-p-q)\, \psi_3 \equiv SU(N)_{k_2^-}  +p\, \psi_1+ (F_1-p)\, \psi_3\ ,
\end{align}
 where $k_2^-= k-p/2$ and $F_2/2-q<k_2^-<F_2$. The phase diagram is then of type $\mathrm{II}$ with the topological theories $\Tp_3$, $\Tp_5$, $\Tp_7$, and $\tilde{\Tp}_8$. The diagonal sigma-model is 
 \begin{align}
 \sigma_{13}^d: Gr(F_2/2+k_2^-,F_2)= \frac{U(F-q)}{U(F/2-k)\times U(F/2-q+k)}\ ,
 \label{sigma13d}
 \end{align}
 the quantum phase $\sigma_{78}$ also exists in this side as in \eq{sigma78} alongside the phase $\sigma_{58}$ that separates $\Tp_5$ and $\tilde{\Tp}_8$ which is  given by
 \begin{align}
\sigma_{58}:Gr(F_2/2-k_2^-,p)=\frac{U(p)}{U(F/2-k)\times U(k-F/2+p)}\equiv \sigma_1^c\ .
\label{sigma58}
\end{align} 
 We should emphasize here that the equivalence between these phases and the cuboid sigma-models is itself a consistency check of our analysis.
 \item $m_3 \rightarrow +\infty$: after integrating $\psi_3$ out the theory is
\begin{align}
SU(N)_{k+\frac{F-p-q}{2}}+p\, \psi_1+ q\, \psi_2 \equiv SU(N)_{k_3^+}  +p\, \psi_1+ (F_3-p)\, \psi_2\ ,
\end{align}
 where $k_3^+= k+(F-p-q)/2$ and $F_3 = p+q$, the relation $k_3^+>F_3/2$ holds, and we have a type $\mathrm{I}$ phase diagram with the topological phases $\Tp_1$, $\Tp_3$, $\Tp_4$, and $\Tp_7$. 
  \item $m_3 \rightarrow -\infty$: integrating $\psi_3$ gives 
\begin{align}
SU(N)_{k-\frac{F-p-q}{2}}+p\, \psi_1+ q\, \psi_2 \equiv SU(N)_{k_3^-}  +p\, \psi_1+ (F_3-p)\, \psi_3\ ,
\end{align}
 where $k_3^-= k-(F-p-q)/2$ and $F_3/2-q<k_3^-<F_3$, but $p>q$ which makes this case cover the range of type $\mathrm{II}$ phase diagram. The phase diagram has the topological phases $\Tp_2$, $\Tp_5$, $\Tp_6$, and $\tilde{\Tp}_8$. The diagonal sigma-model of this side is 
 \begin{align}
 \sigma_{12}^d: Gr(F_3/2+k_3^-,F_3)= \frac{U(p+q)}{U(F/2-k)\times U(k-F/2+p+q)}\ ,
 \label{sigma12d}
 \end{align}
 while the horizontal and vertical quantum regions are $\sigma_{68}$ and $\sigma_{58}$ respectively and given by Eqs.~\eqref{sigma68} and \eqref{sigma58}.
\end{enumerate}

In this range of $k$, we summarize the phase diagram in figures \ref{fig5} and \ref{G2} where the IR theory has the following phases; eight topological phases $(\Tp_{1-7},\tilde{\Tp}_8)$, three cuboid sigma-models $(\sigma_1^c,\sigma_2^c,\sigma_3^c)$, three planes of sigma-models $(\sigma_{12}^d,\sigma_{13}^d,\sigma_{23}^d)$, as well as a line of sigma-model $\sigma$ that appears on the diagonal line $m_1=m_2=m_3$. In the limiting case $q=0$, the phase diagram becomes equivalent to the $k=F/2$ case where all the sigma-models, as well as the topological theories $\Tp_6$ and $\Tp_8$ trivialize. The phase diagram is then reduced to a two-dimensional phase diagram with three topological theories $\Tp_a$, $\Tp_b$, and $\Tp_d$, as well as a trivial theory $SU(N)_0$.
\subsubsection*{\underline{Case 3: $F/2-p\leq k<F/2-q$}}

As in the previous case, the theory has two bosonic dual descriptions. The topological phases are given by Eqs.~\eqref{t3kgf} and \eqref{t3klf1} except for $\Tp_6$, which becomes
\begin{align}
\Tp_6\rightarrow \tilde{\Tp}_6: SU(N)_{k-\frac{F}{2}+q}\longleftrightarrow U(F/2-q-k)_{N}\ .
\label{t3klf2}
\end{align}
$\Tp_8$ is also replaced by $\tilde{\Tp}_8$ as before.

The cuboid quantum regions exist in the following cases:
\begin{itemize}
\item[$\bullet$] $m_2\rightarrow +\infty$ and $m_3\rightarrow-\infty$: the theory is reduced to $SU(N)_{k-F/2+p/2+q}+p\, \psi_1$ which has a sigma-model phase given by
\begin{equation}
\bar{\sigma}_1^c= \frac{U(p)}{U(F/2-q-k)\times U(k-F/2+p+q)}\ .
\label{sigma1cbar}
\end{equation}
\item[$\bullet$] $m_2$, $m_3 \rightarrow -\infty$: the theory has the same cuboid sigma-model as in \eq{sigma1c}.
\item[$\bullet$] $m_2\rightarrow +\infty$ and $m_1\rightarrow-\infty$: the theory is reduced to $SU(N)_{k-p/2+q/2}+(F-p-q)\, \psi_3$ with a sigma-model given by
\begin{equation}
\hat{\sigma}_3^c= \frac{U(F-p-q)}{U(F/2-p+k)\times U(F/2-q-k)}\ .
\label{sigma3chat}
\end{equation}
\item[$\bullet$] $m_1$, $m_2 \rightarrow -\infty$: the theory has the same cuboid sigma-model $\sigma_3^c$ as in \eq{sigma3c}.
\end{itemize}
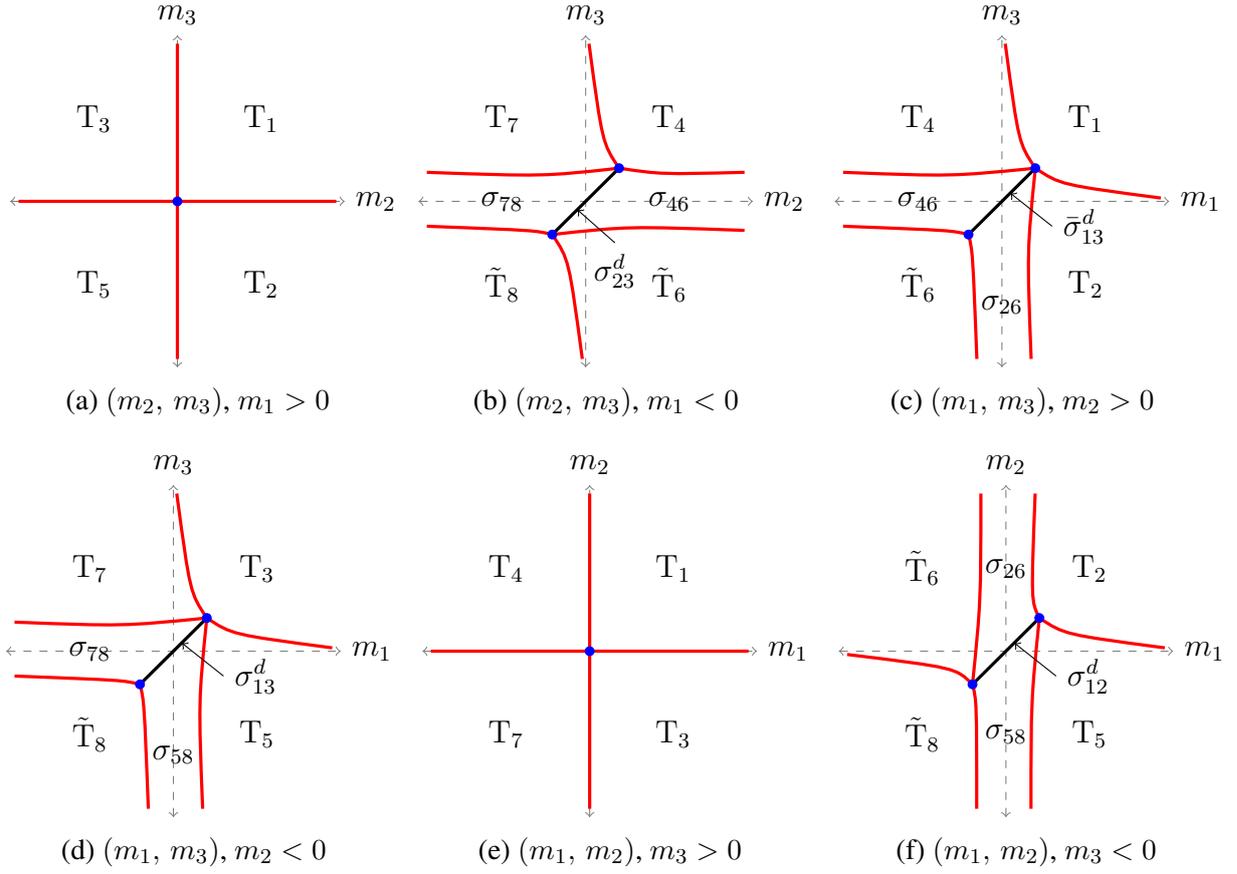
\begin{figure*}[h]
    \centering
    \begin{subfigure}[b]{0.3\textwidth}
\begin{tikzpicture}[scale=.55]
\draw[<->,gray,dashed](-4,0)--(4,0);
\draw (4,0) node[right]{$m_2$};
\draw[<->,gray,dashed](0,-4)--(0,4);
\draw (0,4)node[above]{$m_3$};
\draw (2,2) node{$\Tp_1$};
\draw (-2,2) node{$\Tp_3$};
\draw (2,-2) node{$\Tp_2$};
\draw (-2,-2) node{$\Tp_5$};
\draw[ very thick, red] (-3.8,0)--(3.8,0);
\draw[very thick, red] (0,-3.8)--(0,3.8);
\filldraw[blue] (0,0) circle (3pt);
\end{tikzpicture}
\caption{$(m_2,\,m_3)$, $m_1>0$}        
        \label{fig:6a}
    \end{subfigure}\quad
        \begin{subfigure}[b]{0.3\textwidth}
      \begin{tikzpicture}[scale=.55]
\draw[<->,gray,dashed](-4,0)--(4,0);
\draw (4,0) node[right]{$m_2$};
\draw[<->,gray,dashed](0,-4)--(0,4);
\draw (0,4)node[above]{$m_3$};
\draw (2,2) node{$\Tp_4$};
\draw (-2,2) node{$\Tp_7$};
\draw (2,-2) node{$\tilde{\Tp}_6$};
\draw (-2,-2) node{$\tilde{\Tp}_8$};
\draw[very thick] (-.8,-.8)--(.8,.8);
\draw[very thick, red] (-3.8,-.6)..controls(-1.2,-.7)..(-.8,-.8)..controls (-0.4,-1.4)..(-.08,-3.8);
\draw[very thick, red] (-3.8,.7)..controls(-1.,.6)..(.8,.8)..controls (1.8,.65)..(3.8,.7); 
\draw[very thick, red] (.08,3.8)..controls(.4,1.4)..(.8,.8); 
\draw[very thick, red] (-.8,-.8)..controls(1,-.6)..(3.8,-.7); 
\filldraw[blue] (0.8,0.8) circle (3pt);
\filldraw[blue] (-.8,-.8) circle (3pt);
\draw (-2,0) node{$\sigma_{78}$};
\draw (2,0) node{$\sigma_{46}$};
\draw[->] (.7,-1)node[below]{$\sigma_{23}^d$}--(-.19,-.21);
\end{tikzpicture}
\caption{$(m_2,\,m_3)$, $m_1<0$}        
        \label{fig:6b}
    \end{subfigure} \quad
        \begin{subfigure}[b]{0.3\textwidth}
\begin{tikzpicture}[scale=.55]
\draw[<->,gray,dashed](-4,0)--(4,0);
\draw (4,0) node[right]{$m_1$};
\draw[<->,gray,dashed](0,-4)--(0,4);
\draw (0,4)node[above]{$m_3$};
\draw (2,2) node{$\Tp_1$};
\draw (-2,2) node{$\Tp_4$};
\draw (2,-2) node{$\Tp_2$};
\draw (-2,-2) node{$\tilde{\Tp}_6$};
\draw[very thick] (-.8,-.8)--(.8,.8);
\draw[very thick, red] (-3.8,-.6)..controls(-1.2,-.7)..(-.8,-.8)..controls (-0.7,-1.2)..(-.6,-3.8);
\draw[very thick, red] (-3.8,.7)..controls(-1.2,.6)..(.8,.8)..controls (0.6,-1.2)..(.7,-3.8); 
\draw[very thick, red] (3.8,.08)..controls(1.4,.4)..(.8,.8)..controls(0.4,1.4)..(0.08,3.8); 
\filldraw[blue] (0.8,0.8) circle (3pt);
\filldraw[blue] (-.8,-.8) circle (3pt);
\draw (-2,0) node{$\sigma_{46}$};
\draw (0,-2) node[below]{$\sigma_{26}$};
\draw[->] (1.2,-.6)node[right]{$\bar{\sigma}_{13}^d$}--(.21,.19);
\end{tikzpicture}
\caption{$(m_1,\,m_3)$, $m_2>0$}        
        \label{fig:6c}
    \end{subfigure}\\ \vspace*{.3 cm}
        \begin{subfigure}[b]{0.3\textwidth}
      \begin{tikzpicture}[scale=.55]
\draw[<->,gray,dashed](-4,0)--(4,0);
\draw (4,0) node[right]{$m_1$};
\draw[<->,gray,dashed](0,-4)--(0,4);
\draw (0,4)node[above]{$m_3$};
\draw (2,2) node{$\Tp_3$};
\draw (-2,2) node{$\Tp_7$};
\draw (2,-2) node{$\Tp_5$};
\draw (-2,-2) node{$\tilde{\Tp}_8$};
\draw[very thick] (-.8,-.8)--(.8,.8);
\draw[very thick, red] (-3.8,-.6)..controls(-1.2,-.7)..(-.8,-.8)..controls (-0.7,-1.2)..(-.6,-3.8);
\draw[very thick, red] (-3.8,.7)..controls(-1.2,.6)..(.8,.8)..controls (0.6,-1.2)..(.7,-3.8); 
\draw[very thick, red] (3.8,.08)..controls(1.4,.4)..(.8,.8)..controls(0.4,1.4)..(0.08,3.8); 
\filldraw[blue] (0.8,0.8) circle (3pt);
\filldraw[blue] (-.8,-.8) circle (3pt);
\draw (-2,0) node{$\sigma_{78}$};
\draw (0,-2) node[below]{$\sigma_{58}$};
\draw[->] (1.2,-.6)node[right]{$\sigma_{13}^d$}--(.21,.19);
\end{tikzpicture}
\caption{$(m_1,\,m_3)$, $m_2<0$}        
        \label{fig:6d}
    \end{subfigure}
  \quad
     \begin{subfigure}[b]{0.3\textwidth}
\begin{tikzpicture}[scale=.55]
\draw[<->,gray,dashed](-4,0)--(4,0);
\draw (4,0) node[right]{$m_1$};
\draw[<->,gray,dashed](0,-4)--(0,4);
\draw (0,4)node[above]{$m_2$};
\draw (2,2) node{$\Tp_1$};
\draw (-2,2) node{$\Tp_4$};
\draw (2,-2) node{$\Tp_3$};
\draw (-2,-2) node{$\Tp_7$};
\draw[ very thick, red] (-3.8,0)--(3.8,0);
\draw[very thick, red] (0,-3.8)--(0,3.8);
\filldraw[blue] (0,0) circle (3pt);
\end{tikzpicture}
\caption{$(m_1,\,m_2)$, $m_3>0$}        
        \label{fig:6e}
    \end{subfigure} \quad
        \begin{subfigure}[b]{0.3\textwidth}
      \begin{tikzpicture}[scale=.55]
\draw[<->,gray,dashed](-4,0)--(4,0);
\draw (4,0) node[right]{$m_1$};
\draw[<->,gray,dashed](0,-4)--(0,4);
\draw (0,4)node[above]{$m_2$};
\draw (2,2) node{$\Tp_2$};
\draw (-2,2) node{$\tilde{\Tp}_6$};
\draw (2,-2) node{$\Tp_5$};
\draw (-2,-2) node{$\tilde{\Tp}_8$};
\draw[very thick] (-.8,-.8)--(.8,.8);
\draw[very thick, red] (-3.8,-.08)..controls(-1.2,-.4)..(-.8,-.8)..controls (-0.7,-1.2)..(-.7,-3.8);
\draw[very thick, red] (.8,.8)..controls (0.6,-1.2)..(.6,-3.8);  
\draw[very thick, red](-.8,-.8)..controls (-0.6,1.2)..(-.6,3.8);
\draw[very thick, red] (3.8,.08)..controls(1.4,.4)..(.8,.8)..controls(0.6,1.2)..(.7,3.8); 
\filldraw[blue] (0.8,0.8) circle (3pt);
\filldraw[blue] (-.8,-.8) circle (3pt);
\draw (0,2) node{$\sigma_{26}$};
\draw (0,-2) node{$\sigma_{58}$};
\draw[->] (1.2,-.6)node[right]{$\sigma_{12}^d$}--((.21,.19);
\end{tikzpicture}
\caption{$(m_1,\,m_2)$, $m_3<0$}        
        \label{fig:6f}
    \end{subfigure}
    \caption{Phases of $SU(N)_k+p\, \psi_1+q\, \psi_2+(F-p-q)\,\psi_3$ with $F/2-p\leq k<F/2-q$.}
    \label{fig6}
\end{figure*}
\begin{figure}[h!]
\centering
\includegraphics[scale=.21]{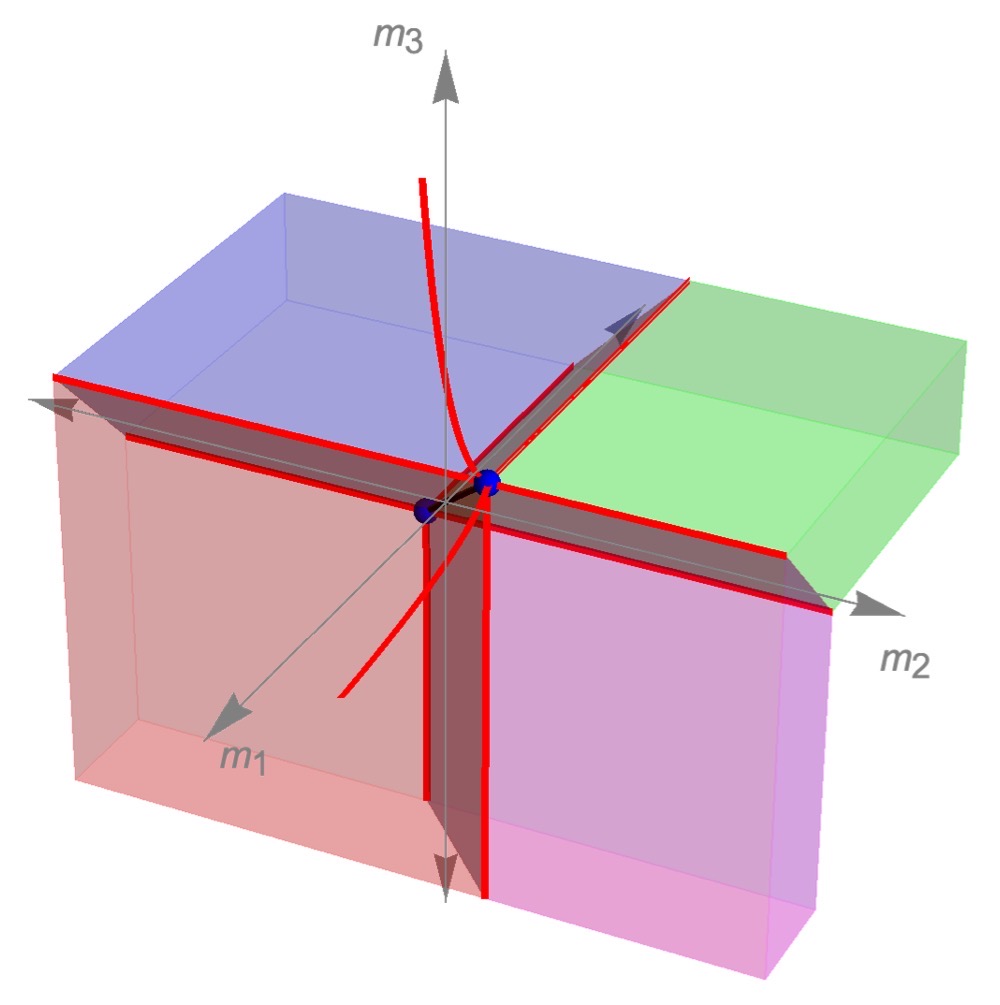}
\caption{The three-dimensional phase diagram of $SU(N)_k+p\,\psi_1+q\,\psi_2+(F-p-q)\,\psi_3$ with $F/2-p\leq k< F/2-q$. $\bar{\sigma}_1^c$ is given by the region in magenta color and $\hat{\sigma}_3^c$ is represented by the green region.}
\label{G3}
\end{figure}
The phases that appear when we take one of the masses to $\pm \infty$ are as follows:
\begin{enumerate}[label=(\roman*)]
\item $m_1\rightarrow +\infty$: this is similar to case 2, where we have a type $\mathrm{I}$ phase diagram with only topological phases.
\item $m_1\rightarrow -\infty$: this limit is different from the previous case, as in this range of $k$ the value of  $k_1^-$ lies between $F_1/2-p$ and $F_1/2 -q$, which gives a type $\mathrm{III}$ phase diagram. The sigma-models $\sigma_{23}^d$ and $\sigma_{78}$ remain the same, while a new phase $\sigma_{46}$ appears between $\Tp_4$ and $\tilde{\Tp}_6$ and is given by
\begin{align}
\sigma_{46}: Gr(F_1/2+k_1^-,F_1-q)= \frac{U(F-p-q)}{U(F/2-p+k)\times U(F/2-q-k)} \equiv \hat{\sigma}_3^c\ .
\label{sigma46}
\end{align}
\item $m_2\rightarrow +\infty$: the range of $k_2^+$ within this range of $k$ is $F_2/2-p\leq k_2^+<F_2/2$ and the phase diagram is of type $\mathrm{II}$ with a new diagonal sigma-model $\bar{\sigma}_{13}^d$ given by
\begin{align}
\bar{\sigma}_{13}^d: Gr(F_2/2+k_2^+,F_2)= \frac{U(F-q)}{U(F/2+k)\times U(F/2-q-k)}\ .
\label{sigma13dbar}
\end{align}
The other two sigma-models are $\sigma_{46}$ as in \eq{sigma46} and a new $\sigma_{26}$ that separates $\Tp_2$ and $\tilde{\Tp}_6$ with a Grassmannian given by
\begin{align}
\sigma_{26}: Gr(F_2/2-k_2^+,p)= \frac{U(p)}{U(F/2-q-k)\times U(k-F/2+p+q)}\equiv \bar{\sigma}_1^c\ .
\label{sigma26}
\end{align}
\item $m_2\rightarrow -\infty$:  in this case, $k_2^-$ is within $F_2/2-p\leq k_2^-<F_2/2$, which is just similar to case 2 with the same quantum phases $\sigma_{13}^d$, $\sigma_{78}$, and $\sigma_{58}$.
\item $m_3\rightarrow +\infty$: $k_3^+>F_3/2$ giving a type $\mathrm{I}$ phase diagram just like in case 2.
\item $m_3\rightarrow -\infty$: it is not clear whether this case is of type $\mathrm{II}$ or $\mathrm{III}$ because $k_3^-$ takes some negative values in this range of $k$. The theory is better understood by rewriting its reduction in the form $SU(N)_{k_3^-}+q\,\psi_2+(F_3-q)\, \psi_1$. Hence it becomes clear that $|k_3^-|<F_3/2-q$ and the theory is type $\mathrm{III}$ with sigma-models appearing only for small $m_1$ both negative and positive sides which are $\sigma_{58}$ and $\sigma_{26}$ respectively, as shown in \fig{fig:6f}.
\end{enumerate}

The phases of case 3 are summarized in figures \ref{fig6} and \ref{G3}. These phases are eight topological field theories $(\Tp_{1-5},\tilde{\Tp}_6,$ $\Tp_7,\tilde{\Tp}_8)$, four cuboid sigma-models $(\sigma_1^c,\bar{\sigma}_1^c, \hat{\sigma}_3^c,\sigma_3^c)$, four planes of sigma-models $(\sigma_{12}^d,\sigma_{13}^d,\bar{\sigma}_{13}^d,\sigma_{23}^d)$, as well as the one-dimensional sigma-model $\sigma$. In the limiting case $q=0$, the Chern-Simons level range becomes $F/2-p\leq k<F/2$ and the phase diagram is reduced to \fig{fig:3b}.

\subsubsection*{\underline{Case 4: $(p+q)-F/2\leq k<F/2-p$}}
Following the same procedure, we found that this case has the same topological phases as in case 3 with an extra change being that  $\Tp_5$ is replaced by $\tilde{\Tp}_5$ as
\begin{align}
\Tp_5\rightarrow \tilde{\Tp}_5: SU(N)_{k-\frac{F}{2}+p}\longleftrightarrow U(F/2-p-k)_{N}\ .
\label{t3klf3}
\end{align}

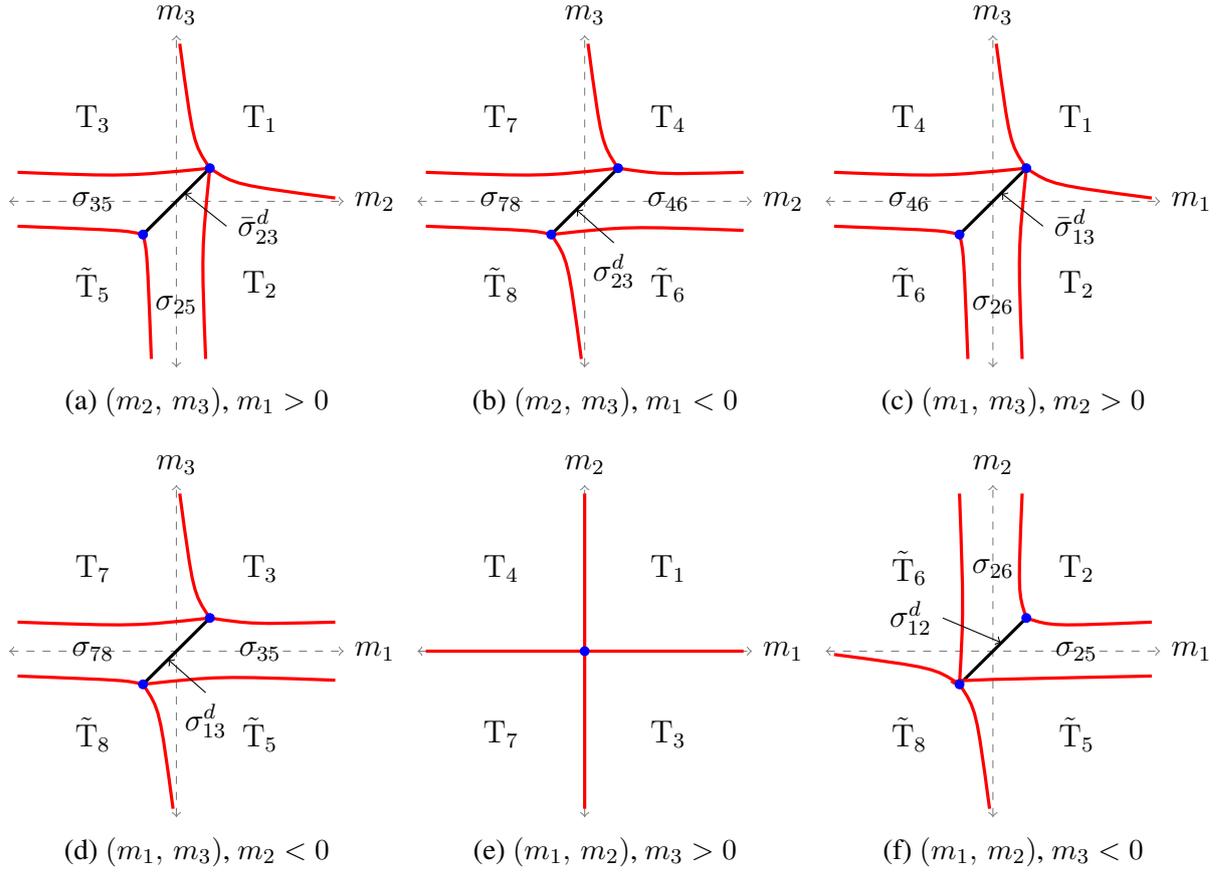
\begin{figure*}[h]
    \centering
   \begin{subfigure}[b]{0.3\textwidth}
\begin{tikzpicture}[scale=.55]
\draw[<->,gray,dashed](-4,0)--(4,0);
\draw (4,0) node[right]{$m_2$};
\draw[<->,gray,dashed](0,-4)--(0,4);
\draw (0,4)node[above]{$m_3$};
\draw (2,2) node{$\Tp_1$};
\draw (-2,2) node{$\Tp_3$};
\draw (2,-2) node{$\Tp_2$};
\draw (-2,-2) node{$\tilde{\Tp}_5$};
\draw[very thick] (-.8,-.8)--(.8,.8);
\draw[very thick, red] (-3.8,-.6)..controls(-1.2,-.7)..(-.8,-.8)..controls (-0.7,-1.2)..(-.6,-3.8);
\draw[very thick, red] (-3.8,.7)..controls(-1.2,.6)..(.8,.8)..controls (0.6,-1.2)..(.7,-3.8); 
\draw[very thick, red] (3.8,.08)..controls(1.4,.4)..(.8,.8)..controls(0.4,1.4)..(0.08,3.8); 
\filldraw[blue] (0.8,0.8) circle (3pt);
\filldraw[blue] (-.8,-.8) circle (3pt);
\draw (-2,0) node{$\sigma_{35}$};
\draw (0,-2) node[below]{$\sigma_{25}$};
\draw[->] (1.2,-.6)node[right]{$\bar{\sigma}_{23}^d$}--((.21,.19);
\end{tikzpicture}
\caption{$(m_2,\,m_3)$, $m_1>0$}        
        \label{fig:7a}
    \end{subfigure}\quad
        \begin{subfigure}[b]{0.3\textwidth}
      \begin{tikzpicture}[scale=.55]
\draw[<->,gray,dashed](-4,0)--(4,0);
\draw (4,0) node[right]{$m_2$};
\draw[<->,gray,dashed](0,-4)--(0,4);
\draw (0,4)node[above]{$m_3$};
\draw (2,2) node{$\Tp_4$};
\draw (-2,2) node{$\Tp_7$};
\draw (2,-2) node{$\tilde{\Tp}_6$};
\draw (-2,-2) node{$\tilde{\Tp}_8$};
\draw[very thick] (-.8,-.8)--(.8,.8);
\draw[very thick, red] (-3.8,-.6)..controls(-1.2,-.7)..(-.8,-.8)..controls (-0.4,-1.4)..(-.08,-3.8);
\draw[very thick, red] (-3.8,.7)..controls(-1.,.6)..(.8,.8)..controls (1.8,.65)..(3.8,.7); 
\draw[very thick, red] (.08,3.8)..controls(.4,1.4)..(.8,.8); 
\draw[very thick, red] (-.8,-.8)..controls(1,-.6)..(3.8,-.7); 
\filldraw[blue] (0.8,0.8) circle (3pt);
\filldraw[blue] (-.8,-.8) circle (3pt);
\draw (-2,0) node{$\sigma_{78}$};
\draw (2,0) node{$\sigma_{46}$};
\draw[->] (.7,-1)node[below]{$\sigma_{23}^d$}--(-.19,-.21);
\end{tikzpicture}
\caption{$(m_2,\,m_3)$, $m_1<0$}        
        \label{fig:7b}
    \end{subfigure}\quad
        \begin{subfigure}[b]{0.3\textwidth}
\begin{tikzpicture}[scale=.55]
\draw[<->,gray,dashed](-4,0)--(4,0);
\draw (4,0) node[right]{$m_1$};
\draw[<->,gray,dashed](0,-4)--(0,4);
\draw (0,4)node[above]{$m_3$};
\draw (2,2) node{$\Tp_1$};
\draw (-2,2) node{$\Tp_4$};
\draw (2,-2) node{$\Tp_2$};
\draw (-2,-2) node{$\tilde{\Tp}_6$};
\draw[very thick] (-.8,-.8)--(.8,.8);
\draw[very thick, red] (-3.8,-.6)..controls(-1.2,-.7)..(-.8,-.8)..controls (-0.7,-1.2)..(-.6,-3.8);
\draw[very thick, red] (-3.8,.7)..controls(-1.2,.6)..(.8,.8)..controls (0.6,-1.2)..(.7,-3.8); 
\draw[very thick, red] (3.8,.08)..controls(1.4,.4)..(.8,.8)..controls(0.4,1.4)..(0.08,3.8); 
\filldraw[blue] (0.8,0.8) circle (3pt);
\filldraw[blue] (-.8,-.8) circle (3pt);
\draw (-2,0) node{$\sigma_{46}$};
\draw (0,-2) node[below]{$\sigma_{26}$};
\draw[->] (1.2,-.6)node[right]{$\bar{\sigma}_{13}^d$}--((.21,.19);
\end{tikzpicture}
\caption{$(m_1,\,m_3)$, $m_2>0$}        
        \label{fig:7c}
    \end{subfigure}\\ \vspace*{.3 cm}
        \begin{subfigure}[b]{0.3\textwidth}
    \begin{tikzpicture}[scale=.55]
\draw[<->,gray,dashed](-4,0)--(4,0);
\draw (4,0) node[right]{$m_1$};
\draw[<->,gray,dashed](0,-4)--(0,4);
\draw (0,4)node[above]{$m_3$};
\draw (2,2) node{$\Tp_3$};
\draw (-2,2) node{$\Tp_7$};
\draw (2,-2) node{$\tilde{\Tp}_5$};
\draw (-2,-2) node{$\tilde{\Tp}_8$};
\draw[very thick] (-.8,-.8)--(.8,.8);
\draw[very thick, red] (-3.8,-.6)..controls(-1.2,-.7)..(-.8,-.8)..controls (-0.4,-1.4)..(-.08,-3.8);
\draw[very thick, red] (-3.8,.7)..controls(-1.,.6)..(.8,.8)..controls (1.8,.65)..(3.8,.7); 
\draw[very thick, red] (.08,3.8)..controls(.4,1.4)..(.8,.8); 
\draw[very thick, red] (-.8,-.8)..controls(1,-.6)..(3.8,-.7); 
\filldraw[blue] (0.8,0.8) circle (3pt);
\filldraw[blue] (-.8,-.8) circle (3pt);
\draw (-2,0) node{$\sigma_{78}$};
\draw (2,0) node{$\sigma_{35}$};
\draw[->] (.7,-1)node[below]{$\sigma_{13}^d$}--(-.19,-.21);
\end{tikzpicture}
\caption{$(m_1,\,m_3)$, $m_2<0$}        
        \label{fig:7d}
    \end{subfigure}\quad
     \begin{subfigure}[b]{0.3\textwidth}
\begin{tikzpicture}[scale=.55]
\draw[<->,gray,dashed](-4,0)--(4,0);
\draw (4,0) node[right]{$m_1$};
\draw[<->,gray,dashed](0,-4)--(0,4);
\draw (0,4)node[above]{$m_2$};
\draw (2,2) node{$\Tp_1$};
\draw (-2,2) node{$\Tp_4$};
\draw (2,-2) node{$\Tp_3$};
\draw (-2,-2) node{$\Tp_7$};
\draw[ very thick, red] (-3.8,0)--(3.8,0);
\draw[very thick, red] (0,-3.8)--(0,3.8);
\filldraw[blue] (0,0) circle (3pt);
\end{tikzpicture}
\caption{$(m_1,\,m_2)$, $m_3>0$}        
        \label{fig:7e}
    \end{subfigure}\quad
        \begin{subfigure}[b]{0.3\textwidth}
\begin{tikzpicture}[scale=.55]
\draw[<->,gray,dashed](-4,0)--(4,0);
\draw (4,0) node[right]{$m_1$};
\draw[<->,gray,dashed](0,-4)--(0,4);
\draw (0,4)node[above]{$m_2$};
\draw (2,2) node{$\Tp_2$};
\draw (-2,2) node{$\tilde{\Tp}_6$};
\draw (2,-2) node{$\tilde{\Tp}_5$};
\draw (-2,-2) node{$\tilde{\Tp}_8$};
\draw[very thick] (-.8,-.8)--(.8,.8);
\draw[very thick, red] (3.8,-.6)..controls(-1.2,-.7)..(-.8,-.8)..controls (-0.7,1)..(-.8,3.8);
\draw[very thick, red] (-3.8,-.08)..controls(-1.4,-.4)..(-.8,-.8)..controls (-0.4,-1.4)..(-.08,-3.8); 
\draw[very thick, red] (3.8,.7)..controls(1.4,.6)..(.8,.8)..controls(0.6,1.4)..(0.7,3.8); 
\filldraw[blue] (0.8,0.8) circle (3pt);
\filldraw[blue] (-.8,-.8) circle (3pt);
\draw (0,2) node{$\sigma_{26}$};
\draw (2,0) node{$\sigma_{25}$};
\draw[->] (-1.2,.8)node[left]{$\sigma_{12}^d$}--((.19,.21);
\end{tikzpicture}
\caption{$(m_1,\,m_2)$, $m_3<0$}        
        \label{fig:7f}
    \end{subfigure}
    \caption{Phases of $SU(N)_k+p\, \psi_1+q\, \psi_2+(F-p-q)\,\psi_3$ with $(p+q)-F/2\leq k<F/2-p$.}
    \label{fig7}
\end{figure*}

The two masses asymptotic limits show that the theory has the following cuboid sigma-models 
\begin{itemize}
\item[$\bullet$] $m_1\rightarrow +\infty$ and $m_3\rightarrow -\infty$: 
\begin{equation}
\bar{\sigma}_2^c= \frac{U(q)}{U(F/2-p-k)\times U(k-F/2+p+q)}\ .
\label{sigma2cbar}
\end{equation}
\item[$\bullet$] $m_1\rightarrow +\infty$ and $m_2\rightarrow -\infty$: 
\begin{equation}
\bar{\sigma}_3^c= \frac{U(F-p-q)}{U(F/2-p-k)\times U(F/2-q+k)}\ ,
\label{sigma3cbar}
\end{equation}
along with $\bar{\sigma}_1^c$, $\hat{\sigma}_3^c$, and $\sigma_3^c$.
\end{itemize}
The six sides of the three-dimensional picture have the following phases:
\begin{enumerate}[label=(\roman*)]
\item $m_1\rightarrow +\infty$: $F_1/2-q\leq k_1^+<F_1/2$ and the theory has a type $\mathrm{II}$ phase diagram with a diagonal sigma-model $\bar{\sigma}_{23}^d$ given by
\begin{align}
\bar{\sigma}_{23}^d: Gr(F_1/2+k_1^+,F_1)= \frac{U(F-p)}{U(F/2+k)\times U(F/2-p-k)}\ .
\label{sigma23dbar}
\end{align}
This diagonal sigma-model separates two other sigma-models given by
\begin{align}
\sigma_{35}: Gr(F_1/2-k_1^+,F_1-q)= \frac{U(F-p-q)}{U(F/2-p-k)\times U(F/2-q+k)}\equiv \bar{\sigma}_3^c \ ,\label{sigma35}
\end{align}
\begin{align}
\sigma_{25}: Gr(F_1/2-k_1^+,q)=  \frac{U(q)}{U(F/2-p-k)\times U(k-F/2+p+q)}\equiv \bar{\sigma}_2^c\ .
\label{sigma25}
\end{align}
\item $m_1\rightarrow -\infty$: this limit is similar to case 3 with $\sigma_{23}^d$, $\sigma_{78}$, and $\sigma_{46}$  appear as quantum phases.
\item $m_2\rightarrow +\infty$: this is also similar to case 3 with $\bar{\sigma}_{13}^d$, $\sigma_{46}$, and $\sigma_{26}$.
\item $m_2\rightarrow -\infty$: we have $ k_2^-<F_2/2-p$ within this range of $k$, which makes this limit to be of type $\mathrm{III}$ with  $\sigma_{13}^d$, $\sigma_{78}$, and $\sigma_{35}$.
\item $m_3\rightarrow +\infty$: this remains of type $\mathrm{I}$ phase diagram just like in cases 2 and 3.
\item $m_3\rightarrow -\infty$: this limit gives a shifted level within the range $F_3/2-p<k_3^-<F_3/2$, which makes this case of type $\mathrm{II}$ phase diagram. However, $k_3^-$ is always negative in this range of $k$, which requires a flip of the masses signs to get the right phase diagram. This allows sigma-models to appear for small $m_2$ but positive $m_1$ ($\sigma_{25}$) and small $m_1$ with positive $m_2$ ($\sigma_{26}$) instead of $\sigma_{68}$ and $\sigma_{58}$, as shown in \fig{fig:7f}. $\sigma_{25}$ and $\sigma_{26}$ are the correct phases to appear in this limit as they are part of $\bar{\sigma}_1^c$ and $\bar{\sigma}_2^c$ while $\sigma_{68}$ and $\sigma_{58}$ are part of $\sigma_1^c$ and $\sigma_2^c$ which do not appear in this range of $k$. 
\end{enumerate}

\begin{figure}[h!]
\centering
\includegraphics[scale=.21]{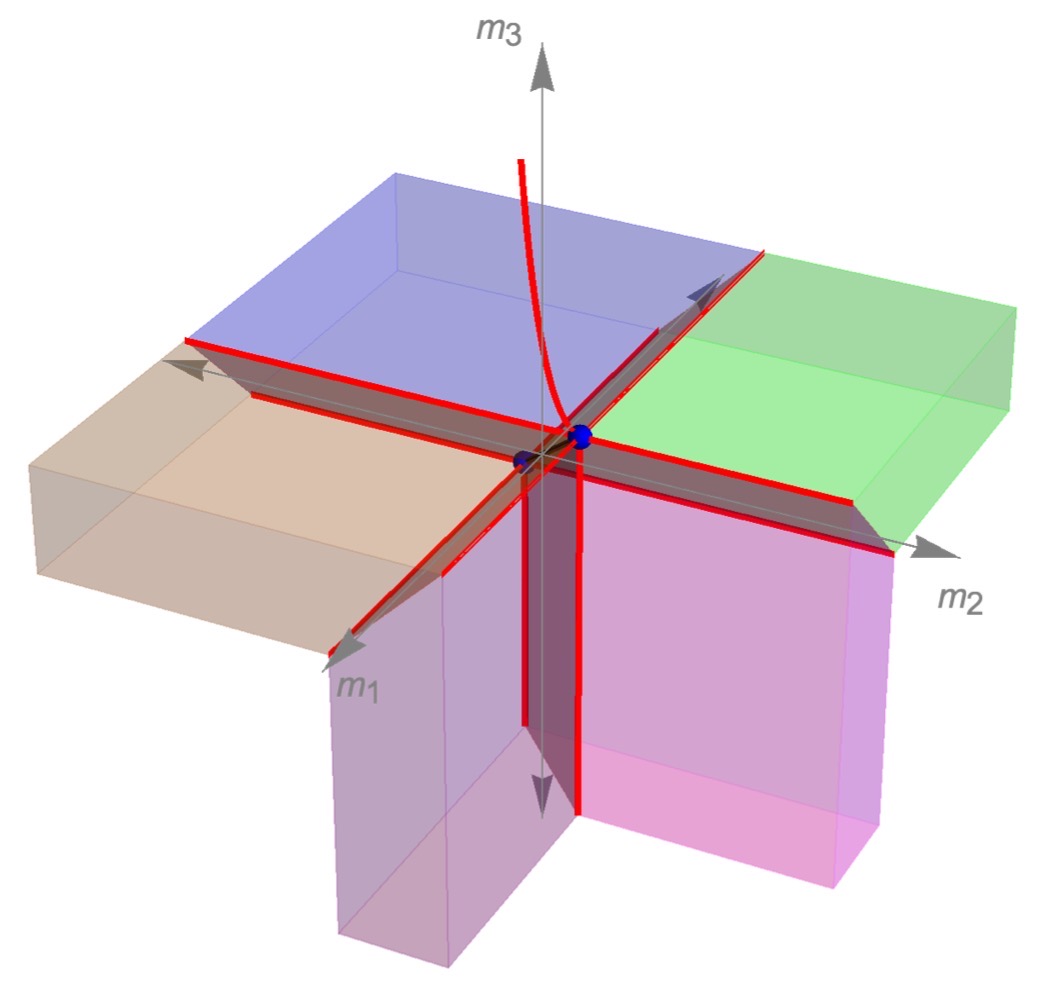}
\caption{The three-dimensional phase diagram of $SU(N)_k+p\,\psi_1+q\,\psi_2+(F-p-q)\,\psi_3$ with $(p+q)-F/2\leq k< F/2-p$. $\bar{\sigma}_2^c$ is the region in purple and $\bar{\sigma}_3^c$ is represented by the light brown region.}
\label{G4}
\end{figure}

Figures \ref{fig7} and \ref{G4} summarize the phases of case 4, which include the following; the eight topological phases $(\Tp_{1-4},$ $\tilde{\Tp}_5,\tilde{\Tp}_6,\Tp_7,\tilde{\Tp}_8)$, five cuboid sigma-models $(\bar{\sigma}_1^c,\bar{\sigma}_2^c,\bar{\sigma}_3^c,\hat{\sigma}_3^c,$ $\sigma_3^c)$, five planes of sigma-models $(\sigma_{12}^d,\sigma_{13}^d$, $\bar{\sigma}_{13}^d,\sigma_{23}^d,\bar{\sigma}_{23}^d)$
together with the sigma-model line $\sigma$. In the limiting case $q=0$, the phase diagram becomes equivalent to a theory with level $k=F/2-p$ where all the sigma-models and the topological theories $\Tp_2$ and $\Tp_5$, trivialize. The phase diagram is then reduced to a two-dimensional phase diagram with three topological theories $\Tp_a$, $\Tp_b$, and $\Tp_c$, as well as a trivial theory $SU(N)_0$.

\subsubsection*{\underline{Case 5: $0\leq k<(p+q)-F/2$}}

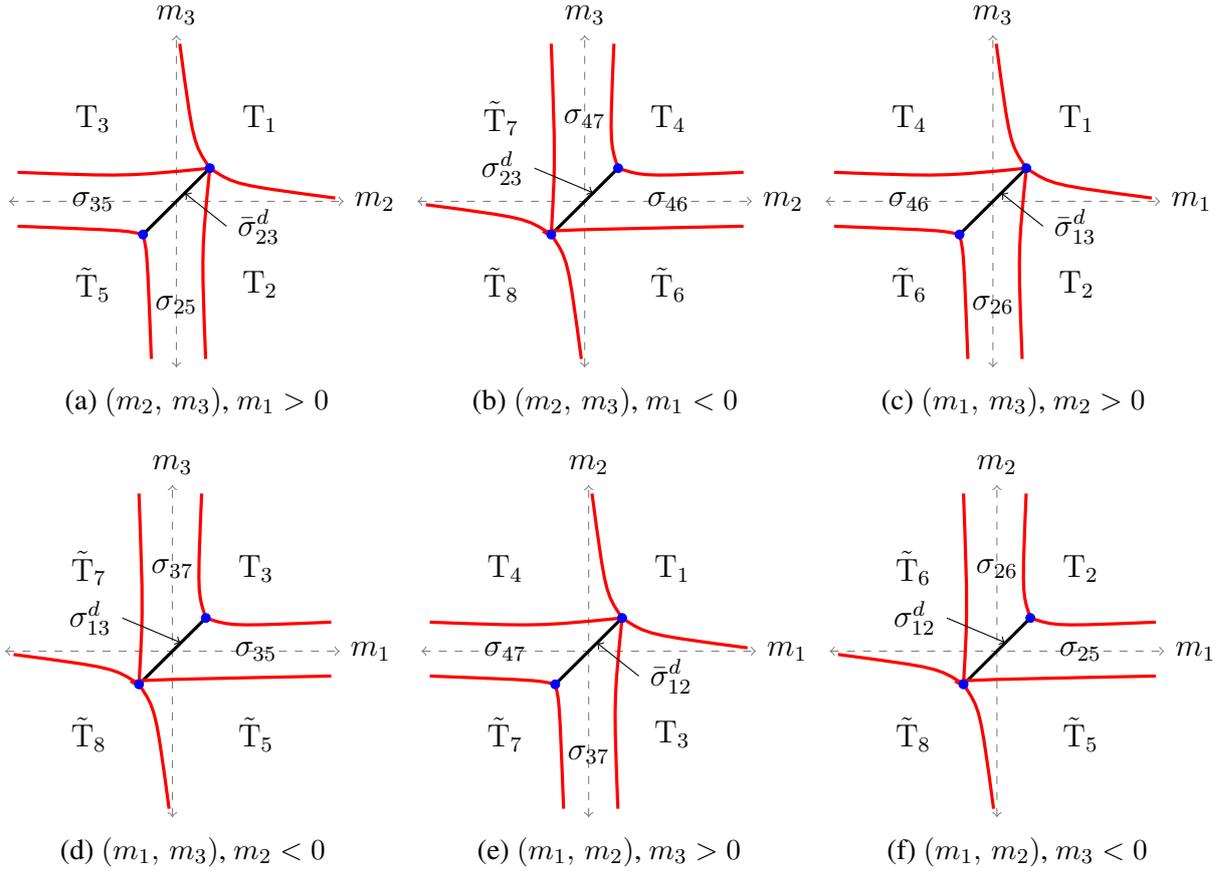
\begin{figure*}[h]
    \centering
    \begin{subfigure}[b]{0.3\textwidth}
\begin{tikzpicture}[scale=.55]
\draw[<->,gray,dashed](-4,0)--(4,0);
\draw (4,0) node[right]{$m_2$};
\draw[<->,gray,dashed](0,-4)--(0,4);
\draw (0,4)node[above]{$m_3$};
\draw (2,2) node{$\Tp_1$};
\draw (-2,2) node{$\Tp_3$};
\draw (2,-2) node{$\Tp_2$};
\draw (-2,-2) node{$\tilde{\Tp}_5$};
\draw[very thick] (-.8,-.8)--(.8,.8);
\draw[very thick, red] (-3.8,-.6)..controls(-1.2,-.7)..(-.8,-.8)..controls (-0.7,-1.2)..(-.6,-3.8);
\draw[very thick, red] (-3.8,.7)..controls(-1.2,.6)..(.8,.8)..controls (0.6,-1.2)..(.7,-3.8); 
\draw[very thick, red] (3.8,.08)..controls(1.4,.4)..(.8,.8)..controls(0.4,1.4)..(0.08,3.8); 
\filldraw[blue] (0.8,0.8) circle (3pt);
\filldraw[blue] (-.8,-.8) circle (3pt);
\draw (-2,0) node{$\sigma_{35}$};
\draw (0,-2) node[below]{$\sigma_{25}$};
\draw[->] (1.2,-.6)node[right]{$\bar{\sigma}_{23}^d$}--((.21,.19);
\end{tikzpicture}
\caption{$(m_2,\,m_3)$, $m_1>0$}        
        \label{fig:8a}
    \end{subfigure}\quad
        \begin{subfigure}[b]{0.3\textwidth}
      \begin{tikzpicture}[scale=.55]
\draw[<->,gray,dashed](-4,0)--(4,0);
\draw (4,0) node[right]{$m_2$};
\draw[<->,gray,dashed](0,-4)--(0,4);
\draw (0,4)node[above]{$m_3$};
\draw (2,2) node{$\Tp_4$};
\draw (-2,2) node{$\tilde{\Tp}_7$};
\draw (2,-2) node{$\tilde{\Tp}_6$};
\draw (-2,-2) node{$\tilde{\Tp}_8$};
\draw[very thick] (-.8,-.8)--(.8,.8);
\draw[very thick, red] (3.8,-.6)..controls(-1.2,-.7)..(-.8,-.8)..controls (-0.7,1)..(-.8,3.8);
\draw[very thick, red] (-3.8,-.08)..controls(-1.4,-.4)..(-.8,-.8)..controls (-0.4,-1.4)..(-.08,-3.8); 
\draw[very thick, red] (3.8,.7)..controls(1.4,.6)..(.8,.8)..controls(0.6,1.4)..(0.7,3.8); 
\filldraw[blue] (0.8,0.8) circle (3pt);
\filldraw[blue] (-.8,-.8) circle (3pt);
\draw (0,2) node{$\sigma_{47}$};
\draw (2,0) node{$\sigma_{46}$};
\draw[->] (-1.2,.8)node[left]{$\sigma_{23}^d$}--((.19,.21);
\end{tikzpicture}
\caption{$(m_2,\,m_3)$, $m_1<0$}        
        \label{fig:8b}
    \end{subfigure}\quad
        \begin{subfigure}[b]{0.3\textwidth}
\begin{tikzpicture}[scale=.55]
\draw[<->,gray,dashed](-4,0)--(4,0);
\draw (4,0) node[right]{$m_1$};
\draw[<->,gray,dashed](0,-4)--(0,4);
\draw (0,4)node[above]{$m_3$};
\draw (2,2) node{$\Tp_1$};
\draw (-2,2) node{$\Tp_4$};
\draw (2,-2) node{$\Tp_2$};
\draw (-2,-2) node{$\tilde{\Tp}_6$};
\draw[very thick] (-.8,-.8)--(.8,.8);
\draw[very thick, red] (-3.8,-.6)..controls(-1.2,-.7)..(-.8,-.8)..controls (-0.7,-1.2)..(-.6,-3.8);
\draw[very thick, red] (-3.8,.7)..controls(-1.2,.6)..(.8,.8)..controls (0.6,-1.2)..(.7,-3.8); 
\draw[very thick, red] (3.8,.08)..controls(1.4,.4)..(.8,.8)..controls(0.4,1.4)..(0.08,3.8); 
\filldraw[blue] (0.8,0.8) circle (3pt);
\filldraw[blue] (-.8,-.8) circle (3pt);
\draw (-2,0) node{$\sigma_{46}$};
\draw (0,-2) node[below]{$\sigma_{26}$};
\draw[->] (1.2,-.6)node[right]{$\bar{\sigma}_{13}^d$}--((.21,.19);
\end{tikzpicture}
\caption{$(m_1,\,m_3)$, $m_2>0$}        
        \label{fig:8c}
    \end{subfigure}\\ \vspace*{.3 cm}
        \begin{subfigure}[b]{0.3\textwidth}
      \begin{tikzpicture}[scale=.55]
\draw[<->,gray,dashed](-4,0)--(4,0);
\draw (4,0) node[right]{$m_1$};
\draw[<->,gray,dashed](0,-4)--(0,4);
\draw (0,4)node[above]{$m_3$};
\draw (2,2) node{$\Tp_3$};
\draw (-2,2) node{$\tilde{\Tp}_7$};
\draw (2,-2) node{$\tilde{\Tp}_5$};
\draw (-2,-2) node{$\tilde{\Tp}_8$};
\draw[very thick] (-.8,-.8)--(.8,.8);
\draw[very thick, red] (3.8,-.6)..controls(-1.2,-.7)..(-.8,-.8)..controls (-0.7,1)..(-.8,3.8);
\draw[very thick, red] (-3.8,-.08)..controls(-1.4,-.4)..(-.8,-.8)..controls (-0.4,-1.4)..(-.08,-3.8); 
\draw[very thick, red] (3.8,.7)..controls(1.4,.6)..(.8,.8)..controls(0.6,1.4)..(0.7,3.8); 
\filldraw[blue] (0.8,0.8) circle (3pt);
\filldraw[blue] (-.8,-.8) circle (3pt);
\draw (0,2) node{$\sigma_{37}$};
\draw (2,0) node{$\sigma_{35}$};
\draw[->] (-1.2,.8)node[left]{$\sigma_{13}^d$}--((.19,.21);
\end{tikzpicture}
\caption{$(m_1,\,m_3)$, $m_2<0$}        
        \label{fig:8d}
    \end{subfigure}   \quad
     \begin{subfigure}[b]{0.3\textwidth}
\begin{tikzpicture}[scale=.55]
\draw[<->,gray,dashed](-4,0)--(4,0);
\draw (4,0) node[right]{$m_1$};
\draw[<->,gray,dashed](0,-4)--(0,4);
\draw (0,4)node[above]{$m_2$};
\draw (2,2) node{$\Tp_1$};
\draw (-2,2) node{$\Tp_4$};
\draw (2,-2) node{$\Tp_3$};
\draw (-2,-2) node{$\tilde{\Tp}_7$};
\draw[very thick] (-.8,-.8)--(.8,.8);
\draw[very thick, red] (-3.8,-.6)..controls(-1.2,-.7)..(-.8,-.8)..controls (-0.7,-1.2)..(-.6,-3.8);
\draw[very thick, red] (-3.8,.7)..controls(-1.2,.6)..(.8,.8)..controls (0.6,-1.2)..(.7,-3.8); 
\draw[very thick, red] (3.8,.08)..controls(1.4,.4)..(.8,.8)..controls(0.4,1.4)..(0.08,3.8); 
\filldraw[blue] (0.8,0.8) circle (3pt);
\filldraw[blue] (-.8,-.8) circle (3pt);
\draw (-2,0) node{$\sigma_{\mathrm{47}}$};
\draw (0,-2) node[below]{$\sigma_{\mathrm{37}}$};
\draw[->] (1.2,-.6)node[right]{$\bar{\sigma}_{12}^d$}--(.21,.19);
\end{tikzpicture}
\caption{$(m_1,\,m_2)$, $m_3>0$}        
        \label{fig:8e}
    \end{subfigure}\quad
        \begin{subfigure}[b]{0.3\textwidth}
      \begin{tikzpicture}[scale=.55]
\draw[<->,gray,dashed](-4,0)--(4,0);
\draw (4,0) node[right]{$m_1$};
\draw[<->,gray,dashed](0,-4)--(0,4);
\draw (0,4)node[above]{$m_2$};
\draw (2,2) node{$\Tp_2$};
\draw (-2,2) node{$\tilde{\Tp}_6$};
\draw (2,-2) node{$\tilde{\Tp}_5$};
\draw (-2,-2) node{$\tilde{\Tp}_8$};
\draw[very thick] (-.8,-.8)--(.8,.8);
\draw[very thick, red] (3.8,-.6)..controls(-1.2,-.7)..(-.8,-.8)..controls (-0.7,1)..(-.8,3.8);
\draw[very thick, red] (-3.8,-.08)..controls(-1.4,-.4)..(-.8,-.8)..controls (-0.4,-1.4)..(-.08,-3.8); 
\draw[very thick, red] (3.8,.7)..controls(1.4,.6)..(.8,.8)..controls(0.6,1.4)..(0.7,3.8); 
\filldraw[blue] (0.8,0.8) circle (3pt);
\filldraw[blue] (-.8,-.8) circle (3pt);
\draw (0,2) node{$\sigma_{26}$};
\draw (2,0) node{$\sigma_{25}$};
\draw[->] (-1.2,.8)node[left]{$\sigma_{12}^d$}--((.19,.21);
\end{tikzpicture}
\caption{$(m_1,\,m_2)$, $m_3<0$}        
        \label{fig:8f}
    \end{subfigure}
    \caption{Phases of $SU(N)_k+p\, \psi_1+q\, \psi_2+(F-p-q)\,\psi_3$ with $0\leq k<(p+q)-F/2$.}
    \label{fig8}
\end{figure*}

This is the last possible range of $k$ has a three-dimensional phase diagram with the same topological phases as in case 4 except that $\Tp_7$ is now
\begin{align}
\Tp_7\rightarrow \tilde{\Tp}_7: SU(N)_{k+\frac{F}{2}-p-q}\longleftrightarrow U(p+q-F/2-k)_{N}\ .
\label{t7til}
\end{align}
The topological theories come along with $\bar{\sigma}_1^c$, $\bar{\sigma}_2^c$, $\bar{\sigma}_3^c$, $\hat{\sigma}_3^c$, and the following new cuboid quantum regions
\begin{itemize}
\item[$\bullet$] $ m_2\rightarrow -\infty,\, m_3\rightarrow+\infty$:
\begin{align}
\hat{\sigma}_1^c= \frac{U(p)}{U(F/2+k-q)\times U(p+q-F/2-k)}\ .
\label{sigma1chat}
\end{align}
\item[$\bullet$] $ m_1\rightarrow -\infty,\, m_3\rightarrow+\infty$:
\begin{align}
\hat{\sigma}_2^c= \frac{U(q)}{U(F/2+k-p)\times U(p+q-F/2-k)}\ .
\label{sigma2chat}
\end{align}
\end{itemize}
The one mass asymptotic limits are now
\begin{enumerate}[label=(\roman*)]
\item $m_1\rightarrow +\infty$: $F_1/2-q\leq k_1^+<F_1/2$ which gives a type $\mathrm{II}$ phase diagram with the quantum regions $\bar{\sigma}_{23}^d$, $\sigma_{35}$, and $\sigma_{25}$
\item $m_1\rightarrow -\infty$: $F_1/2-q\leq |k_1^-|<F_1/2$ with negative $k_1^-$ which leads to a time-reversed type $\mathrm{II}$ phase diagram as shown in \ref{fig:8b} with the phases $\sigma_{23}^d$, $\sigma_{46}$, and $\sigma_{47}$, where
\begin{align}
\sigma_{47}: Gr(F_1/2+k_1^-,q)= \frac{U(q)}{U(F/2+k-p)\times U(p+q-F/2-k)}\equiv \hat{\sigma}_2^c\ .
\label{sigma47}
\end{align}
\item $m_2\rightarrow +\infty$: $F_2/2-p<k_2^+<F_2/2$ and the phase diagram is of type $\mathrm{II}$ with the phases $\bar{\sigma}_{13}^d$, $\sigma_{46}$, and $\sigma_{26}$.
\item $m_2\rightarrow -\infty$: $F_2/2-p<|k_2^-|<F_2/2$, with $k_2^-<0$, hence we have a time-reversed type $\mathrm{II}$ phase diagram with $\sigma_{13}^d$, $\sigma_{35}$, and $\sigma_{37}$, where
\begin{align}
\sigma_{37}: Gr(F_2/2+k_2^-,p)= \frac{U(p)}{U(F/2+k-q)\times U(p+q-F/2-k)}\equiv \hat{\sigma}_1^c\ .
\label{sigma37}
\end{align}
\item $m_3\rightarrow +\infty$: $F_3/2-p\leq |k_3^+|<F_3/2$ which gives a type $\mathrm{II}$ phase diagram with $\sigma_{37}$, $\sigma_{47}$, and a new diagonal sigma model $\bar{\sigma}_{12}^d$ given by
 \begin{align}
 \bar{\sigma}_{12}^d: Gr(F_3/2+k_3^+,F_3)= \frac{U(p+q)}{U(F/2+k)\times U(p+q-F/2-k)}\ ,
 \label{sigma12dbar}
 \end{align}
\item $m_3\rightarrow -\infty$: $F_3/2-p\leq |k_3^-|<F_3/2$ this limit has a time-reversed type $\mathrm{II}$ phase diagram with $\sigma_{12}^d$, $\sigma_{25}$, and $\sigma_{26}$. 
\end{enumerate}

\begin{figure}[h!]
\centering
\includegraphics[scale=.2]{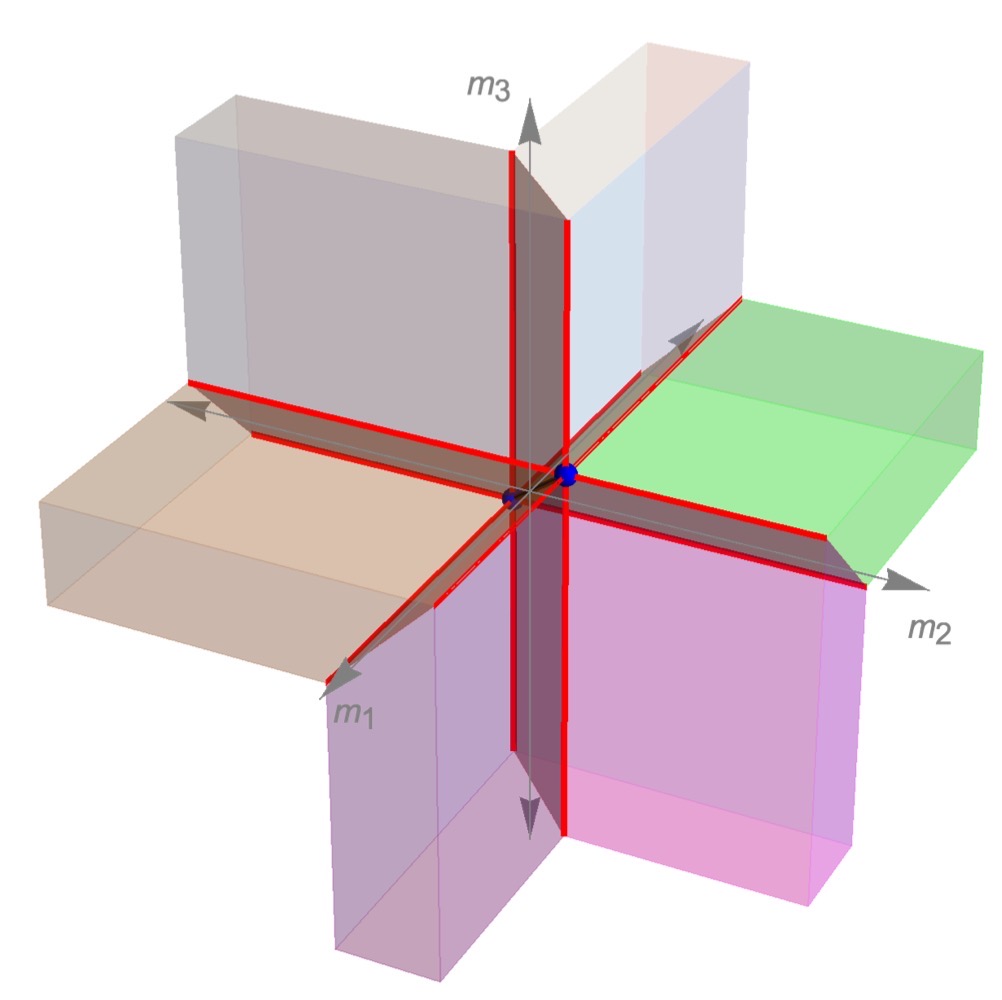}
\caption{The three-dimensional phase diagram of $SU(N)_k+p\,\psi_1+q\,\psi_2+(F-p-q)\,\psi_3$ with $0\leq k< (p+q)-F/2$. $\hat{\sigma}_1^c$ is represented by the region in gray, while $\hat{\sigma}_2^c$ is the white region.}
\label{G5}
\end{figure}
We summarize the phases of case 5 in figures \ref{fig8} and \ref{G5} where these are: eight topological field theories $(\Tp_1,\Tp_2,$ $\Tp_3,\Tp_4,\tilde{\Tp}_5,\tilde{\Tp}_6,\tilde{\Tp}_7,\tilde{\Tp}_8)$, six cuboid sigma-models $(\bar{\sigma}_1^c,\hat{\sigma}_1^c,\bar{\sigma}_2^c,\hat{\sigma}_2^c,\bar{\sigma}_3^c,$ $\hat{\sigma}_3^c)$, six planes of sigma-models $(\sigma_{12}^d,\bar{\sigma}_{12}^d,\sigma_{13}^d,\bar{\sigma}_{13}^d,\sigma_{23}^d,\bar{\sigma}_{23}^d)$, as well as $\sigma$. The limiting case $q=0$ reproduces the phase diagram in \fig{fig:3c}.
\subsection{$F-p-q>F/2$ Scenario}\label{sec:2b}
In this scenario we consider the choice of $p$ and $q$ such that $0<q\leq p\leq p+q \leq F/2$. The range of $k$ diagram is divided into: $k\geq F/2$, $F/2-q\leq k<F/2$, $F/2-p\leq k<F/2-q$, $F/2-(p+q)\leq k<F/2-p$, and $0\leq k<F/2-(p+q)$, we label these cases by $\bar{1}$, $\bar{2}$, $\bar{3}$, $\bar{4}$, and $\bar{5}$, respectively. We notice that the cases $\bar{1}$ to $\bar{3}$ are precisely similar to cases 1 to 3 of the first scenario, and the phase diagrams of these cases are equivalent. Case $\bar{4}$ is identical to case 4 in the first scenario since we have $k>|F/2-(p+q)|$ in both scenarios. The only difference then would be in the $0\leq k<F/2-(p+q)$ case.

\subsubsection*{\underline{Case $\bar{5}$: $0\leq k<F/2-(p+q)$}}
This range of $k$ under the constraint of this scenario has a three-dimensional phase diagram with the same topological phases as in case 4 of the first scenario except that $\Tp_2$ is now
\begin{align}
\Tp_2\rightarrow \tilde{\Tp}_2: SU(N)_{k-\frac{F}{2}+p+q}\longleftrightarrow U(F/2-p-q-k)_{N}\ .
\label{t2til}
\end{align}
The topological theories come along with $\sigma_3^c$, $\bar{\sigma}_3^c$, $\hat{\sigma}_3^c$, and a new cuboid region when we send $m_1,\, m_2\rightarrow +\infty$ given by
\begin{align}
\tilde{\sigma}_3^c= \frac{U(F-p-q)}{U(F/2+k)\times U(F/2-p-q-k)}\ .
\label{sigma3ctil}
\end{align}
\begin{figure*}[h]
    \centering
    \begin{subfigure}[b]{0.3\textwidth}
\begin{tikzpicture}[scale=.55]
\draw[<->,gray,dashed](-4,0)--(4,0);
\draw (4,0) node[right]{$m_2$};
\draw[<->,gray,dashed](0,-4)--(0,4);
\draw (0,4)node[above]{$m_3$};
\draw (2,2) node{$\Tp_1$};
\draw (-2,2) node{$\Tp_3$};
\draw (2,-2) node{$\tilde{\Tp}_2$};
\draw (-2,-2) node{$\tilde{\Tp}_5$};
\draw[very thick] (-.8,-.8)--(.8,.8);
\draw[very thick, red] (-3.8,-.6)..controls(-1.2,-.7)..(-.8,-.8)..controls (-0.4,-1.4)..(-.08,-3.8);
\draw[very thick, red] (-3.8,.7)..controls(-1.,.6)..(.8,.8)..controls (1.8,.65)..(3.8,.7); 
\draw[very thick, red] (.08,3.8)..controls(.4,1.4)..(.8,.8); 
\draw[very thick, red] (-.8,-.8)..controls(1,-.6)..(3.8,-.7); 
\filldraw[blue] (0.8,0.8) circle (3pt);
\filldraw[blue] (-.8,-.8) circle (3pt);
\draw (-2,0) node{$\sigma_{35}$};
\draw (2,0) node{$\sigma_{12}$};
\draw[->] (.7,-1)node[below]{$\bar{\sigma}_{23}^d$}--(-.19,-.21);
\end{tikzpicture}
\caption{$(m_2,\,m_3)$, $m_1>0$}        
        \label{fig:8abar}
    \end{subfigure}\quad
        \begin{subfigure}[b]{0.3\textwidth}
  \begin{tikzpicture}[scale=.55]
\draw[<->,gray,dashed](-4,0)--(4,0);
\draw (4,0) node[right]{$m_2$};
\draw[<->,gray,dashed](0,-4)--(0,4);
\draw (0,4)node[above]{$m_3$};
\draw (2,2) node{$\Tp_4$};
\draw (-2,2) node{$\Tp_7$};
\draw (2,-2) node{$\tilde{\Tp}_6$};
\draw (-2,-2) node{$\tilde{\Tp}_8$};
\draw[very thick] (-.8,-.8)--(.8,.8);
\draw[very thick, red] (-3.8,-.6)..controls(-1.2,-.7)..(-.8,-.8)..controls (-0.4,-1.4)..(-.08,-3.8);
\draw[very thick, red] (-3.8,.7)..controls(-1.,.6)..(.8,.8)..controls (1.8,.65)..(3.8,.7); 
\draw[very thick, red] (.08,3.8)..controls(.4,1.4)..(.8,.8); 
\draw[very thick, red] (-.8,-.8)..controls(1,-.6)..(3.8,-.7); 
\filldraw[blue] (0.8,0.8) circle (3pt);
\filldraw[blue] (-.8,-.8) circle (3pt);
\draw (-2,0) node{$\sigma_{78}$};
\draw (2,0) node{$\sigma_{46}$};
\draw[->] (.7,-1)node[below]{$\sigma_{23}^d$}--(-.19,-.21);
\end{tikzpicture}
\caption{$(m_2,\,m_3)$, $m_1<0$}        
        \label{fig:8bbar}
    \end{subfigure}\quad
        \begin{subfigure}[b]{0.3\textwidth}
    \begin{tikzpicture}[scale=.55]
\draw[<->,gray,dashed](-4,0)--(4,0);
\draw (4,0) node[right]{$m_1$};
\draw[<->,gray,dashed](0,-4)--(0,4);
\draw (0,4)node[above]{$m_3$};
\draw (2,2) node{$\Tp_1$};
\draw (-2,2) node{$\Tp_4$};
\draw (2,-2) node{$\tilde{\Tp}_2$};
\draw (-2,-2) node{$\tilde{\Tp}_6$};
\draw[very thick] (-.8,-.8)--(.8,.8);
\draw[very thick, red] (-3.8,-.6)..controls(-1.2,-.7)..(-.8,-.8)..controls (-0.4,-1.4)..(-.08,-3.8);
\draw[very thick, red] (-3.8,.7)..controls(-1.,.6)..(.8,.8)..controls (1.8,.65)..(3.8,.7); 
\draw[very thick, red] (.08,3.8)..controls(.4,1.4)..(.8,.8); 
\draw[very thick, red] (-.8,-.8)..controls(1,-.6)..(3.8,-.7); 
\filldraw[blue] (0.8,0.8) circle (3pt);
\filldraw[blue] (-.8,-.8) circle (3pt);
\draw (-2,0) node{$\sigma_{46}$};
\draw (2,0) node{$\sigma_{12}$};
\draw[->] (.7,-1)node[below]{$\bar{\sigma}_{13}^d$}--(-.19,-.21);
\end{tikzpicture}
\caption{$(m_1,\,m_3)$, $m_2>0$}        
        \label{fig:8cbar}
    \end{subfigure}\\ \vspace*{.3 cm}
        \begin{subfigure}[b]{0.3\textwidth}
    \begin{tikzpicture}[scale=.55]
\draw[<->,gray,dashed](-4,0)--(4,0);
\draw (4,0) node[right]{$m_1$};
\draw[<->,gray,dashed](0,-4)--(0,4);
\draw (0,4)node[above]{$m_3$};
\draw (2,2) node{$\Tp_3$};
\draw (-2,2) node{$\Tp_7$};
\draw (2,-2) node{$\tilde{\Tp}_5$};
\draw (-2,-2) node{$\tilde{\Tp}_8$};
\draw[very thick] (-.8,-.8)--(.8,.8);
\draw[very thick, red] (-3.8,-.6)..controls(-1.2,-.7)..(-.8,-.8)..controls (-0.4,-1.4)..(-.08,-3.8);
\draw[very thick, red] (-3.8,.7)..controls(-1.,.6)..(.8,.8)..controls (1.8,.65)..(3.8,.7); 
\draw[very thick, red] (.08,3.8)..controls(.4,1.4)..(.8,.8); 
\draw[very thick, red] (-.8,-.8)..controls(1,-.6)..(3.8,-.7); 
\filldraw[blue] (0.8,0.8) circle (3pt);
\filldraw[blue] (-.8,-.8) circle (3pt);
\draw (-2,0) node{$\sigma_{78}$};
\draw (2,0) node{$\sigma_{35}$};
\draw[->] (.7,-1)node[below]{$\sigma_{13}^d$}--(-.19,-.21);
\end{tikzpicture}
\caption{$(m_1,\,m_3)$, $m_2<0$}        
        \label{fig:8dbar}
    \end{subfigure}\quad
   \begin{subfigure}[b]{0.3\textwidth}
\begin{tikzpicture}[scale=.55]
\draw[<->,gray,dashed](-4,0)--(4,0);
\draw (4,0) node[right]{$m_1$};
\draw[<->,gray,dashed](0,-4)--(0,4);
\draw (0,4)node[above]{$m_2$};
\draw (2,2) node{$\Tp_1$};
\draw (-2,2) node{$\tilde{\Tp}_4$};
\draw (2,-2) node{$\Tp_3$};
\draw (-2,-2) node{$\Tp_7$};
\draw[ very thick, red] (-3.8,0)--(3.8,0);
\draw[very thick, red] (0,-3.8)--(0,3.8);
\filldraw[blue] (0,0) circle (3pt);
\end{tikzpicture}
\caption{$(m_1,\,m_2)$, $m_3>0$}        
        \label{fig:8ebar}
    \end{subfigure} \quad
        \begin{subfigure}[b]{0.3\textwidth}
  \begin{tikzpicture}[scale=.55]
\draw[<->,gray,dashed](-4,0)--(4,0);
\draw (4,0) node[right]{$m_1$};
\draw[<->,gray,dashed](0,-4)--(0,4);
\draw (0,4)node[above]{$m_2$};
\draw (2,2) node{$\tilde{\Tp}_2$};
\draw (-2,2) node{$\tilde{\Tp}_6$};
\draw (2,-2) node{$\tilde{\Tp}_5$};
\draw (-2,-2) node{$\Tp_8$};
\draw[ very thick, red] (-3.8,0)--(3.8,0);
\draw[very thick, red] (0,-3.8)--(0,3.8);
\filldraw[blue] (0,0) circle (3pt);
\end{tikzpicture}
\caption{$(m_1,\,m_2)$, $m_3<0$}        
        \label{fig:8fbar}
    \end{subfigure}
    \caption{Phases of $SU(N)_k+p\, \psi_1+q\, \psi_2+(F-p-q)\,\psi_3$ with $0\leq k<F/2-(p+q)$.}
    \label{fig8bar}
\end{figure*}
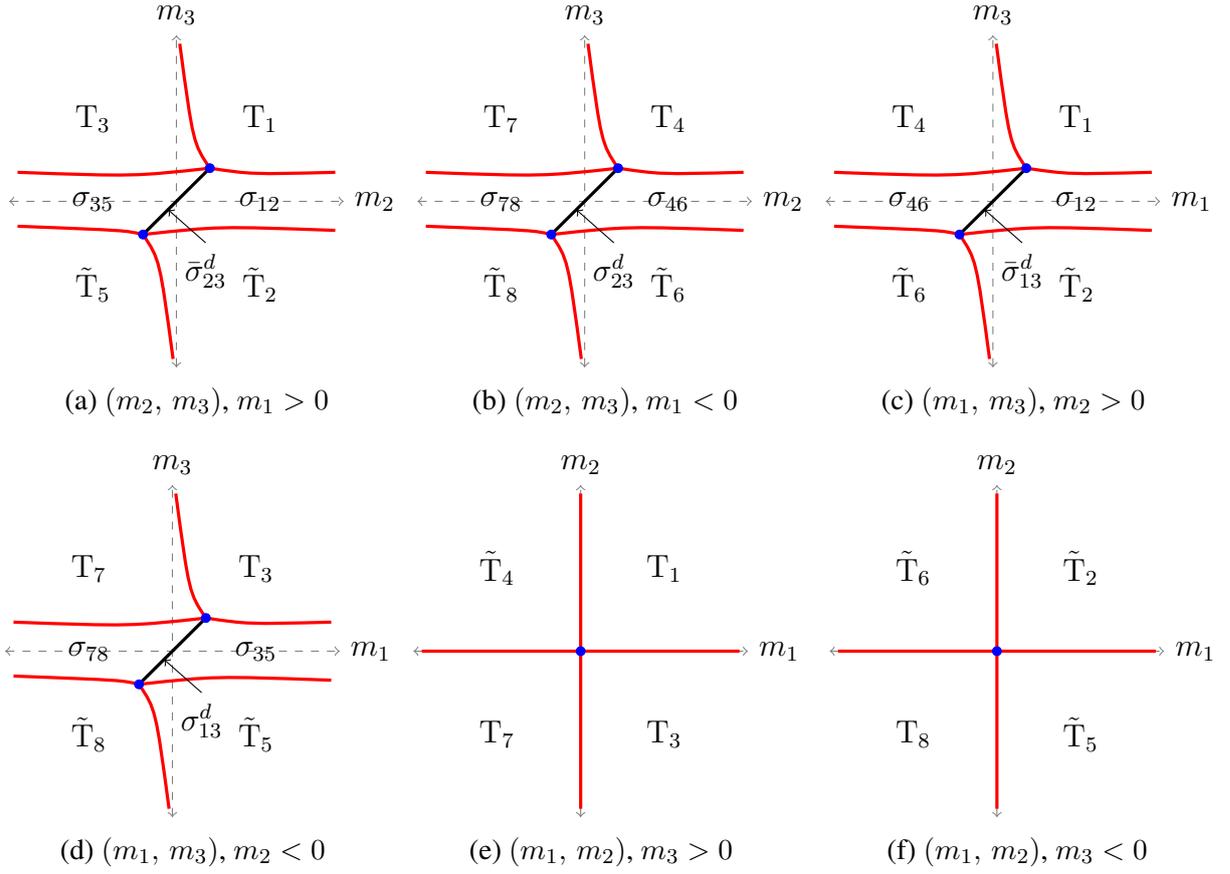
\begin{figure}[h!]
\centering
\includegraphics[scale=.2]{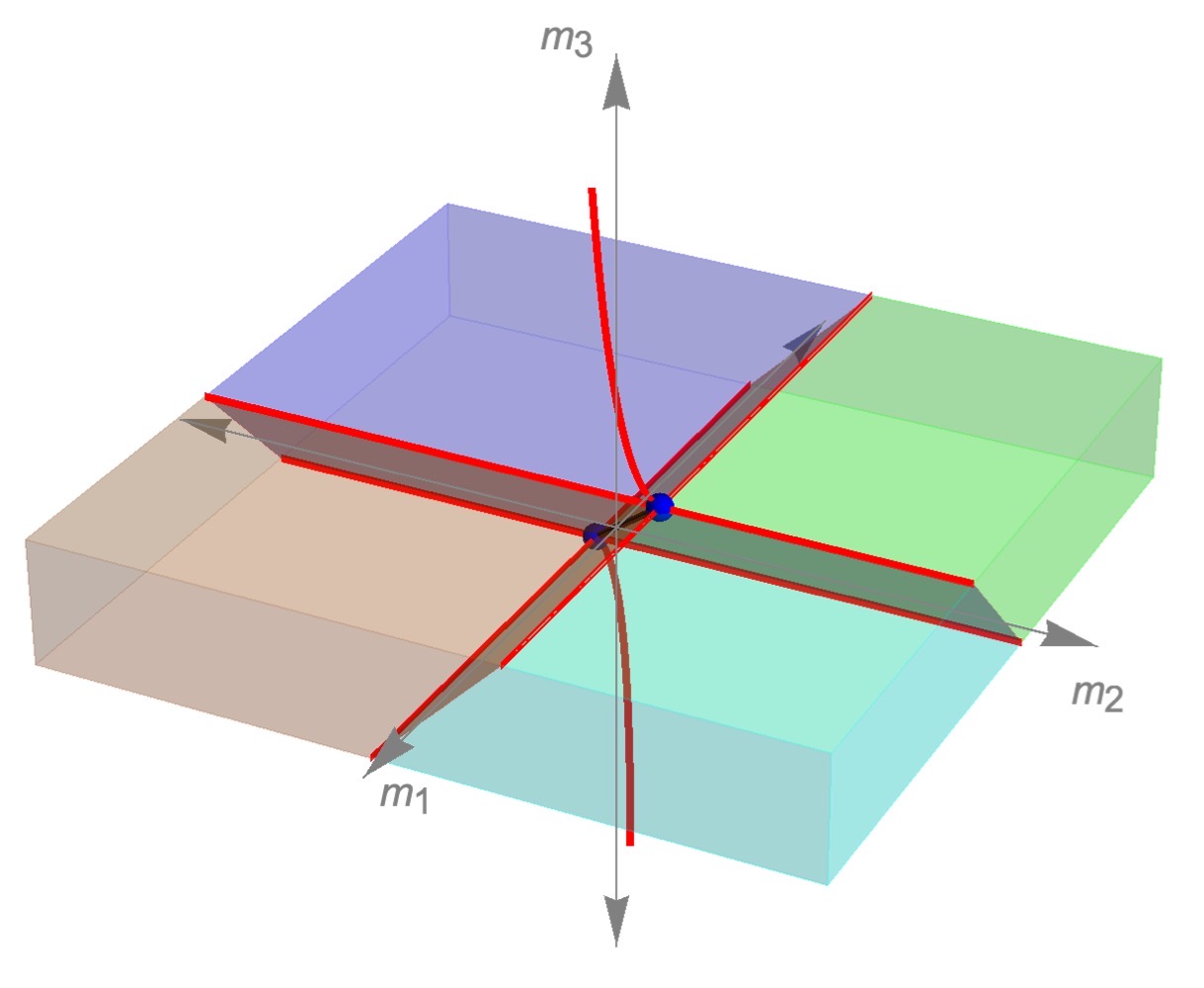}
\caption{The three-dimensional phase diagram of $SU(N)_k+p\,\psi_1+q\,\psi_2+(F-p-q)\,\psi_3$ with $0\leq k< F/2-(p+q)$. $\tilde{\sigma}_3^c$ is represented by the region in cyan.}
\label{G5bar}
\end{figure}

The one mass asymptotic limits are now
\begin{enumerate}[label=(\roman*)]
\item $m_1\rightarrow +\infty$: $p/2\leq k_1^+<F_1/2-q$ which gives a type $\mathrm{III}$ phase diagram with the quantum regions $\bar{\sigma}_{23}^d$, $\sigma_{35}$ as well as a new region for small $m_3$ but positive $m_2$:
\begin{align}
\sigma_{12}: Gr(F_1/2+k_1^+,F_1-q)= \frac{U(F-p-q)}{U(F/2+k)\times U(F/2-p-q-k)}\equiv \tilde{\sigma}_3^c\ .
\label{sigma12}
\end{align}
\item $m_1\rightarrow -\infty$: $|k_1^-|<F_1/2-q$ giving a type $\mathrm{III}$ phase diagram as in case 4 with the phases $\sigma_{23}^d$, $\sigma_{78}$, and $\sigma_{46}$.
\item $m_2\rightarrow +\infty$: $q/2<k_2^+<F_2/2-p$ and the phase diagram is of type $\mathrm{III}$ with the phases $\bar{\sigma}_{13}^d$, $\sigma_{46}$, and $\sigma_{12}$.
\item $m_2\rightarrow -\infty$:  this is a type $\mathrm{III}$ phase diagram with $\sigma_{13}^d$, $\sigma_{78}$, and $\sigma_{35}$.
\item $m_3\rightarrow +\infty$: type $\mathrm{I}$ phase diagram.
\item $m_3\rightarrow -\infty$: this limit has a type $\mathrm{I}$ phase diagram with no quantum regions as it satisfies $|k_3^-|>F_3/2$. 
\end{enumerate}

We summarize the phases of case $\bar{5}$ in figures \ref{fig8bar} and \ref{G5bar} where these are: eight topological field theories $(\Tp_1,\tilde{\Tp}_2,$ $\Tp_3,\Tp_4,\tilde{\Tp}_5,\tilde{\Tp}_6,\Tp_7,\tilde{\Tp}_8)$, four cuboid sigma-models $(\tilde{\sigma}_3^c,$ $\bar{\sigma}_3^c,$ $\hat{\sigma}_3^c,\sigma_3^c)$, four planes of sigma-models $(\sigma_{13}^d,\bar{\sigma}_{13}^d,\sigma_{23}^d,\bar{\sigma}_{23}^d)$, and the one-dimensional sigma-model $\sigma$. The limiting case $q=0$ reproduces the phase diagram in \fig{fig:3c}.

We see that this scenario does not include $\bar{\sigma}_{12}^d$, $\hat{\sigma}_1^c$, and $\hat{\sigma}_2^c$ as in the first scenario. However, this scenario includes $\tilde{\sigma}_3^c$, which was missing in the first scenario. Hence there is no single analysis that discusses the full phases of the three-family case; one should choose a scenario for the analysis based on the number $F-p-q$. The difference appears in some new and other missing quantum phases for each scenario. 
\section{Consistency checks }\label{sec:3}
In this section, we discuss a few ways to check our analysis for the three-family case. We mostly zoom on the region around the blue critical points on our figures, when we use the boson/fermion duality adapted to our three-family situation. We generalize the procedure used in \cite{Argurio:2019tvw} to our model of the three-family theory. The discussion for this section is mainly for the first scenario discussed in section \ref{sec:2b}, and we will mention the possible changes to the analysis when we use the second scenario accordingly.
\subsection{Planar sigma-models}

Before we start looking at the bosonic phases, we give an alternative way of the reduction to the two-family case by looking at the planes where two of the masses are equal. This reduction gives more insights into the nature of the diagonal sigma-models that appear in the three-family theory. 
\begin{enumerate}
\item \underline{$(m_1=m_2,m_3)$ plane:}\\
The theory is reduced to $SU(N)_k+\bar{p}\, \psi_1+(F-\bar{p})\,\psi_3$ with $\bar{p}=p+q$ fermions of mass $m_1=m_2$ and $F-\bar{p}$ of mass $m_3$. However, in this case the choice of $\bar{p}$ is such that $0\leq F-\bar{p}\leq F/2$, which makes the quantum regions appear for small $m_1=m_2$ in the type $\mathrm{III}$ phase diagram. The phase diagram is then of type $\mathrm{I}$ in case 1, type $\mathrm{II}$ in cases 2, 3, and 4, and type $\mathrm{III}$ in case 5 of section \ref{sec:2a}. The phase diagrams are now reduced to the phases in \fig{fig9}. In \fig{fig:9b}, the quantum phases are $\sigma_{17}$ and $\sigma_{28}$ where
\begin{align}
\sigma_{17}: Gr(F/2+k,\bar{p}) =  \frac{p+q}{U(F/2+k)\times U(p+q-F/2-k)}\equiv \bar{\sigma}_{12}^d\ ,
\label{sigma17}
\end{align}
\begin{align}
\sigma_{28}: Gr(F/2-k,\bar{p}) =  \frac{p+q}{U(F/2-k)\times U(k-F/2+p+q)}\equiv \sigma_{12}^d\ .
\label{sigma28}
\end{align}
We note that $\sigma_{28}$ is equivalent to $\sigma_{12}^d$, which clearly shows that $\sigma_{12}^d$ is not a line of quantum phase but rather a plane, and it appears in cases 2, 3, and 4. The quantum regions in \fig{fig:9c} are $\sigma_{78}$ and $\sigma_{17}$, which are given by Eqs.~\eqref{sigma78} and \eqref{sigma17}.

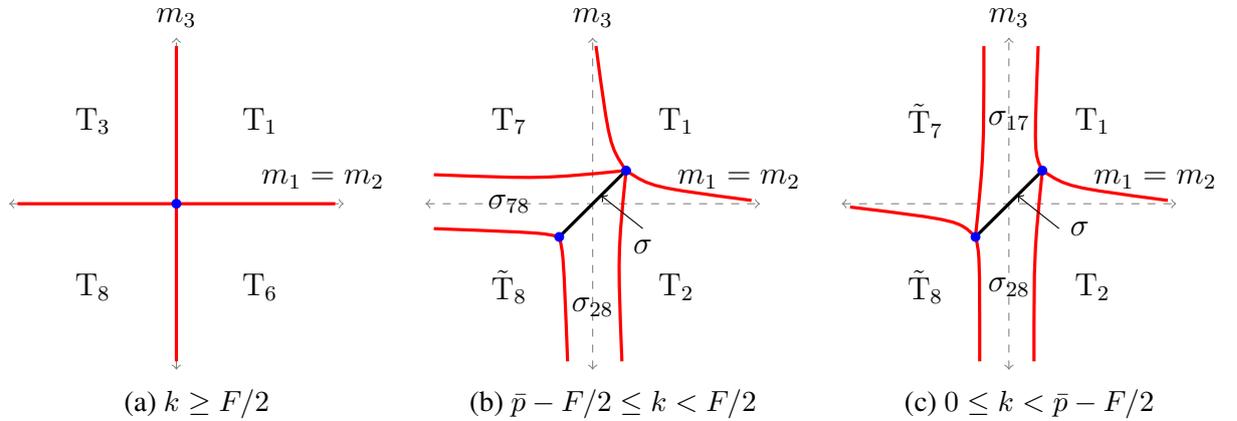
\begin{figure*}[h!]
    \centering
    \begin{subfigure}[b]{0.3\textwidth}
\begin{tikzpicture}[scale=.55]
\draw[<->,gray,dashed](-4,0)--(4,0);
\draw (3.5,0.1) node[above,align=left]{$m_1=m_2$};
\draw[<->,gray,dashed](0,-4)--(0,4);
\draw (0,4)node[above]{$m_3$};
\draw (2,2) node{$\Tp_1$};
\draw (-2,2) node{$\Tp_3$};
\draw (2,-2) node{$\Tp_6$};
\draw (-2,-2) node{$\Tp_8$};
\draw[ very thick, red] (-3.8,0)--(3.8,0);
\draw[very thick, red] (0,-3.8)--(0,3.8);
\filldraw[blue] (0,0) circle (3pt);
\end{tikzpicture}
\caption{$k\geq F/2$}        
        \label{fig:9a}
    \end{subfigure}
  \quad
    \begin{subfigure}[b]{0.3\textwidth}
      \begin{tikzpicture}[scale=.55]
\draw[<->,gray,dashed](-4,0)--(4,0);
\draw (3.5,0.1) node[above,align=left]{$m_1=m_2$};
\draw[<->,gray,dashed](0,-4)--(0,4);
\draw (0,4)node[above]{$m_3$};
\draw (2,2) node{$\Tp_1$};
\draw (-2,2) node{$\Tp_7$};
\draw (2,-2) node{$\Tp_2$};
\draw (-2,-2) node{$\tilde{\Tp}_8$};
\draw[very thick] (-.8,-.8)--(.8,.8);
\draw[very thick, red] (-3.8,-.6)..controls(-1.2,-.7)..(-.8,-.8)..controls (-0.7,-1.2)..(-.6,-3.8);
\draw[very thick, red] (-3.8,.7)..controls(-1.2,.6)..(.8,.8)..controls (0.6,-1.2)..(.7,-3.8); 
\draw[very thick, red] (3.8,.08)..controls(1.4,.4)..(.8,.8)..controls(0.4,1.4)..(0.08,3.8); 
\filldraw[blue] (0.8,0.8) circle (3pt);
\filldraw[blue] (-.8,-.8) circle (3pt);
\draw (-2,0) node{$\sigma_{78}$};
\draw (0,-2) node[below]{$\sigma_{28}$};
\draw[->] (1.2,-.6)node[below]{$\sigma$}--((.21,.19);
\end{tikzpicture}
        \caption{$\bar{p}-F/2\leq k<F/2$}
        \label{fig:9b}
    \end{subfigure}
   \quad
    \begin{subfigure}[b]{0.3\textwidth}
  \begin{tikzpicture}[scale=.55]
\draw[<->,gray,dashed](-4,0)--(4,0);
\draw (3.5,0.1) node[above]{$m_1=m_2$};
\draw[<->,gray,dashed](0,-4)--(0,4);
\draw (0,4)node[above]{$m_3$};
\draw (2,2) node{$\Tp_1$};
\draw (-2,2) node{$\tilde{\Tp}_7$};
\draw (2,-2) node{$\Tp_2$};
\draw (-2,-2) node{$\tilde{\Tp}_8$};
\draw[very thick] (-.8,-.8)--(.8,.8);
\draw[very thick, red] (-3.8,-.08)..controls(-1.2,-.4)..(-.8,-.8)..controls (-0.7,-1.2)..(-.7,-3.8);
\draw[very thick, red] (.8,.8)..controls (0.6,-1.2)..(.6,-3.8);  
\draw[very thick, red](-.8,-.8)..controls (-0.6,1.2)..(-.6,3.8);
\draw[very thick, red] (3.8,.08)..controls(1.4,.4)..(.8,.8)..controls(0.6,1.2)..(.7,3.8); 
\filldraw[blue] (0.8,0.8) circle (3pt);
\filldraw[blue] (-.8,-.8) circle (3pt);
\draw (0,2) node{$\sigma_{17}$};
\draw (0,-2) node{$\sigma_{28}$};
\draw[->] (1.2,-.6)node[right]{$\sigma$}--((.21,.19);
\end{tikzpicture}
        \caption{$0\leq k<\bar{p}-F/2$}
        \label{fig:9c}
    \end{subfigure}
    \caption{Phase diagrams in the limiting case $m_1=m_2$.}\label{fig9}
\end{figure*}

The only difference between the first and second scenarios is that the phase diagram in figure \ref{fig:9c} becomes a type $\mathrm{III}$ with quantum regions for small $m_3$ instead of small $m_1=m_2$. This makes the diagonal sigma model $\bar{\sigma}_{12}^d$ disappear as we expected from the discussion of case $\bar{5}$. 
\item \underline{$(m_1=m_3,m_2)$ plane:}\\
The theory is reduced to $SU(N)_k+q\, \psi_2+(F-q)\,\psi_3$  with $q$ fermions of mass $m_1 = m_3$ and $F-q$ fermions of mass $m_2$. The phase diagram is now of type $\mathrm{I}$ in case 1, type $\mathrm{II}$ in case 2, and type $\mathrm{III}$ in cases 3, 4, and 5. The phase diagram is summarized in \fig{fig10} with the following quantum regions: in \fig{fig:10b} we have $\sigma_{68}$ and $\sigma_{38}$ where
\begin{align}
\sigma_{38}: Gr(F/2-k,F-q) =  \frac{F-q}{U(F/2-k)\times U(F/2-q+k)}\equiv \sigma_{13}^d\ .
\label{sigma38}
\end{align}
In \fig{fig:10c} the quantum phases are $\sigma_{38}$ and $\sigma_{16}$ where
\begin{align}
\sigma_{16}: Gr(F/2+k,F-q) =\frac{F-q}{U(F/2+k)\times U(F/2-q-k)}\equiv \bar{\sigma}_{13}^d\ .
\label{sigma16}
\end{align}
We conclude that the diagonal sigma-model $\sigma_{13}^d$ appears in all the cases except case 1 while $\bar{\sigma}_{13}^d$ appears only in cases 3, 4, and 5 of section \ref{sec:2a}. The analysis is the same for the second scenario.

\begin{figure*}[h!]
    \centering
    \begin{subfigure}[b]{0.3\textwidth}
\begin{tikzpicture}[scale=.55]
\draw[<->,gray,dashed](-4,0)--(4,0);
\draw (4,0) node[right]{$m_2$};
\draw[<->,gray,dashed](0,-4)--(0,4);
\draw (0,4)node[above]{$m_3=m_1$};
\draw (2,2) node{$\Tp_1$};
\draw (-2,2) node{$\Tp_3$};
\draw (2,-2) node{$\Tp_6$};
\draw (-2,-2) node{$\Tp_8$};
\draw[ very thick, red] (-3.8,0)--(3.8,0);
\draw[very thick, red] (0,-3.8)--(0,3.8);
\filldraw[blue] (0,0) circle (3pt);
\end{tikzpicture}
\caption{ $k\geq F/2$}        
        \label{fig:10a}
    \end{subfigure}
    \quad
    \begin{subfigure}[b]{0.3\textwidth}
      \begin{tikzpicture}[scale=.55]
\draw[<->,gray,dashed](-4,0)--(4,0);
\draw (4,0) node[right]{$m_2$};
\draw[<->,gray,dashed](0,-4)--(0,4);
\draw (0,4)node[above]{$m_3=m_1$};
\draw (2,2) node{$\Tp_1$};
\draw (-2,2) node{$\Tp_3$};
\draw (2,-2) node{$\Tp_6$};
\draw (-2,-2) node{$\tilde{\Tp}_8$};
\draw[very thick] (-.8,-.8)--(.8,.8);
\draw[very thick, red] (-3.8,-.6)..controls(-1.2,-.7)..(-.8,-.8)..controls (-0.7,-1.2)..(-.6,-3.8);
\draw[very thick, red] (-3.8,.7)..controls(-1.2,.6)..(.8,.8)..controls (0.6,-1.2)..(.7,-3.8); 
\draw[very thick, red] (3.8,.08)..controls(1.4,.4)..(.8,.8)..controls(0.4,1.4)..(0.08,3.8); 
\filldraw[blue] (0.8,0.8) circle (3pt);
\filldraw[blue] (-.8,-.8) circle (3pt);
\draw (-2,0) node{$\sigma_{38}$};
\draw (0,-2) node[below]{$\sigma_{68}$};
\draw[->] (1.2,-.6)node[below]{$\sigma$}--((.21,.19);
\end{tikzpicture}
        \caption{$F/2-q\leq k<F/2$}
        \label{fig:10b}
    \end{subfigure}
    \quad
    \begin{subfigure}[b]{0.3\textwidth}
      \begin{tikzpicture}[scale=.55]
\draw[<->,gray,dashed](-4,0)--(4,0);
\draw (4,0) node[right]{$m_2$};
\draw[<->,gray,dashed](0,-4)--(0,4);
\draw (0,4)node[above]{$m_3=m_1$};
\draw (2,2) node{$\Tp_1$};
\draw (-2,2) node{$\Tp_3$};
\draw (2,-2) node{$\tilde{\Tp}_6$};
\draw (-2,-2) node{$\tilde{\Tp}_8$};
\draw[very thick] (-.8,-.8)--(.8,.8);
\draw[very thick, red] (-3.8,-.6)..controls(-1.2,-.7)..(-.8,-.8)..controls (-0.4,-1.4)..(-.08,-3.8);
\draw[very thick, red] (-3.8,.7)..controls(-1.,.6)..(.8,.8)..controls (1.8,.65)..(3.8,.7); 
\draw[very thick, red] (.08,3.8)..controls(.4,1.4)..(.8,.8); 
\draw[very thick, red] (-.8,-.8)..controls(1,-.6)..(3.8,-.7); 
\filldraw[blue] (0.8,0.8) circle (3pt);
\filldraw[blue] (-.8,-.8) circle (3pt);
\draw (-2,0) node{$\sigma_{38}$};
\draw (2,0) node{$\sigma_{16}$};
\draw[->] (1.,-1)node[below]{$\sigma$}--(-.19,-.21);
\end{tikzpicture}
        \caption{$0\leq k<F/2-q$}
        \label{fig:10c}
    \end{subfigure}
    \caption{Phase diagrams in the limiting case $m_1=m_3$}\label{fig10}
\end{figure*}
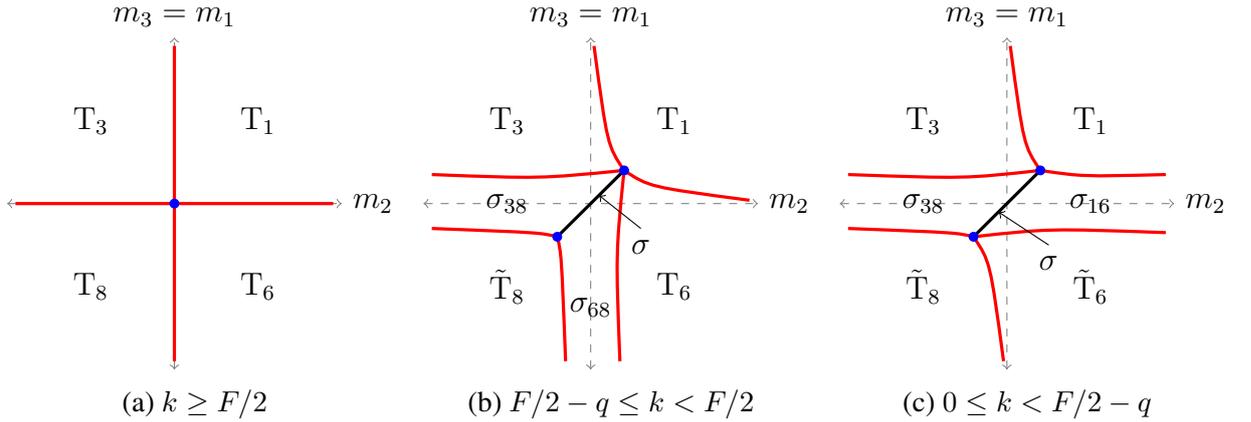
\item \underline{$(m_1,m_2=m_3)$ plane:}\\
The theory is reduced to $SU(N)_k+p\, \psi_1+(F-p)\,\psi_3$ where $\psi_2=\psi_3$. The phase diagram is of type $\mathrm{I}$ in case 1, type $\mathrm{II}$ only in cases 2 and 3, and type $\mathrm{III}$ in cases 4 and 5. The phase diagram is summarized in \fig{fig11}, wherein \fig{fig:11b}, the quantum regions are $\sigma_{58}$ and $\sigma_{48}$ with
\begin{align}
\sigma_{48}: Gr(F/2-k,F-p) =  \frac{F-p}{U(F/2-k)\times U(F/2-p+k)}\equiv \sigma_{23}^d\ .
\label{sigma48}
\end{align}
In \fig{fig:11c} the quantum phases are $\sigma_{48}$ and $\sigma_{15}$ where
\begin{align}
\sigma_{15}: Gr(F/2+k,F-p) = \frac{F-p}{U(F/2+k)\times U(F/2-p-k)}\equiv \bar{\sigma}_{23}^d\ ,
\label{sigma15}
\end{align}
which shows that the diagonal sigma-model $\sigma_{23}^d$ appears cases 2 and 3 while $\bar{\sigma}_{23}^d$ appears in cases 4 and 5. The analysis is also the same for the second scenario in this limiting case.

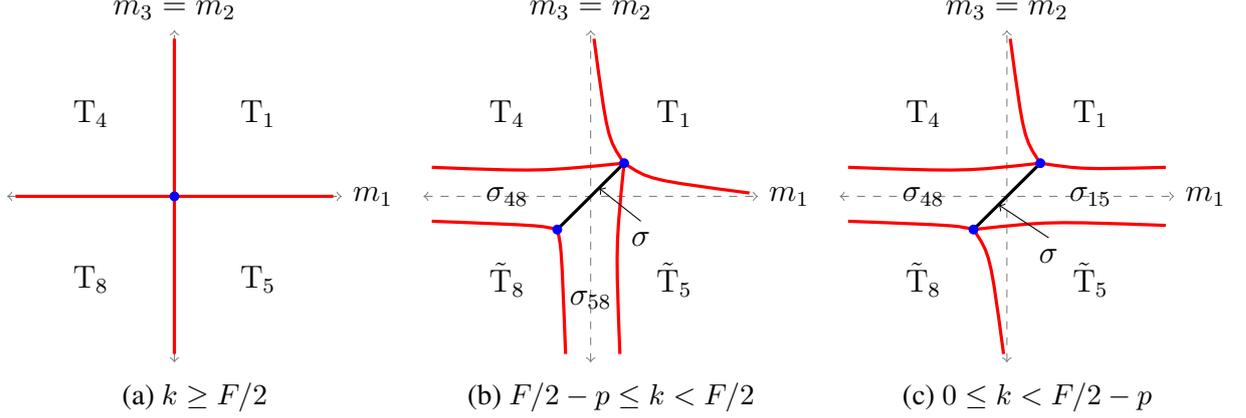
\begin{figure*}[h!]
    \centering
    \begin{subfigure}[b]{0.3\textwidth}
\begin{tikzpicture}[scale=.55]
\draw[<->,gray,dashed](-4,0)--(4,0);
\draw (4,0) node[right]{$m_1$};
\draw[<->,gray,dashed](0,-4)--(0,4);
\draw (0,4)node[above]{$m_3=m_2$};
\draw (2,2) node{$\Tp_1$};
\draw (-2,2) node{$\Tp_4$};
\draw (2,-2) node{$\Tp_5$};
\draw (-2,-2) node{$\Tp_8$};
\draw[ very thick, red] (-3.8,0)--(3.8,0);
\draw[very thick, red] (0,-3.8)--(0,3.8);
\filldraw[blue] (0,0) circle (3pt);
\end{tikzpicture}
\caption{ $k\geq F/2$}        
        \label{fig:11a}
    \end{subfigure}
    \quad
    \begin{subfigure}[b]{0.3\textwidth}
      \begin{tikzpicture}[scale=.55]
\draw[<->,gray,dashed](-4,0)--(4,0);
\draw (4,0) node[right]{$m_1$};
\draw[<->,gray,dashed](0,-4)--(0,4);
\draw (0,4)node[above]{$m_3=m_2$};
\draw (2,2) node{$\Tp_1$};
\draw (-2,2) node{$\Tp_4$};
\draw (2,-2) node{$\tilde{\Tp}_5$};
\draw (-2,-2) node{$\tilde{\Tp}_8$};
\draw[very thick] (-.8,-.8)--(.8,.8);
\draw[very thick, red] (-3.8,-.6)..controls(-1.2,-.7)..(-.8,-.8)..controls (-0.7,-1.2)..(-.6,-3.8);
\draw[very thick, red] (-3.8,.7)..controls(-1.2,.6)..(.8,.8)..controls (0.6,-1.2)..(.7,-3.8); 
\draw[very thick, red] (3.8,.08)..controls(1.4,.4)..(.8,.8)..controls(0.4,1.4)..(0.08,3.8); 
\filldraw[blue] (0.8,0.8) circle (3pt);
\filldraw[blue] (-.8,-.8) circle (3pt);
\draw (-2,0) node{$\sigma_{48}$};
\draw (0,-2) node[below]{$\sigma_{58}$};
\draw[->] (1.2,-.6)node[below]{$\sigma$}--((.21,.19);
\end{tikzpicture}
        \caption{$F/2-p\leq k<F/2$}
        \label{fig:11b}
    \end{subfigure}
    \quad
    \begin{subfigure}[b]{0.3\textwidth}
      \begin{tikzpicture}[scale=.55]
\draw[<->,gray,dashed](-4,0)--(4,0);
\draw (4,0) node[right]{$m_1$};
\draw[<->,gray,dashed](0,-4)--(0,4);
\draw (0,4)node[above]{$m_3=m_2$};
\draw (2,2) node{$\Tp_1$};
\draw (-2,2) node{$\Tp_4$};
\draw (2,-2) node{$\tilde{\Tp}_5$};
\draw (-2,-2) node{$\tilde{\Tp}_8$};
\draw[very thick] (-.8,-.8)--(.8,.8);
\draw[very thick, red] (-3.8,-.6)..controls(-1.2,-.7)..(-.8,-.8)..controls (-0.4,-1.4)..(-.08,-3.8);
\draw[very thick, red] (-3.8,.7)..controls(-1.,.6)..(.8,.8)..controls (1.8,.65)..(3.8,.7); 
\draw[very thick, red] (.08,3.8)..controls(.4,1.4)..(.8,.8); 
\draw[very thick, red] (-.8,-.8)..controls(1,-.6)..(3.8,-.7); 
\filldraw[blue] (0.8,0.8) circle (3pt);
\filldraw[blue] (-.8,-.8) circle (3pt);
\draw (-2,0) node{$\sigma_{48}$};
\draw (2,0) node{$\sigma_{15}$};
\draw[->] (1.,-1)node[below]{$\sigma$}--(-.19,-.21);
\end{tikzpicture}
        \caption{$0\leq k<F/2-p$   }
        \label{fig:11c}
    \end{subfigure}
        \caption{Phase diagrams in the limiting case $m_2=m_3$}\label{fig11}
\end{figure*}
\end{enumerate}
\subsection{Matching the bosonic phases}
Near the critical points, the fermionic theory with three families of fermions is conjectured to have a bosonic dual description with gauge group $U(n)_{l}$ and three sets of scalar fields in the fundamental representation of the gauge group which are $p\ \phi_1$, $q\, \phi_2$ and $(F-p-q)\, \phi_3$. In $U(n)_{l}$, $n$ and $l$ take the value $F/2\pm k$ and $\pm N$, respectively. We have split the $F$ scalars into three sets which can acquire independent mass deformations to be denoted $M_i^2,\, i = 1,2,3$. 

This bosonic theory has six gauge invariants operators which can be written in terms of the three scalars as 
\begin{equation}
\begin{aligned}
&X=\phi_1\, \phi_1^{\dagger},\ Y=\phi_2\, \phi_2^{\dagger},\ Z=\phi_3\, \phi_3^{\dagger}\\
&U=\phi_1\, \phi_2^{\dagger},\ W=\phi_1\, \phi_3^{\dagger},\ T=\phi_2\, \phi_3^{\dagger}\ ,
\end{aligned}
\label{gio}
\end{equation}
where $X$, $Y$, and $Z$ are positive semidefinite diagonal Hermitian matrices of dimensions $p$, $q$, and $F-p-q$, respectively. We consider a scalar potential for the critical theory including up to quartic order in the scalar field, which is further deformed by symmetry breaking mass operators. Written in terms of the six gauge invariants operators, this is
\begin{multline}
V=  M_1^2 \text{Tr}X + M_2^2 \text{Tr}Y+ M_3^2 \text{Tr}Z+\lambda(\text{Tr}^2X+\text{Tr}^2Y\\+\text{Tr}^2Z+2\text{Tr}X\text{Tr}Y+2\text{Tr}X\text{Tr}Z+2\text{Tr}Y\text{Tr}Z)+\\
\mu(\text{Tr}X^2+\text{Tr}Y^2+\text{Tr}Z^2+2\text{Tr}UU^{\dagger}+2\text{Tr}WW^{\dagger}+2\text{Tr}TT^{\dagger})\ ,
\label{v}
\end{multline}
where $\lambda$ and $\mu$ are the coupling constants for the quartic terms. The quartic couplings are chosen such that the full $U(F)$ flavor symmetry is preserved. We choose $\mu\geq 0$, which requires $\mu+\text{min}(n,F)\lambda>0$ for the potential to be bounded from below. 

Consider that $X$, $Y$, and $Z$ have $r_1$, $r_2$, and $r_3$ degenerate eigenvalues $x$, $y$, and $z$ respectively such that 
\begin{equation}
\text{Tr}X = r_1x,\ \text{Tr}Y = r_2y,\ \text{Tr}Z = r_3z,\ .
\end{equation}
The gauge group $U(n)$ is never Higgsed if the squared mass of $X$, $Y$, and $Z$ are non-negative. In this case, all the six gauge invariants operators vanish on-shell, so there is no scalar condensation, all matter fields are integrated out due to being massive, and one obtains a topological $U(n)_l$ theory in the infrared. 

On the other hand, if at least one of the scalars has a negative mass squared, the minimum of the potential can be found by solving the equations of motion 
\begin{align}
&M_1^2+2\lambda (\text{Tr}X+\text{Tr}Y+\text{Tr}Z)+2\mu X=0\ ,\label{eof1}\\
&M_2^2+2\lambda (\text{Tr}X+\text{Tr}Y+\text{Tr}Z)+2\mu Y=0\ ,\label{eof2}\\
&M_3^2+2\lambda (\text{Tr}X+\text{Tr}Y+\text{Tr}Z)+2\mu Z=0\ .
\label{eof3}
\end{align}
It also implies that $U=W=T=0$. Solving the equations of motion gives the following eigenvalues 

\begin{align}
x= \frac{\lambda(M_2^2r_2+M_3^2 r_3)-M_1^2[\mu +\lambda(r_2+r_3)]}{2\mu[\mu+\lambda(r_1+r_2+r_3)]}\ ,\label{eigx}\\
y= \frac{\lambda(M_1^2r_1+M_3^2 r_3)-M_2^2[\mu +\lambda(r_1+r_3)]}{2\mu[\mu+\lambda(r_1+r_2+r_3)]}\ ,\label{eigy}\\
z= \frac{\lambda(M_1^2r_1+M_2^2 r_2)-M_3^2[\mu +\lambda(r_1+r_2)]}{2\mu[\mu+\lambda(r_1+r_2+r_3)]}\ .\label{eigz}
\end{align}
It can be easily seen that minimizing the potential always requires maximization of $r_1+r_2+r_3$. The ranks $r_1$, $r_2$, and $r_3$ are non-negative integers satisfying the following conditions 
\begin{equation}
\begin{aligned}
r_1\leq \text{min}(n,p)&,\ r_2\leq \text{min}(n,q),\ r_3\leq \text{min}(n,F-p-q),\\& r_1+r_2+r_3\leq \text{min}(n,F)\ .
\end{aligned}
\label{rcons}
\end{equation}

The constraints in \eq{rcons} and the sign of the mass squared of each gauge invariant operator defines the phases that appear in the bosonic theory. The bosonic theory experiences Higgsing of the gauge group or Higgsing plus spontaneous symmetry breaking except when $M_1^2$, $M_2^2$, and $M_3^2$ are all non-negative, as discussed above. 

The phase diagram of the bosonic theory can be divided into five cases:

\begin{enumerate}

\item \underline{$q\leq p\leq F-p-q\leq F<n$:} $F<n$ does not allow any spontaneous symmetry breaking for  the flavor symmetry $U(F)$. We expect to have eight different regions to describe the phase diagram in this range. Region $\mathbb{A}$ describes the theory when all the masses squared are non-negative with no scalar condensation. The regions $\mathbb{B}$, $\mathbb{C}$, and $\mathbb{D}$ are reached when only one scalar mass squared is negative, allowing a condensation for $\phi_1$, $\phi_2$, or $\phi_3$, respectively. There are also three regions $\mathbb{E}$, $\mathbb{F}$, and $\mathbb{G}$ where two of the scalars condense before integrating them out. The last region, $\mathbb{H}$, describes a phase when the three scalars condense simultaneously. 
\begin{table*}[h!]
\centering
\resizebox{.9\textwidth}{!}{%
\begin{tabular}{ |c|l|l|l|l|c|c|c|c|}
\hline
\multirow{2}{*}{Region} & \multirow{2}{*}{$r_1$} & \multirow{2}{*}{$r_2$ }& \multirow{2}{*}{$r_3$}& \multirow{2}{*}{Phase}&\multicolumn{2}{c|}{Scenario 1} & \multicolumn{2}{c|}{Scenario 2}  \\ 
\cline{6-9}
 & & & & & $n=F/2+k$& $n=F/2-k$&$n=F/2+k$&$n=F/2-k$ \\
\hline
$\mathbb{A}$ &$0$&$0$&$0$&$U(n)_l$& $\Tp_1$ & $\tilde{\Tp}_8$&$\Tp_1$&$\tilde{\Tp}_8$\\
\hline
$\mathbb{B}$&$p$&$0$&$0$&$U(n-p)_l$& $\Tp_4$& $\tilde{\Tp}_5$&$\Tp_4$&$\tilde{\Tp}_5$\\
\hline
$\mathbb{C}$&$0$&$q$&$0$&$U(n-q)$&$\Tp_3$& $\tilde{\Tp}_6$&$\Tp_3$&$\tilde{\Tp}_6$\\
\hline
$\mathbb{D}$&$0$&$0$&$F-p-q$&$U(n-F+p+q)_l$&$\Tp_2$& $\tilde{\Tp}_7$&$\Tp_2$&N/A\\
\hline
$\mathbb{E}$&$p$&$q$&$0$&$U(n-p-q)_l$&$\Tp_7$&N/A&$\Tp_7$&$\tilde{\Tp}_2$\\
\hline
$\mathbb{F}$&$p$&$0$&$F-p-q$&$U(n-F+q)_l$&$\Tp_6$&N/A&$\Tp_6$&N/A\\
\hline
$\mathbb{G}$&$0$&$q$&$F-p-q$&$U(n-F+p)_l$&$\Tp_5$&N/A&$\Tp_5$&N/A\\
\hline
$\mathbb{H}$&$p$&$q$&$F-p-q$&$U(n-F)_l$&$\Tp_8$&N/A&$\Tp_8$&N/A\\
\hline
\end{tabular}}
\caption{Phases of the bosonic theory with $q\leq p\leq F-p-q\leq F<n$.}
\label{table1}
\end{table*}

The phases of the bosonic theory in this range are summarized in table \ref{table1}. For $n=F/2+k$, the phases reproduce the topological theories of \eq{t3kgf}, which match the phases of case 1 in the fermionic description. For $n=F/2-k$, the remaining topological phases from cases 2 to 5 appear.

For the second scenario, the bosonic phases are similar except that, for $n=F/2-k$, region $\mathbb{D}$ will not be allowed, and region $\mathbb{E}$ will be described by $\tilde{\Tp}_2$. The bosonic phases then match the topological phases of the fermionic theory for cases $\bar{1}$ to $\bar{5}$.

\item \underline{$q\leq p\leq F-p-q\leq n<F$:} In this range, there is a possibility of spontaneous symmetry breaking, which allows sigma-models to appear in the bosonic phases. The sigma-models appear when there is a condensation of more than one scalar. The region $\mathbb{E}$, where $\phi_1$ and $\phi_2$ condense, splits into two regions: $\mathbb{E}_1$ where only the constraint on $r_1$ is saturated and $\mathbb{E}_2$ where the constraint on $r_2$ is saturated. The same scenario occurs for regions $\mathbb{F}$ and $\mathbb{G}$, while region $\mathbb{H}$ splits into three subregions, each of them is described when one of the constraints on $r_1$, $r_2$, or $r_3$ is saturated. 

The phases of the bosonic theory in this range are summarized in table \ref{table2}. For $n=F/2+k$, the quantum phases are $(\sigma_1^c,\, \bar{\sigma}_1^c,\, \hat{\sigma}_1^c,\, \sigma_2^c,\, \bar{\sigma}_2^c,\, \hat{\sigma}_2^c,\, \sigma_3^c,\, \bar{\sigma}_3^c,\, \hat{\sigma}_3^c)$, which are equivalent to all the cuboid quantum phases of the fermionic theory in cases 2, 3, 4, and 5 of section \ref{sec:2a}. For $n=F/2-k$, the quantum phases are $ \bar{\sigma}_1^c,\, \hat{\sigma}_1^c,\, \bar{\sigma}_2^c,\, \hat{\sigma}_2^c,\, \bar{\sigma}_3^c,\, \hat{\sigma}_3^c)$, which match the fermionic phases from case 5.

For the second scenario, the only allowed substitution is $n=F/2+k$ with slightly changed phases where the regions $\mathbb{E}_1$ and $\mathbb{E}_2$ will not be allowed under the constraint $F-p-q>F/2$. This makes the quantum phases $\hat{\sigma}_1^c$ and $\hat{\sigma}_2^c$ disappear, and the bosonic phases match correctly the fermionic phases in cases $\bar{2}$, $\bar{3}$, and $\bar{4}$.

\begin{table*}[h!]
\centering
\resizebox{.9\textwidth}{!}{%
\begin{tabular}{ |c|l|l|l|l|c|c|c|}
\hline
\multirow{2}{*}{Region} & \multirow{2}{*}{$r_1$} & \multirow{2}{*}{$r_2$ }& \multirow{2}{*}{$r_3$}& \multirow{2}{*}{Phase}&\multicolumn{2}{c|}{Scenario 1} &  Scenario 2\\ 
\cline{6-8}
 & & & & & $n=F/2+k$& $n=F/2-k$&$n=F/2+k$\\
\hline
$\mathbb{A}$ &$0$&$0$&$0$&$U(n)_l$&$\Tp_1$&$\tilde{\Tp}_8$&$\Tp_1$\\
\hline
$\mathbb{B}$&$p$&$0$&$0$&$U(n-p)_l$&$\Tp_4$&$\tilde{\Tp}_5$&$\Tp_4$\\
\hline
$\mathbb{C}$&$0$&$q$&$0$&$U(n-q)$&$\Tp_3$&$\tilde{\Tp}_6$&$\Tp_3$\\
\hline
$\mathbb{D}$&$0$&$0$&$F-p-q$&$U(n-F+p+q)_l$&$\Tp_2$&$\tilde{\Tp}_7$&$\Tp_2$\\
\hline
$\mathbb{E}_1$&$p$&$n-p$&$0$&$Gr(n-p,q)$&$\hat{\sigma}_2$&$\bar{\sigma}_2^c$&N/A\\
\hline
$\mathbb{E}_2$&$n-q$&$q$&$0$&$Gr(n-q,p)$&$\hat{\sigma}_1$&$\bar{\sigma}_1^c$&N/A\\
\hline
$\mathbb{F}_1$&$p$&$0$&$n-p$&$Gr(n-p,F-p-q)$&$\hat{\sigma}_3^c$&$\bar{\sigma}_3^c$&$\hat{\sigma}_3^c$\\
\hline
$\mathbb{F}_2$&$n-F+p+q$&$0$&$F-p-q$&$Gr(n-F+p+q,p)$&$\bar{\sigma}_1^c$&$\hat{\sigma}_1^c$&$\bar{\sigma}_1^c$\\
\hline
$\mathbb{G}_1$&$0$&$q$&$n-q$&$Gr(n-q,F-p-q)$&$\bar{\sigma}_3^c$&$\hat{\sigma}_3^c$&$\bar{\sigma}_3^c$\\
\hline
$\mathbb{G}_2$&$0$&$n-F+p+q$&$F-p-q$&$Gr(n-F+p+q,q)$&$\bar{\sigma}_2^c$&$\hat{\sigma}_2^c$&$\bar{\sigma}_2^c$\\
\hline
$\mathbb{H}_1$&$p$&$q$&$n-p-q$&$Gr(n-p-q,F-p-q)$&$\sigma_3^c$&N/A&$\sigma_3^c$\\
\hline
$\mathbb{H}_2$&$p$&$n-F+q$&$F-p-q$&$Gr(n-F+q,q)$&$\sigma_2^c$&N/A&$\sigma_2^c$\\
\hline
$\mathbb{H}_3$&$n-F+p$&$q$&$F-p-q$&$Gr(n-F+p,p)$&$\sigma_1^c$&N/A&$\sigma_1^c$\\
\hline
\end{tabular}}
\caption{Phases of the bosonic theory with $q\leq p\leq F-p-q\leq n< F$.}
\label{table2}
\end{table*}

\item \underline{$q\leq p \leq n< F-p-q\leq F$:} In this range, the regions $\mathbb{A}$, $\mathbb{B}$, and $\mathbb{C}$ are similar to the previous cases. Since $n<F-p-q$, $r_1$ is saturated to $n$ and region $\mathbb{D}$ now shrinks to a smaller region $\mathbb{D}_1$ with a sigma-model phase. The regions $\mathbb{E}_1$ and $\mathbb{E}_2$ remain the same as in the previous case while only $\mathbb{F}_1$, $\mathbb{G}_1$, and $\mathbb{H}_1$ subregions appear in this case. Each of the remaining subregions shares the same phase as in one of the other regions (e.\,g.\,the subregion $\mathbb{F}_2$ has the same sigma-model as in $\mathbb{D}_1$). 

The phases of this case are summarized in table \ref{table3}. Only $n=F/2-k$ is allowed for the first scenario which gives the phases $(\bar{\sigma}_1^c,\, \bar{\sigma}_2^c,\, \sigma_3^c,\, \hat{\sigma}_3^c,\, \bar{\sigma}_3^c)$ . These quantum phases are equivalent to the phases of the fermionic theory in case 4 of section \ref{sec:2a}. 

For the second scenario, when $n=F/2+k$, the phases are $(\bar{\sigma}_1^c,\,  \bar{\sigma}_2^c,\, \bar{\sigma}_3^c,\, \hat{\sigma}_3^c,\, \tilde{\sigma}_3^c,\, \sigma_3^c)$, which are equivalent to the phases of the fermionic theory in cases $\bar{4}$ and $\bar{5}$. For $n=F/2-k$, the bosonic phases are $(\sigma_3^c,\, \bar{\sigma}_3^c,\, \hat{\sigma}_3^c,\, \tilde{\sigma}_3^c)$ which match the fermionic phases in case $\bar{5}$ of section \ref{sec:2b}.
\begin{table*}[h!]
\centering
\resizebox{.9\textwidth}{!}{%
\begin{tabular}{ |c|l|l|l|l|c|c|c| }
\hline
\multirow{2}{*}{Region} & \multirow{2}{*}{$r_1$} & \multirow{2}{*}{$r_2$ }& \multirow{2}{*}{$r_3$}& \multirow{2}{*}{Phase}&Scenario 1& \multicolumn{2}{c|}{Scenario 2}  \\ 
\cline{6-8}
 & & & & & $n=F/2-k$&$n=F/2+k$&$n=F/2-k$ \\
 \hline
$\mathbb{A}$ &$0$&$0$&$0$&$U(n)_l$&$\tilde{\Tp}_8$&$\Tp_1$&$\tilde{\Tp}_8$\\
\hline
$\mathbb{B}$&$p$&$0$&$0$&$U(n-p)_l$&$\tilde{\Tp}_5$&$\Tp_4$&$\tilde{\Tp}_5$\\
\hline
$\mathbb{C}$&$0$&$q$&$0$&$U(n-q)$&$\tilde{\Tp}_6$&$\Tp_3$&$\tilde{\Tp}_6$\\
\hline
$\mathbb{D}_1$&$0$&$0$&$n$&$Gr(n,F-p-q)$&$\sigma_3^c$&$\tilde{\sigma}_3^c$&$\sigma_3^c$\\
\hline
$\mathbb{E}_1$&$p$&$n-p$&$0$&$Gr(n-p,q)$&$\bar{\sigma}_2^c$&N/A&$\bar{\sigma}_2^c$\\
\hline
$\mathbb{E}_2$&$n-q$&$q$&$0$&$Gr(n-q,p)$&$\bar{\sigma}_1^c$&N/A&$\bar{\sigma}_1^c$\\
\hline
$\mathbb{F}_1$&$p$&$0$&$n-p$&$Gr(n-p,F-p-q)$&$\bar{\sigma}_3^c$&$\hat{\sigma}_3^c$&$\bar{\sigma}_3^c$\\
\hline
$\mathbb{G}_1$&$0$&$q$&$n-q$&$Gr(n-q,F-p-q)$&$\hat{\sigma}_3^c$&$\bar{\sigma}_3^c$&$\hat{\sigma}_3^c$\\
\hline
$\mathbb{H}_1$&$p$&$q$&$n-p-q$&$Gr(n-p-q,F-p-q)$&N/A&$\sigma_3^c$&$\tilde{\sigma}_3^c$\\
\hline
\end{tabular}}
\caption{Phases of the bosonic theory with $q\leq p \leq n< F-p-q\leq F$.}
\label{table3}
\end{table*}
\item \underline{$q\leq n< p\leq F-p-q\leq F$:} In this case, the regions $\mathbb{A}$ and $\mathbb{C}$ do not experience any spontaneous symmetry breaking. Since $n<(p,F-p-q)$, both $r_1$ and $r_3$ are saturated to $n$ in regions $\mathbb{B}$ and $\mathbb{D}$, which shrink to smaller regions $\mathbb{B}_1$ and $\mathbb{D}_1$ with sigma-model phases. The subregions $\mathbb{F}_1$ and $\mathbb{H}_1$ now join the subregion $\mathbb{B}_1$ to form a broader region sharing the same sigma-model, and the same happens for $\mathbb{F}_2$ and $\mathbb{H}_2$ which join the subregion $\mathbb{D}_1$. 

The phases of this case are summarized in table \ref{table4}. Only $n=F/2-k$ is allowed in this case which gives $( \sigma_1^c,\, \bar{\sigma}_1^c,\, \sigma_3^c,\, \hat{\sigma}_3^c)$, which match the phases of the fermionic theory in case 3 of section \ref{sec:2a}. The analysis is precisely the same for the second scenario, where case 3 and case $\bar{3}$ are identical.

\begin{table}[h!]
\centering

\begin{tabular}{ |c|l|l|l|l|c|}
\hline
Region & $r_1$ & $r_2$ & $r_3$& Phase&$n=F/2-k$  \\ 
\hline
$\mathbb{A}$ &$0$&$0$&$0$&$U(n)_l$&$\tilde{\Tp}_8$\\
\hline
$\mathbb{B}_1$&$n$&$0$&$0$&$U(n-q)_l$&$\tilde{\Tp}_6$\\
\hline
$\mathbb{C}$&$0$&$q$&$0$&$Gr(n,p)$&$\sigma_1^c$\\
\hline
$\mathbb{D}_1$&$0$&$0$&$n$&$Gr(n,F-p-q)$&$\sigma_3^c$\\
\hline
$\mathbb{E}_2$&$n-q$&$q$&$0$&$Gr(n-q,p)$&$\bar{\sigma}_1^c$\\
\hline
$\mathbb{G}_1$&$0$&$q$&$n-q$&$Gr(n-q,F-p-q)$&$\hat{\sigma}_3^c$\\
\hline
\end{tabular}
\caption{Phases of the bosonic theory with $q\leq n< p\leq F-p-q\leq F$.}
\label{table4}
\end{table}
\item \underline{$n<q\leq p\leq F-p-q\leq F$:} In this range, all the single condensation cases experience spontaneous symmetry breaking scenario where each of the corresponding ranks is saturated to $n$ producing a sigma-model. The double and triple condensation cases share the same sigma-model as in the single condensation case. 

The phases are now reduced to include only regions $\mathbb{A}$, $\mathbb{B}_1$, $\mathbb{C}_1$, and $\mathbb{D}_1$, as shown in table \ref{table5}. For $n=F/2-k$, the phases are $(\sigma_1^c,\, \sigma_2^c,\, \sigma_3^c)$, matching the fermionic phases in case 2 of section \ref{sec:2a}. The analysis is also the same for the second scenario, where case 2 and case $\bar{2}$ are identical.
\begin{table}[h!]
\centering
\begin{tabular}{ |c|l|l|l|l|c| }
\hline
Region & $r_1$ & $r_2$ & $r_3$& Phase &$n=F/2-k$ \\ 
\hline
$\mathbb{A}$ &$0$&$0$&$0$&$U(n)_l$&$\tilde{\Tp}_8$\\
\hline
$\mathbb{B}_1$&$n$&$0$&$0$&$Gr(n,p)$&$\sigma_1^c$\\
\hline
$\mathbb{C}_1$&$0$&$n$&$0$&$Gr(n,q)$&$\sigma_2^c$\\
\hline
$\mathbb{D}_1$&$0$&$0$&$n$&$Gr(n,F-p-q)$&$\sigma_3^c$\\
\hline
\end{tabular}
\caption{Phases of the bosonic theory with $n<q\leq p\leq F-p-q\leq F$.}
\label{table5}
\end{table}
\end{enumerate}
An additional and straightforward consistency check is to reduce the bosonic theory to the two-family case by putting $q=0$ where the tables \ref{table1}, \ref{table2}, \ref{table3}, \ref{table4}, and \ref{table5} reduce to the tables in \cite{Argurio:2019tvw}.
\subsection{Perturbing the lower dimension sigma-models}
We saw in the previous subsection how to match the phases of the bosonic and the fermionic theories around the critical points by considering perturbations in the bosonic dual descriptions. We now want to perturb the diagonal sigma-models in both two-dimensional and three-dimensional pictures. We do this by adding a mass term which explicitly breaks the flavor symmetry $U(F)$, as considered in \cite{Argurio:2019tvw} for the two-family case. 
The target space of the sigma model $\sigma$ is 
\begin{equation}
\sigma: Gr(n,F)= \frac{U(F)}{U(n)\times U(F-n)}\ ,
\label{sigmam}
\end{equation}
where again $n$ can be either $F/2+k$ or $F/2 - k$. $\sigma$ appears on the diagonal line of the three different limiting cases discussed in the previous subsection, which show that there exist three different possibilities of the mass deformation corresponding to deforming the mass of each of the scalars independently.

For $(m_1,m_2=m_3)$ plane, the theory has $p\, \phi_1+(F-p)\, \phi_2$ scalar. This allows us to perturb $\sigma$ by deforming $\phi_1$ or $\phi_2$ where the result is independent of the choice of the scalar set that we deform so let us say that we deform $\phi_1$  by adding an infinitesimal mass squared $\delta M_1^2$ to $M_1^2$. Hence we have four possibilities: 
\begin{itemize} 
\item[$\bullet$] If $\delta M_1^2>0$ and $F-p>n$, $\phi_3$ condenses first Higgsing the gauge group $U(n)$, then one can integrate $\phi_1$ out and the resulting sigma-model has a Grassmannian $Gr(n,F-p)$. 
\item[$\bullet$] If $\delta M_1^2>0$ but $F-p<n$ the condensation of $\phi_3$ partially Higgs the gauge group down to $U(n-F+p)$ and then can be integrated out followed by integrating out $\phi_1$. This gives a sigma-model with a Grassmannian $Gr(F-p,p)$. 
\item[$\bullet$] For $\delta M_1^2<0$ and $p>n$, $\phi_1$ condenses first with a complete Higgsing of the gauge group which leads to a sigma-model with a Grassmannian $Gr(n,p)$. 
\item[$\bullet$] For $\delta M_1^2<0$ and $p<n$, the theory has a sigma-model with a Grassmannian $Gr(F-n,F-p)$.
\end{itemize}
Substituting $n=F/2\pm k$ gives Grassmannians describing the sigma-models $\sigma_{23}^d$, $\sigma_1^c$, and $\bar{\sigma}_{23}^d$, which match the theories around $\sigma$ in \fig{fig:11c}.

Similarly for the $(m_2,m_1=m_3)$ plane, where we deform the mass of $\phi_1$ to perturb sigma leading to the Grassmannians $Gr(n,F-q)$, $Gr(F-n,q)$, $Gr(n,q)$, and $Gr(F-n,F-q)$. These Grassmanians correspond to the sigma-models $\sigma_{13}^d$, $\sigma_2^c$, and $\bar{\sigma}_{13}^d$ matching the phases around $\sigma$ in \fig{fig:10c}. Lastly, the mass deformation of $\phi_3$ in the $(m_1=m_2,m_3)$ plane leads to the Grassmannians $Gr(n,F-p-q)$, $Gr(F-n,p+q)$, $Gr(n,p+q)$, and $Gr(F-n,F-p-q)$ describing the target space of the sigma-models $\sigma_3^c$, $\sigma_{12}^d$, and $\tilde{\sigma}_3^c$ which match the phases around $\sigma$ in the fermionic theory, as shown in \fig{fig:9c}. 

We now move to perturb the other diagonal sigma-models when we simultaneously deform the mass of two scalars. We consider the perturbation of pairs of diagonal sigma-models as follows:
\begin{enumerate}
\item \underline{Perturbing $\sigma_{23}^d$ and $\bar{\sigma}_{23}^d$:} we rewrite both theories in the general form $Gr(n,F-p)$ where $\sigma_{23}^d$ can be found by substituting $n=F/2-k$ while $\bar{\sigma}_{23}^d$ is found by substituting $n=F/2+k$. This can be obtained by deforming the mass of $\phi_1$ with $\delta M_1^2>0$ and $F-p>n$ on the $(m_1,m_2=m_3)$ plane. In addition, we deform the mass of $\phi_2$ and check the four possibilities. 

Now we have a perturbation of $Gr(n,F_1)$  $\delta M_2^2>0$ or $\delta M_2^2<0$.  As in the previous discussion, this gives sigma-models with Grassmannians $Gr(n,F-p-q)$, $Gr(F-p-n,q)$, $Gr(n,q)$, and $Gr(F-p-n,F-p-q)$. For $n=F/2-k$, these Grassmannians correspond to $\sigma_3^c$, $\hat{\sigma}_3^c$, $\sigma_2^c$, and $\hat{\sigma}_2^c$ matching the fermionic phases around $\sigma_{23}^d$, as shown in Figs.~\ref{fig:5b}, \ref{fig:6b}, \ref{fig:7b}, and \ref{fig:8b}. For $n=F/2+k$, these Grassmannians correspond to $\bar{\sigma}_3^c$ and $\bar{\sigma}_2^c$ matching the fermionic phases around $\bar{\sigma}_{23}^d$, as shown in Figs.~\ref{fig:7a} and \ref{fig:8a}. 

We should clarify that not all the Grassmannians are allowed when we make the substitution $n=F/2\pm k$ where they are subject to $k$ being non-negative and the constraint of the first deformation, which is $F-p>n$ in this case.

In the second scenario, $Gr(F-p-n,F-p-q)$ is not allowed for $n=F/2-k$ and $Gr(n,F-p-q)$ provides $\tilde{\sigma}_3^c$ for $n= F/2+k$. Then we have only $\sigma_3^c$, $\hat{\sigma}_3^c$, and $\sigma_2^c$ that surround $\sigma_{23}^d$, as shown in Figs.~\ref{fig:5b}, \ref{fig:6b}, \ref{fig:7b}, and \ref{fig:8bbar}. $\tilde{\sigma}_3^c$, $\bar{\sigma}_3^c$ and $\bar{\sigma}_2^c$ are the phases around $\bar{\sigma}_{23}^d$ matching the fermionic description, as shown in Figs.~\ref{fig:7a}, \ref{fig:8a}, and \ref{fig:8abar}.

\item \underline{Perturbing $\sigma_{13}^d$ and $\bar{\sigma}_{13}^d$:} these two theories have Grassmannians written in a single form $Gr(n,F-q)$ which is obtained by deforming the mass of $\phi_2$ with $\delta M_2^2>0$ and $F-q>n$ on the $(m_2,m_1=m_3)$ plane. By adding a deformation to the mass of $\phi_1$, the resulting sigma-models have Grassmannians $Gr(n,F-p-q)$, $Gr(F-q-n,p)$, $Gr(n,p)$, and $Gr(F-q-n,F-p-q)$. For $n=F/2-k$, these Grassmannians correspond to $\sigma_3^c$, $\sigma_1^c$, $\bar{\sigma}_3^c$, and $\hat{\sigma}_1^c$ which match the phases around $\sigma_{13}^d$.  For $n=F/2+k$, the Grassmanians correspond to $\bar{\sigma}_1^c$ and $\hat{\sigma}_3^c$, which match the phases around $\bar{\sigma}_{13}^d$ in the fermionic picture. The analysis is similar for the second scenario except that the Grassmannian $Gr(F-q-n,F-p-q)$ is no longer allowed for $n=F/2-k$.

\item \underline{Perturbing $\sigma_{12}^d$ and $\bar{\sigma}_{12}^d$:} we start from $Gr(n,p+q)$ which can be read from deforming the mass of $\phi_3$ with $\delta M_3^2<0$ and $p+q>n$ on the $(m_1=m_2,m_3)$ plane. An extra deformation of the mass of $\phi_1$ gives sigma-models. For $n=F/2-k$, the only allowed sigma-models have Grassmannians $Gr(F/2-k,p)$, $Gr(p+q-F/2+k,p)$, and $Gr(p+q-F/2+k,q)$. These Grassmannians correspond to $\sigma_1^c$, $\bar{\sigma}_1^c$, and $\bar{\sigma}_2^c$, which are the phases that appear around $\sigma_{12}^d$. For $n=F/2+k$, the allowed sigma-models are $\hat{\sigma}_1^c$ and $\hat{\sigma}_2^c$, which are the only quantum phases that appear around $\bar{\sigma}_{12}^d$. The substitution $n=F/2+k$ is not allowed for the second scenario; the analysis remains the same elsewhere.

\end{enumerate}
\section{Conclusion}\label{sec:4}
We investigated the IR behaviour of $SU(N)$ gauge theory coupled to three-families of flavors in the fundamental representation, extending the previous work for one family \cite{Komargodski:2017keh} and two families \cite{Argurio:2019tvw, Baumgartner:2019frr}. Our description covers the full phase diagram in all the possible ranges of the Chern-Simons level. Our analysis leads to a three-dimensional phase diagram which is well described semiclassically by topological and gapped phases for $k\geq F/2$ as in the one and two-family cases. In addition to the topological phases, we encounter one-dimensional sigma-models, planar, and cuboid sigma-models which are all quantum phases. The cuboid sigma-models are an intrinsic feature of the three-dimensional phase diagram which appear when one of the fermion masses become small.

We also provided consistency checks such as matching the phases of the bosonic dual descriptions with the fermionic ones and perturbing the diagonal sigma-model via mass deformations to match the off-diagonal lines phases.  The reduction to the two-family case by describing the various planes when two of the masses are equal reproduces the results of  \cite{Argurio:2019tvw, Baumgartner:2019frr}.

The order of the phase transitions that appear in the phase diagrams is only known in some limits to be second-order phase transitions \cite{Aharony:2012nh, GurAri:2012is, Jain:2013py, Jain:2013gza}. It would be interesting to study such transitions in the two and three-dimensional phase diagrams when more than one fermion is massless and decipher the type of phase transition. We leave this for future work. 
\bibliographystyle{JHEP}
\bibliography{Refs}

\end{document}